\newif \ifarxiv
\arxivtrue
\documentclass[a4paper,english]{lipics-v2018}
\pdfoutput=1

\ifarxiv
  \def \copyrightline {\vbox{
    \vskip5mm \small \par
    A version of this article without the Appendix
    is being submitted for publication in the post-proceedings of the
    \emph{24th International Conference on Types for Proofs and Programs}
    (\emph{TYPES 2018}).
  }}
\fi

\usepackage{microtype}

\title{Semantic subtyping for non-strict languages}

\author%
  {Tommaso Petrucciani}%
  {DIBRIS, Università di Genova, Italy %
    \& IRIF, Université Paris Diderot, France}{}{}{}%

\author%
  {Giuseppe Castagna}%
  {CNRS, IRIF, Université Paris Diderot, France}{}{}{}%

\author%
  {Davide Ancona and Elena Zucca}%
  {DIBRIS, Università di Genova, Italy}{}{}{}%

\authorrunning{T. Petrucciani, G. Castagna, D. Ancona, and E. Zucca}

\Copyright{Tommaso Petrucciani, Giuseppe Castagna, Davide Ancona, and Elena Zucca}

\subjclass{Dummy classification}


\category{}

\relatedversion{}

\supplement{}

\funding{}

\EventShortTitle{TYPES 2018}
\nolinenumbers 

\newif \ifshowproofs
\showproofsfalse
\showproofstrue

\usepackage{setup}

\begin{document}

\maketitle

\begin{abstract}
  Semantic subtyping is an approach to define subtyping relations
  for type systems featuring union and intersection type connectives.
  It has been studied only for strict languages,
  and it is unsound for non-strict semantics.
  In this work, we study how to adapt this approach to non-strict languages:
  in particular, we define a type system using semantic subtyping
  for a functional language with a call-by-need semantics.
  We do so by introducing an explicit representation for divergence in the types,
  so that the type system distinguishes expressions that are results
  from those which are computations that might diverge.
 \end{abstract}

\section{Introduction}

Semantic subtyping is a powerful framework
which allows language designers to define subtyping relations
for rich languages of types -- including union and intersection types --
that can express precise properties of programs.
However, it has been developed for languages with call-by-value semantics
and, in its current form, it is unsound for non-strict languages.
We show how to design a type system
which keeps the advantages of semantic subtyping
while being sound for non-strict languages
(more specifically, for call-by-need semantics).

\subsection{Semantic subtyping}

Union and intersection types can be used to type
several language constructs
-- from branching and pattern matching to function overloading --
very precisely.
However, they make it challenging to define
a subtyping relation that behaves precisely and intuitively.

\emph{Semantic subtyping} is a technique to do so,
studied by Frisch, Castagna, and Benzaken~\cite{Frisch2008}
for types given by:
\[
  t \Coloneqq
         b \mid t \to t \mid t \times t
    \mid t \lor t \mid t \land t \mid \lnot t \mid \Empty \mid \Any
  \qquad
  \text{where }
  b \Coloneqq \Int \mid \Bool \mid \cdots
\]
Types include constructors
-- basic types $ b $, arrows, and products --
plus
union $ t \lor t $, intersection $ t \land t $,
negation (or complementation) $ \lnot t $,
and the bottom and top types $ \Empty $ and $ \Any $ (actually, $ t_1 \land t_2 $ and $ \Any $ can be defined respectively as
$ \lnot (\lnot t_1 \lor \lnot t_2) $ and $ \lnot \Empty $). 
The grammar above is interpreted \emph{coinductively} rather than inductively,
thus allowing infinite terms that correspond to recursive types.
Subtyping is defined
by giving an interpretation $ \Inter{} $ of types as sets
and defining $ t_1 \leq t_2 $
as the inclusion of the interpretations, that is, $ t_1 \leq t_2 $
is defined as $ \Inter{t_1} \subseteq \Inter{t_2} $.
Intuitively, we can see $ \Inter{t} $
as the set of values that inhabit $ t $ in the language.
By interpreting union, intersection, and negation
as the corresponding operations on sets
and by giving appropriate interpretations to the other constructors,
we ensure that
subtyping will satisfy all commutative and distributive laws we expect:
for example, $
  (t_1 \times t_2) \lor (t'_1 \times t'_2) \leq
  (t_1 \lor t_1') \times (t_2 \lor t_2')
$ or $
  (t \to t_1) \land (t \to t_2) \leq t \to (t_1 \land t_2)
$.

This relation is used in \cite{Frisch2008} to type a call-by-value language
featuring higher-order functions,
data constructors and destructors (pairs),
and a typecase construct
which models runtime type dispatch and acts as a form of pattern matching.
Functions can be recursive
and are explicitly typed;
their type can be an intersection of arrow types,
describing overloaded behaviour.
A simple example of an overloaded function is\\[1mm]
\centerline{
\texttt{let f x = if (x is Int) then (x + 1) else not(x)}}\\[1mm]
which tests whether its argument \texttt{x} is of type $ \Int $
and in this case returns its successor, its negation otherwise.
This function can
be given the type $
  (\Int \to \Int) \land (\Bool \to \Bool)
$, which signifies
that it has both type $ \Int \to \Int $ and type $ \Bool \to \Bool $:
the two types define its two possible behaviours
depending on the outcome of the test (and, thus, on the type of the input).
This is done in~\cite{Frisch2008} by explicitly annotating the whole function definition.
Using notation for typecases from \cite{Frisch2008}:
\(
  \Let{f : (\Int \to \Int) \land (\Bool \to \Bool)
    \,\texttt{=}\, \Abstr{x. \Case{(x \IN \Int) ? (x + \K{1}) : (\textsf{not } x)}}}
\).
The type deduced for this function is  $(\Int \to \Int) \land (\Bool \to \Bool)$, but it can also be given the type $
  (\Int \lor \Bool) \to (\Int \lor \Bool)
$:
the latter type states that the function can be applied to both integers and booleans
and that its result is either an integer or a boolean.
This latter type is less precise than the intersection,
since it loses the correlation between the types of the argument and of the result. Accordingly, the semantic definition of subtyping ensures $
  (\Int \to \Int) \land (\Bool \to \Bool)
  \leq (\Int \lor \Bool) \to (\Int \lor \Bool)
$.

The work of \citet{Frisch2008}
has been extended to treat more language features,
including parametric polymorphism
\citep{Castagna2011,Castagna2014,Castagna2015},
type inference \citep{Castagna2016},
and gradual typing \citep{Castagna2017} and adapted to SMT solvers~\cite{Bie10}. It has been used to type object-oriented languages~\cite{DGV13,AC16}, XML queries~\cite{CHNB15}, NoSQL languages~\cite{BCNS13}, and scientific languages~\cite{julia18}.
It is also at the basis of the definition of \CDuce,
an XML-processing functional programming language with union and intersection types~\cite{BCF03}.
However, only strict evaluation had been considered, until now.

\subsection{Semantic subtyping in lazy languages}

Our work started as an attempt to design
a type system for the Nix Expression Language~\citep{Dolstra2008},
an untyped, purely functional, and lazily evaluated language
for Unix/Linux package management.
Since Nix is untyped,
some programming idioms it encourages
require advanced type system features to be analyzed properly.
Notably,
the possibility of writing functions
that use type tests to have an overloaded-like behaviour
made intersection types and semantic subtyping a good fit for the language.
However,
existing semantic subtyping relations are unsound for non-strict semantics;
this was already observed in \citet{Frisch2008}
and no adaptation has been proposed later.
Here we describe our solution
to define a type system based on semantic subtyping
which is sound for a non-strict language.
In particular,
we consider a call-by-need variant
of the language studied in \citet{Frisch2008}.

Current semantic subtyping systems are unsound for non-strict semantics
because of the way they deal with the bottom type $ \Empty $,
which corresponds to the empty set of values
($ \Inter{\Empty} = \emptyset $).
The intuition is that a reducible expression $e$ can be safely given a type $t$ only if all results (i.e., values) it can return
are of type $ t $.
Accordingly,
$ \Empty $ can only be assigned
to expressions that are statically known to diverge (i.e., that never return a result).
For example,
the ML expression \textsf{let rec f x = f x in f ()} can be given type $ \Empty $.
Let us use $ \bar e $ to denote any diverging expression that, like this, can be given type $\Empty$.
Consider the following typing derivations,
which are valid in current semantic subtyping systems
($ \pi_2 $ projects the second component of a pair).
\[
  \Infer
    {
      \Infer[$ \simeq $]
        { \Typing{|- (\bar e, \K{3}): \Empty \times \Int} }
        { \Typing{|- (\bar e, \K{3}): \Empty \times \Bool} }
        {}
    }
    { \Typing{|- \ProjSnd{(\bar e, \K{3})}: \Bool} }
    {}
  \qquad
  \Infer
    {
      \Infer[$ \simeq $]
        { \Typing{|- \Abstr{x. \K{3}}: \Empty \to \Int} }
        { \Typing{|- \Abstr{x. \K{3}}: \Empty \to \Bool} }
        {}
      \\
      \Typing{|- \bar e: \Empty}
    }
    { \Typing{|- \Appl{(\Abstr{x. \K{3}})}{\bar e}: \Bool} }
    {}
\]

Note that both $ \ProjSnd{(\bar e, \K{3})} $ and $ \Appl{(\Abstr{x. \K{3}})}{\bar e} $
diverge in call-by-value semantics
(since $ \bar e $ must be evaluated first),
while they both reduce to $ \K{3} $ in call-by-name or call-by-need.
The derivations are therefore sound for call-by-value,
while they are clearly unsound with non-strict evaluation.

Why are these derivations valid?
The crucial steps are those marked with $ [\simeq] $,
which convert between types
that have the same interpretation;
$ \simeq $ denotes this equivalence relation.
With semantic subtyping, $ \Empty \times \Int \simeq \Empty \times \Bool $ holds
because all types of the form $ \Empty \times t $ are equivalent to $ \Empty $ itself:
none of these types contains any value
(indeed, product types are interpreted as Cartesian products and therefore the product with the empty set is itself empty).
It can appear more surprising
that $ \Empty \to \Int \simeq \Empty \to \Bool $ holds.
We interpret a type $ t_1 \to t_2 $
as the set of functions which,
on arguments of type $ t_1 $, either diverge or return results in type $ t_2 $.
Since there is no argument of type $ \Empty $
(because, in call-by-value, arguments are always values),
all types of the form $ \Empty \to t $ are equivalent (they all contain every well-typed function).

\subsection{Our approach}

The intuition behind our solution is that, with non-strict semantics,
it is not appropriate to see a type as the set of the values that have that type.
In a call-by-value language,
operations like application or projection
occur on values:
thus, we can identify two types (and, in some sense, the expressions they type) if they contain (and their expressions may produce) the same values.
In non-strict languages,
though,
operations also occur on partially evaluated results:
these, like $ (\bar{e}, \K{3}) $ in our example,
can contain diverging sub-expressions below their top-level constructor.

As a result,
it is unsound, for example,
to type $ (\bar{e}, \K{3}) $ as $ \Empty \times \Int $,
since we have that $ \Empty \times \Int $ and $ \Empty \times \Bool $ are equivalent.
It is also unsound to have subtyping rules for functions
which assume implicitly that every argument will eventually be a value.

One approach to solve this problem would be
to change the interpretation of $ \Empty $ so that it is non-empty.
However, the existence of types with an empty interpretation
is important for the internal machinery of semantic subtyping.
Notably, the decision procedure for subtyping relies on them
(checking whether $ t_1 \leq t_2 $ holds is reduced to checking
whether the type $ t_1 \land \lnot t_2 $ is empty).
Therefore, we keep the interpretation $
  \Inter{\Empty} = \emptyset
$, but we change the type system so that this type is \emph{never} derivable,
not even for diverging expressions.
We keep it as a purely ``internal'' type useful to describe subtyping,
but never used to type expressions.

We introduce instead a separate type $ \bot $
as the type of diverging expressions.
This type is non-empty
but disjoint from the types of constants, functions, and pairs:
$ \Inter{\bot} $ is a singleton
whose unique element represents divergence.
Introducing the $ \bot $ type
means that we track termination in types.
In particular, we distinguish two classes of types:
those that are disjoint from $ \bot $
(for example, $ \Int $, $ \Int \to \Bool $, or $ \Int \times \Bool $)
and those that include $ \bot $
(since the interpretation of $ \bot $ is a singleton,
no type can contain a proper subset of it).
Intuitively, the former correspond to computations that are guaranteed to terminate:
for example, $ \Int $ is the type of terminating expressions
producing an integer result.
Conversely, the types of diverging expressions must always contain $ \bot $ and,
as a result, they can always be written
in the form $ t \lor \bot $, for some type $ t $.
Subtyping verifies $ t \leq t \lor \bot $ for any $ t $:
this ensures that a terminating expression
can always be used when a possibly diverging one is expected.
This subdivision of types suggests
that $\bot$ is used to approximate the set of diverging  well-typed expressions: an expression whose type contains $ \bot $ is an expression that \emph{may} diverge.
Actually, the type system we propose performs a rather gross approximation.
We derive ``terminating types'' (i.e., subtypes of $ \lnot \bot $)
only for expressions that are already results and cannot be reduced: constants, functions, or pairs.
Applications and projections, instead,
are always typed by assuming that they might diverge. The typing rules are written to handle and
\begin{floatingfigure}[r]{5.4cm}
 \hspace*{-1.2cm}\vspace{-4mm}\begin{minipage}{7.5cm}\vspace{-3mm} \[
  \Infer
    {
      \Typing{\Gamma |- e_1: (t' \to t) \lor \bot} \\
      \Typing{\Gamma |- e_2: t'}
    }
    {
      \Typing{\Gamma |- \Appl{e_1}{e_2}: t \lor \bot}
    }
    {}
\]
 \end{minipage}\vspace{1mm}
\end{floatingfigure}
\noindent propagate the $ \bot $ type. For example,
we type applications using the rule on the right.
This rule allows the expression $ e_1 $ to be possibly diverging:
we require it to have the type $ (t' \to t) \lor \bot $
instead of the usual $ t' \to t $ (but an expression with the latter type can always be subsumed to have the former type).
We type the whole application as $ t \lor \bot $
to signify that it can diverge
even if the codomain $ t $ does not include $ \bot $,
since $ e_1 $ can diverge.

This system avoids the problems we have seen with semantic subtyping:
no expression can be assigned the empty type,
which was the type on which subtyping had incorrect behaviour.
The new type $ \bot $ does not cause the same problems
because $ \Inter{\bot} $ is non-empty.
For example, the type of expressions like $ (\bar{e}, \K{3}) $
-- where $ \bar{e} $ is diverging --
is now $ \bot \times \Int $.
This type is not equivalent to $ \bot \times \Bool $:
indeed, the two interpretations are different because
the interpretation of types includes an element ($ \Inter{\bot} $)
to represent divergence.

Typing all applications as possibly diverging
-- even very simple ones like $
  \Appl{(\Abstr{x. \K{3}})}{e}
$ --
is a very coarse approximation
which can seem unsatisfactory.
We could try to amend the rule to say that
if $ e_1 $ has type $ t' \to t $,
then $ \Appl{e_1}{e_2} $ has type $ t $ instead of $ t \lor \bot $.
However, we prefer to keep the simpler rules
since they achieve our goal of giving a sound type system
that still enjoys most benefits of semantic subtyping.

An advantage of the simpler system is that it allows us to treat
$ \bot $ as an internal type
that does not need to be written explicitly by programmers.
Since the language is explicitly typed,
if $ \bot $ were to be treated more precisely,
programmers would presumably need to include it or exclude it explicitly
from function signatures.
This would make the type system significantly different from conventional ones
where divergence is not explicitly expressed in the types.
In the present system, instead,
we can assume that programmers annotate programs using standard set-theoretic types
and $ \bot $ is introduced only behind the scenes and, thus, is transparent to programmers.

We define this type system
for a call-by-need variant of the language
studied in \cite{Frisch2008},
and we prove its soundness
in terms of progress and subject reduction.

The choice of call-by-need rather than call-by-name
stems from the behaviour of semantic subtyping on intersections of arrow types.
Our type system would actually be unsound for call-by-name
if the language were extended
with constructs that can reduce non-deterministically to different answers.
For example, the expression $ \mathsf{rnd}(t) $ of \citet{Frisch2008}
that returns a random value of type $ t $
could not be added while keeping soundness.
This is because in call-by-name,
if such an expression is duplicated, each occurrence could reduce differently;
in call-by-need, instead, its evaluation would be shared.
Intersection and union types make the type system precise enough
to expose this difference.
In the absence of such non-deterministic constructs,
call-by-name and call-by-need can be shown to be observationally equivalent,
so that soundness should hold for both;
however, call-by-need also simplifies the technical work to prove soundness.

We show an example of this,
though we will return on this point later.
Consider the following derivation,
where $ \bar{e} $ is an expression of type $ \Int \lor \Bool $.
\[
  \Infer
  {
    \Infer[$ \leq $]
    {
      \Infer
      {
        \Typing{x\colon \Int |- (x, x): \Int \times \Int} \\
        \Typing{x\colon \Bool |- (x, x): \Bool \times \Bool}
      }
      {
        \Typing{|- \Abstr{x. (x, x)}:
          (\Int \to \Int \times \Int) \land (\Bool \to \Bool \times \Bool)}
      }
      {}
    }
    {
      \Typing{|- \Abstr{x. (x, x)}:
        \Int \lor \Bool \to (\Int \times \Int) \lor (\Bool \times \Bool)}
    }
    {}
    \\
    \Typing{|- \bar{e}: \Int \lor \Bool}
  }
  {
    \Typing{|- \Appl{(\Abstr{x. (x, x)})}{\bar{e}}:
      (\Int \times \Int) \lor (\Bool \times \Bool)}
  }
  {}
\]
In a system with intersection types,
the function $ \Abstr{x. (x, x)} $ can be given the type $
  (\Int \to \Int \times \Int) \land (\Bool \to \Bool \times \Bool)
$ because it has both arrow types
(in practice, the function will have to be annotated with the intersection).
Then, the step marked with \Rule{$ \leq $}
is allowed because, in semantic subtyping, $
  (\Int \to \Int \times \Int) \land (\Bool \to \Bool \times \Bool)
$ is a subtype of $
  (\Int \lor \Bool) \to ((\Int \times \Int) \lor (\Bool \times \Bool))
$
(in general, $
  (t_1 \to t_1') \land (t_2 \to t_2') \leq
  t_1 \lor t_2 \to t_1' \lor t_2'
$).
Therefore, the application $ \Appl{(\Abstr{x. (x, x)})}{\bar{e}} $
is well-typed with type $
  (\Int \times \Int) \lor (\Bool \times \Bool)
$.
In call-by-name, it reduces
to $ (\bar{e}, \bar{e}) $:
therefore, for the system to satisfy subject reduction,
we must be able to type $ (\bar{e}, \bar{e}) $ with the type $
  (\Int \times \Int) \lor (\Bool \times \Bool)
$ too.
But this type is intuitively unsound for $ (\bar{e}, \bar{e}) $
if each occurrence of $ \bar{e} $ could reduce
independently and non-deterministically either to an integer or to a boolean.
Using a typecase we can actually exhibit a term that breaks subject reduction.

There are several ways to approach this problem.
We could change the type system or the subtyping relation
so that $ \Abstr{x. (x, x)} $ cannot be given the type $
  (\Int \lor \Bool) \to ((\Int \times \Int) \lor (\Bool \times \Bool))
$.
However, this would curtail the expressive power of intersection types
as used in the semantic subtyping approach.
We could instead assume explicitly that the semantics is deterministic.
In this case, the typing would not be unsound intuitively,
but a proof of subject reduction would be difficult:
we should give a complex union disjunction rule to type $ (\bar{e}, \bar{e}) $.
We choose instead to consider a call-by-need semantics
because it solves both problems.
With call-by-need, non-determinism poses no difficulty because of sharing.
We still need a union disjunction rule,
but it is simpler to state since we only need it to type the \K{let} bindings
which represent shared computations.

\subsection{Contributions}

The main contribution of this work is the development of a type system
for non-strict languages based on semantic subtyping;
to our knowledge, this had not been studied before.

Although the idea of our solution is simple -- to track divergence -- its technical development is far from trivial.
Our work highlights how a type system featuring union and intersection types
is sensitive to the difference between strict and non-strict semantics
and also,
in the presence of non-determinism,
to that between call-by-name and call-by-need.
This shows once more
how union and intersection types can express very fine properties of programs.
Our main technical contribution is the description of sound typing
for \K{let} bindings
-- a construct peculiar to most of the formalizations of call-by-need semantics
-- in the presence of union types.
%
%
%
Finally, our work shows how to integrate the $ \bot $ type,
which is an explicit representation for divergence,
in a semantic subtyping system.
It can thus also be seen as a first step
towards the definition of a type system based on semantic subtyping
that performs a non-trivial form of termination analysis.

\subsection{Related work}

Previous work on semantic subtyping does not discuss non-strict semantics.
Castagna and Frisch~\citet{Castagna2005}
describe how to add a type constructor $ \mathsf{lazy}(t) $
to semantic subtyping systems,
but this is meant just to have lazily constructed expressions
within a call-by-value language.

Many type systems for functional languages
-- like the simply-typed \textlambda-calculus or Hindley-Milner typing --
are sound for both strict and non-strict semantics.
However, difficulties similar to ours are found in work on refinement types.
Vazou et al.~\citet{Vazou2014}
study how to adapt refinement types for Haskell.
Their types contain logical predicates as refinements:
e.g., the type of positive integers is $
  \Setc{ v\colon \Int \given v > \K{0} }
$.
They observe that the standard approach to typechecking in these systems
-- checking implication between predicates with an SMT solver --
is unsound for non-strict semantics.
In their system, a type like $ \Setc{ v\colon \Int \given \textsf{false} } $
is analogous to $ \Empty $ in our system
insofar as it is not inhabited by any value.
These types can be given to diverging expressions,
and their introduction into the environment causes unsoundness.
To avoid this problem, they stratify types,
with types divided in diverging and non-diverging ones.
This corresponds in a way to our use of a type $ \bot $
in types of possibly diverging expressions.
As for ours, their type system can track termination to a certain extent.
Partial correctness properties
can be verified even without precise termination analysis.
However,
with their kind of analysis
(which goes beyond what is expressible with set-theoretic types)
there is a significant practical benefit to tracking termination more precisely.
Hence,
they also study how to check termination of recursive functions.

The notion of a stratification of types
to keep track of divergence
can also be found in work of a more theoretical strain.
For instance, in \citet{Constable1987} it is used
to model partial functions in constructive type theory.
This stratification can be understood as a monad for partiality,
as it is treated in \citet{Capretta2005}.
Our type system can also be seen, intuitively,
as following this monadic structure.
Notably, the rule for applications
in a sense lifts the usual rule for application in this partiality monad.
Injection in this monad is performed implicitly by subtyping
via the judgment $ t \leq t \lor \bot $.
However, we have not developed this intuition formally.

The fact that a type system with union and intersection types
can require changes to account for non-strict semantics
is also remarked in work on refinement types.
Dunfield and Pfenning~\cite[p. 8, footnote 3]{Dunfield2003}
notice how a union elimination rule
cannot be used to eliminate unions in function arguments
if arguments are passed by name:
this is analogous to the aforementioned difficulties
which led to our choice of call-by-need
(their system uses a dedicated typing rule
for what our system handles by subtyping).
Dunfield~\cite[Section 8.1.5]{Dunfield2007}
proposes as future work to adapt a subset of the type system he considers
(of refinement types for a call-by-value effectful language)
to call-by-name.
He notes some of the difficulties and advocates studying call-by-need
as a possible way to face them.
In our work we show, indeed, that a call-by-need semantics
can be used to have the type system
handle union and intersection types expressively
without requiring complex rules.

\subsection{Outline}

Our presentation proceeds as follows.
In Section~\ref{sec:types},
we define the types and the subtyping relation which we use in our type system.
In Section~\ref{sec:language},
we define the language we study,
its syntax and its operational semantics.
In Section~\ref{sec:typing},
we present the type system;
we state the result of soundness for it
and outline the main lemmas required to prove it;
we also complete the discussion about why we chose a call-by-need semantics.
In Section~\ref{sec:changing-subtyping},
we study the relation
between the interpretation of types used to define subtyping
and the expressions that are definable in the language;
we show how we can look for a more precise interpretation.
In Section~\ref{sec:conclusion}
we conclude and point out more directions for future work.

For space reasons, some auxiliary definitions and results,
as well as the proofs of the results we state,
are omitted and can all be found
\ifarxiv
  in the Appendix.
\else
  in the extended version available online~\cite{extendedversion}.
\fi

\section{Types and subtyping}
\label{sec:types}

We begin by describing in more detail
the types and the subtyping relation of our system.

In order to define types,
we first fix two countable sets:
a set $ \Constants $ of \emph{language constants}
(ranged over by $ c $)
and a set $ \BasicTypes $ of \emph{basic types}
(ranged over by $ b $).
For example, we can take constants to be booleans and integers: $
  \Constants = \Set{\K{true}, \K{false}, \K{0}, \K{1}, \K{-1}, \dots}
$.
$ \BasicTypes $ might then contain $ \Bool $ and $ \Int $;
however, we also assume that, for every constant $ c $,
there is a ``singleton'' basic type which corresponds to that constant alone
(for example, a type for $ \K{true} $, which will be a subtype of $ \Bool $).
We assume that a function $
  \ConstantsInBasicType : \BasicTypes \to \Pset(\Constants)
$ assigns to each basic type the set of constants of that type
and that a function $
  \BasicTypeOfConstant{(\cdot)} : \Constants \to \BasicTypes
$ assigns to each constant $ c $ a basic type $ \BasicTypeOfConstant{c} $
such that $ \ConstantsInBasicType(\BasicTypeOfConstant{c}) = \Set{c} $.

\begin{Definition}[Types]
  The set $ \Types $ of \emph{types} is the set of terms $ t $
  coinductively produced by the following grammar\\[-.5mm]\centerline{
  \(
    t \Coloneqq \bot \mid b \mid t \times t \mid t \to t \mid
    t \lor t \mid \lnot t \mid \Empty
 \vspace{-1.5mm}
  \)}\\[1.5mm]
  and which satisfy two additional constraints: $(1)$ \emph{regularity}:
          the term must have a finite number of different sub-terms; $(2)$ \emph{contractivity}:
          every infinite branch must contain an infinite number
          of occurrences of the product or arrow type constructors.\vspace{1.2mm}
\end{Definition}

We introduce the abbreviations
$
  t_1 \land t_2 \eqdef \lnot (\lnot t_1 \lor \lnot t_2)
$, $
  t_1 \setminus t_2 \eqdef t_1 \land (\lnot t_2)
$, and $
  \Any \eqdef \lnot \Empty
$.
We refer to $ b $, $ \times $, and $ \to $ as \emph{type constructors},
and to $ \lor $, $ \lnot $, $ \land $, and $ \setminus $
as \emph{type connectives.}

The contractivity condition is crucial since it bars out ill-formed types
such as $ t = t \lor t $
(which does not carry any information about the set denoted by the type)
or $ t = \lnot t $ (which cannot represent any set).
It also ensures that the binary relation $ \triangleright \subseteq \Types^2 $
defined by $ t_1 \lor t_2 \triangleright t_i $, $ \lnot t \triangleright t $
is Noetherian (that is, strongly normalizing).
This gives an induction principle on $ \Types $ that we will use
without further reference to the relation. The regularity condition is necessary only to ensure the decidability of the subtyping relation.

In the semantic subtyping approach
we give an interpretation of types as sets;
this interpretation is used to define the subtyping relation
in terms of set containment.
We want to see a type as the set of the values of the language that have that type.
However, this set of values cannot be used directly to define the interpretation,
because of a problem of circularity.
Indeed, in a higher-order language,
values include well-typed \textlambda-abstractions;
hence to know which values inhabit a type
we need to have already defined the type system (to type $\lambda$-abstractions),
which depends on the subtyping relation,
which in turn depends on the interpretation of types.
To break this circularity,
types are actually interpreted as subsets of a set $ \Domain $,
an \emph{interpretation domain},
which is not the set of values,
though it corresponds to it intuitively
(in \citet{Frisch2008}, a correspondence is also shown formally:
we return to this in Section~\ref{sec:changing-subtyping}).
We use the following domain
which includes an explicit representation for divergence.

\begin{Definition}[Interpretation domain]
The \emph{interpretation domain} $ \Domain $ is the set of finite terms $ d $
produced inductively by the following grammar
\begin{align*}
  d & \Coloneqq \bot \mid c \mid (d, d) \mid \Set{(d, d_\Omega), \dots, (d, d_\Omega)}
    &
    d_\Omega & \Coloneqq d \mid \Omega
\end{align*}
where $ c $ ranges over the set $ \Constants $ of constants
and where $ \Omega $ is such that $ \Omega \notin \Domain $.
\end{Definition}

The elements of $ \Domain $ correspond, intuitively,
to the results of the evaluation of expressions.
The element $ \bot $ stands for divergence.
Expressions can produce as results constants or pairs of results,
so we include both in $ \Domain $.
For example, a result can be a pair of
a terminating computation returning \K{true}
and a diverging computation:
we represent this by $ (\K{true}, \bot) $.
Finally, in a higher-order language,
the result of a computation can be a function.
Functions are represented in this model by finite relations
of the form $ \Set{(d^1, d_\Omega^1), \dots, (d^n, d_\Omega^n)} $,
where $ \Omega $ (which is not in $ \Domain $)
can appear in second components to signify
that the functions fails (i.e., evaluation is stuck) on the corresponding input.
The restriction to \emph{finite} relations
is standard in semantic subtyping~\cite{Frisch2008};
we say more about it in Section~\ref{sec:changing-subtyping}.

We define the interpretation $ \Inter{t} $ of a type $ t $
so that it satisfies the following equalities,
where $ \Domain_\Omega = \Domain \cup \Set{\Omega} $
and where $ \PsetFin $ denotes the restriction of the powerset to finite subsets:
\begin{align*}
  \Inter{\bot} & = \Set{\bot} &
  \Inter{b} & = \ConstantsInBasicType(b) &
  \Inter{t_1 \times t_2} & = \Inter{t_1} \times \Inter{t_2} \\[-.3mm]
  \Inter{t_1 \to t_2} & =
    \Setc[\Big]{R \in \PsetFin(\Domain \times \Domain_\Omega) \given
      \forall (d, d') \in R. \: d \in \Inter{t_1} \implies d' \in \Inter{t_2}}
    \span\span\span\span \\[-.3mm]
  \Inter{t_1 \lor t_2} & = \Inter{t_1} \cup \Inter{t_2} &
  \Inter{\lnot t} & = \Domain \setminus \Inter{t} &
  \Inter{\Empty} & = \emptyset
\end{align*}

We cannot take the equations above
directly as an inductive definition of $ \Inter{} $
because types are not defined inductively but coinductively.
Therefore we give the following definition,
which validates these equalities
and which uses the aforementioned induction principle on types
and structural induction on $ \Domain $.

\begin{Definition}[Set-theoretic interpretation of types]\label{def:interpretation-of-types}
We define a binary predicate $ (d_\Omega : t) $
(``the element $ d_\Omega $ belongs to the type $ t $''),
where $ d_\Omega \in \Domain \cup \Set{\Omega} $ and $ t \in \Types $,
by induction on the pair $ (d_\Omega, t) $ ordered lexicographically.
The predicate is defined as follows:
\begin{align*}
  (\bot : \bot) & = \mathsf{true} \\
  (c : b) & = c \in \ConstantsInBasicType(b) \\
  ((d_1, d_2) : t_1 \times t_2 ) & =
    (d_1 : t_1) \mathrel{\mathsf{and}} (d_2 : t_2) \\
  (\Set{(d^1, d^1_\Omega), \dots, (d^n, d^n_\Omega)} : t_1 \to t_2) & =
    \forall i \in \Set{1, \dots, n} . \:
    \mathsf{if} \: (d^i : t_1) \mathrel{\mathsf{then}} (d^i_\Omega : t_2) \\
  (d : t_1 \lor t_2) & = (d : t_1) \mathrel{\mathsf{or}} (d : t_2) \\
  (d : \lnot t) & = \mathsf{not} \: (d : t) \\
  (d_\Omega : t) & = \mathsf{false} & \text{otherwise}
\end{align*}

We define the \emph{set-theoretic interpretation}
$ \Inter{} : \Types \to \Pset(\Domain) $
as $ \Inter{t} = \Setc{d \in \Domain \given (d : t)} $.
\end{Definition}

Finally,
we define the subtyping preorder and its associated equivalence relation
as follows.

\begin{Definition}[Subtyping relation]\label{def:subtyping}
  We define the \emph{subtyping} relation $ \leq $
  and the \emph{subtyping equivalence} relation $ \simeq $
  as
  \(
    t_1 \leq t_2 \iffdef \Inter{t_1} \subseteq \Inter{t_2}\) and   
  \(t_1 \simeq t_2 \iffdef (t_1 \leq t_2) \mathrel{\mathsf{and}} (t_2 \leq t_1)
    \: .
  \)
\end{Definition}

\section{Language syntax and semantics}
\label{sec:language}

We consider a language based on that studied in \citet{Frisch2008}:
a \textlambda-calculus
with recursive explicitly annotated functions,
pair constructors and destructors,
and a typecase construct.
We actually define two languages:
a \emph{source language} in which programs will be written
and a slightly different \emph{internal language},
on which we define the semantics.
The internal language adds a \K{let} construct;
this is a form of explicit substitution
used to model call-by-need semantics in a small-step operational style,
following a standard approach
\citep{Ariola1995,Ariola1997a,Maraist1998}.
Typecases are also defined slightly differently in the two languages
(to simplify the semantics),
so we show how to compile source programs to the internal language.

First,
we give some auxiliary definitions on types.
We introduce the abbreviations:
\(
  \WithBot{t} \eqdef t \lor \bot\); \(
  t_1 \toBot t_2 \eqdef \WithBot{t_1} \to \WithBot{t_2}\); and \(
  t_1 \timesBot t_2 \eqdef \WithBot{t_1} \times \WithBot{t_2}
  \: .
\)
These are compact notations for types including $ \bot $.
The first, $ \WithBot{t} $,
is an abbreviated way to write the type of possibly diverging expressions
whose result has type $ t $.
The latter two are used in type annotations.
The intent is that programmers never write $ \bot $ explicitly.
Rather, they use the $ \toBot $ and $ \timesBot $ constructors
instead of $ \to $ and $ \times $
so that $ \bot $ is introduced implicitly.
The $ \to $ and $ \times $ constructors are never written directly in program.
We define the following restricted grammars of types
\label{pag:types-big-t-and-tau}\vspace{-2.8mm}
\begin{align*}
  \InterfaceType & \Coloneqq b \mid \InterfaceType \timesBot \InterfaceType
    \mid \InterfaceType \toBot \InterfaceType
    \mid \InterfaceType \lor \InterfaceType \mid \lnot \InterfaceType
    \mid \Empty
  &
  \TypecaseType & \Coloneqq
    b \mid \TypecaseType \timesBot \TypecaseType \mid \Empty \to \Any
    \mid \TypecaseType \lor \TypecaseType \mid \lnot \TypecaseType \mid \Empty\\[-8mm]
\end{align*}
both of which are interpreted coinductively,
with the same restrictions of regularity and contractivity
as in the definition of types.
The types defined by these grammars will be the only ones
which appear in programs:
neither includes $ \bot $ explicitly.

In particular,
functions will be annotated with $ T $ types,
where the $ \timesBot $ and $ \toBot $ forms are used
to ensure that every type below a constructor is of the form $ t \lor \bot $.

Typecases, instead, will check $ \TypecaseType $ types.
The only arrow type that can appear in them is $ \Empty \to \Any $,
which is the top type of functions (every well-typed function has this type).
This restriction means that typecases
will not be able to test the types of functions,
but only, at most, whether a value is a function or not.
This restriction is not imposed in \cite{Frisch2008},
and actually it could be lifted here without difficulty.
We include it because the purpose of typecases in our language is, to some extent,
the modelling of pattern matching,
which cannot test the type of functions.
Restricting typecases on arrow types
also facilitates the extension of the system with polymorphism and type inference.

\subsection{Source language}

The \emph{source language expressions} are the terms $ \SourceExpr $
produced inductively by the grammar\vspace{-2.8mm}
\begin{align*}
  \SourceExpr & \Coloneqq x \mid c
    \mid \RAAbstr{f: \Interface. x. \SourceExpr}
    \mid \Appl{\SourceExpr}{\SourceExpr}
    \mid (\SourceExpr, \SourceExpr) \mid \ProjIth{\SourceExpr}
    \mid \Case{(x = \SourceExpr) \IN \TypecaseType ? \SourceExpr : \SourceExpr}
  \\
  \Interface & \Coloneqq \textstyle\bigwedge_{i \in I}
    \InterfaceType_i' \toBot \InterfaceType_i
    & |I| > 0\\[-8mm]
\end{align*}
where $ f $ and $ x $ range over a  set $ \XVars $
of \emph{expression variables},
$ c $ over the set $ \Constants $ of constants,
$ i $ in $ \ProjIth{\SourceExpr} $ over $ \Set{1, 2} $,
and where $ \TypecaseType $ in $
  \Case{(x = \SourceExpr) \IN \TypecaseType ? \SourceExpr : \SourceExpr}
$ is such that $ \TypecaseType \not\simeq \Empty $
and $ \TypecaseType \not\simeq \Any $.

Source language expressions include variables, constants,
\textlambda-abstractions, applications,
pairs constructors $ (\SourceExpr, \SourceExpr) $
and destructors $ \ProjFst{\SourceExpr} $ and $ \ProjSnd{\SourceExpr} $,
plus the typecase $
  \Case{(x = \SourceExpr) \IN \TypecaseType ? \SourceExpr : \SourceExpr}
$.

A \textlambda-abstraction $ \RAAbstr{f: \Interface. x. \SourceExpr} $
is a possibly recursive function,
with recursion parameter $ f $ and argument $ x $,
both of which are bound in the body;
the function is explicitly annotated with its type $ \Interface $,
which is a finite intersection of types of the form
$ \InterfaceType' \toBot \InterfaceType $.

A typecase expression $
  \Case{(x = \SourceExpr_0) \IN \TypecaseType ? \SourceExpr_1 : \SourceExpr_2}
$ has the following intended semantics:
$ \SourceExpr_0 $ is evaluated until it can be determined
whether it has type $ \TypecaseType $ or not,
then the selected branch
($ \SourceExpr_1 $ if the result of $ \SourceExpr_0 $ has type $ \TypecaseType $,
$ \SourceExpr_2 $ if it has type $ \lnot \TypecaseType $:
one of the two cases always occurs)
is evaluated in an environment where
$ x $ is bound to the result of $ \SourceExpr_0 $.
Actually, to simplify the presentation,
we will give a non-deterministic semantics
in which we allow to evaluate $ \SourceExpr_0 $ more than what is needed to ascertain
whether it has type $ \TypecaseType $.

In the syntax definition above we have restricted
the types $ \TypecaseType $ in typecases
asking both $ \TypecaseType \not\simeq \Any $
and $ \TypecaseType \not\simeq \Empty $.
A typecase checking the type $ \Any $ is useless:
since all expressions have type $ \Any $,
it immediately reduces to its first branch.
Likewise, a typecase checking the type $ \Empty $
reduces directly to the second branch.
Therefore, the two cases are uninteresting to consider.
We forbid them
because this allows us to give a simpler typing rule for typecases.
Allowing them is just a matter of adding two (trivial) typing rules
specific to these cases, as we show later.

As customary, we consider expressions
up to renaming of bound variables.
In $ \RAAbstr{f: \Interface. x. \SourceExpr} $,
$ f $ and $ x $ are bound in $ \SourceExpr $.
In $
  \Case{(x = \SourceExpr_0) \IN \TypecaseType ? \SourceExpr_1 : \SourceExpr_2}
$, $ x $ is bound in $ \SourceExpr_1 $ and $ \SourceExpr_2 $.

We do not provide mechanisms to define cyclic data structures.
For example, we do not have a direct syntactic construct to define
the infinitely nested pair $ (\K{1}, (\K{1}, \dots)) $.
We can define it by writing a fixpoint operator
(which can be typed in our system since types can be recursive)
or by defining and applying a recursive function which constructs the pair.
A general \K{letrec} construct as in \cite{Ariola1997a}
might be useful in practice
(for efficiency or to provide greater sharing)
but we omit it here since we are only concerned with typing.

\subsection{Internal language}

The \emph{internal language expressions} are the terms $ e $
produced inductively by the grammar\vspace{-2.8mm}
\begin{align*}
  e & \Coloneqq x \mid c
    \mid \RAAbstr{f: \Interface. x. e} \mid \Appl{e}{e}
    \mid (e, e) \mid \ProjIth{e}
    \mid \Case{(x = \TypecaseExpr) \IN \TypecaseType ? e : e}
    \mid \Let{x = e \IN e} \\
  \TypecaseExpr & \Coloneqq x \mid c \mid
    \RAAbstr{f: \Interface. x. e} \mid (\TypecaseExpr, \TypecaseExpr)\\[-8mm]
\end{align*}
where metavariables and conventions are as in the source language.
There are two differences with respect to the source language.
One is the introduction of the construct $ \Let{x = e_1 \IN e_2} $,
which is a binder used to model call-by-need semantics
(in $ \Let{x = e_1 \IN e_2} $, $ x $ is bound in $ e_2 $).
The other difference is that typecases cannot check arbitrary expressions,
but only expressions of the restricted form given by $ \TypecaseExpr $.

A source language expression $ \SourceExpr $ can be compiled
to an internal language expression $ \Compile{\SourceExpr} $
as follows.
Compilation is straightforward for all expressions apart from typecases:\vspace{-2mm}
\begin{align*}
  \Compile{x} & = x
  &
  \Compile{c} & = c
  &
  \Compile{\RAAbstr{f: \Interface. x. \SourceExpr}} & =
    \RAAbstr{f: \Interface. x. \Compile{\SourceExpr}}
  \\[-1mm]
  \Compile{\Appl{\SourceExpr_1}{\SourceExpr_2}} & =
    \Appl{\Compile{\SourceExpr_1}}{\Compile{\SourceExpr_2}}
  &
  \Compile{(\SourceExpr_1, \SourceExpr_2)} & =
    (\Compile{\SourceExpr_1}, \Compile{\SourceExpr_2})
  &
  \Compile{\ProjIth{\SourceExpr}} & = \ProjIth{\Compile{\SourceExpr}}\\[-7mm]
\end{align*}
and for typecases it introduces a \K{let} binder
to ensure that the checked expression is a variable:\vspace{-1.9mm}
\begin{align*}
  \Compile{
    \Case{(x = \SourceExpr_0) \IN \TypecaseType ? \SourceExpr_1 :
      \SourceExpr_2}}
    & =
    \Let{y = \Compile{\SourceExpr_0} \IN \Case{
      (x = y) \IN \TypecaseType ?
        \Compile{\SourceExpr_1} : \Compile{\SourceExpr_2}}}\\[-7mm]
\end{align*}
where $ y $ is chosen not free in $ \SourceExpr_1 $ and $ \SourceExpr_2 $.
(The other forms for $ \TypecaseExpr $ appear during reduction.)

\subsection{Semantics}

We define the operational semantics of the internal language
as a small-step reduction relation using call-by-need.
The semantics of the source language is then given indirectly
through the translation.
The choice of call-by-need rather than call-by-name
was briefly motivated in the Introduction
and will be discussed more extensively in Section~\ref{sec:typing}.

We first define the sets of \emph{answers} (ranged over by $a$) and of \emph{values}
(ranged over by $v$) as the subsets of expressions produced by the following grammars:\vspace{-1.8mm}
\begin{align*}
  a & \Coloneqq c \mid \RAAbstr{f: \Interface . x. e}
    \mid (e, e) \mid \Let{x = e \IN a}
  &
  v & \Coloneqq c \mid \RAAbstr{f: \Interface . x. e}\\[-7mm]
\end{align*}
Answers are the results of evaluation.
They correspond to expressions
which are fully evaluated up to their top-level constructor
(constant, function, or pair)
but which may include arbitrary expressions below that constructor
(so we have $ (e, e) $ rather than $ (a, a) $).
Since they also include \K{let} bindings,
they represent closures
in which variables can be bound to arbitrary expressions.
Values are a subset of answers treated specially in a reduction rule.

The semantics uses evaluation contexts to direct the order of evaluation.
A \emph{context} $ C $ is an expression with a hole (written $ \Hole{} $) in it.
We write $ C\Hole{e} $ for the expression obtained
by replacing the hole in $ C $ with $ e $.
We write $ C\HoleUnbound{e} $ for $ C\Hole{e} $
when the free variables of $ e $ are not bound by $ C $:
for example,
$ \Let{x = e_1 \IN x} $ is of the form $ C\Hole{x} $
-- with $ C \equiv (\Let{x = e_1 \IN \Hole{}}) $ --
but not of the form $ C\HoleUnbound{x} $;
conversely, $ \Let{x = e_1 \IN y} $
is both of the form $ C \Hole{y} $ and $ C \HoleUnbound{y} $.

\emph{Evaluation contexts} $ E $ are the subset of contexts
generated by the following grammar:\vspace{-1.8mm}
\begin{align*}
  E & \Coloneqq \Hole{} \mid \Appl{E}{e} \mid \ProjIth{E}
    \mid \Case{(x = F) \IN \TypecaseType ? e : e}
    \mid \Let{x = e \IN E} \mid \Let{x = E \IN E\HoleUnbound{x}} \\
  F & \Coloneqq \Hole{} \mid (F, \TypecaseExpr) \mid (\TypecaseExpr, F)\\[-7.5mm]
\end{align*}
Evaluation contexts allow reduction to occur on the left of applications
and below projections, but not
on the right of applications and below pairs.
For typecases alone, the contexts allow reduction also below pairs,
since this reduction might be necessary to be able to determine
whether the expression has type $ \TypecaseType $ or not.
This is analogous to the behaviour of pattern matching in lazy languages,
which can force evaluation below constructors.
The contexts for \K{let} are from standard presentations of call-by-need
\citep{Ariola1997a,Maraist1998}.
They allow reduction of the body of the \K{let},
while they only allow reductions of the bound expression
when it is required to continue evaluating the body:
this is enforced by requiring the body to have the form $ E \HoleUnbound{x} $.


\begin{figure}\vspace{-5mm}
  \input{figures/semantics}\vspace{-4mm}
  \caption{Operational semantics.\vspace{-3mm}}
  \label{fig-semantics}
\end{figure}

Figure~\ref{fig-semantics} presents the reduction rules.
They rely on the $ \Typeof $ function,
which assigns types to expressions of the form $ \TypecaseExpr $.
It is defined as follows:\vspace{-2.8mm}
\begin{align*}
  \Typeof(x) & = \Any
  &
  \Typeof(\RAAbstr{f: \Interface . x. e}) & =
    \Empty \to \Any
  \\[-1mm]
  \Typeof(c) & = b_c
  &
  \Typeof((\TypecaseExpr_1, \TypecaseExpr_2)) & =
    \Typeof(\TypecaseExpr_1) \times \Typeof(\TypecaseExpr_2)\\[-8mm]
\end{align*}
\Rule{Appl} is the standard application rule for call-by-need:
the application $ \Appl{( \RAAbstr{f: \Interface. x. e} )}{e'} $
reduces to $ e $
prefixed by two \K{let} bindings that bind
the recursion variable $ f $ to the function itself
and the parameter $ x $ to the argument $ e' $.
\Rule{ApplL} instead deals with applications
with a \K{let} expression in function position:
it moves the application below the \K{let}.
The rule is necessary to prevent loss of sharing:
substituting the binding of $ x $ to $ e $ in $ a $
would duplicate $ e $.
Symmetrically,
there are two rules for pair projections, \Rule{Proj} and \Rule{ProjL}.

There are three rules for \K{let} expressions.
They rewrite expressions of the form $ \Let{x = a \IN E \HoleUnbound{x}} $:
that is, \K{let} bindings
where the bound expression is an answer
and the body is an expression whose evaluation requires the evaluation of $ x $.
If $ a $ is a value $ v $, \Rule{LetV} applies
and the expression is reduced by just replacing $ v $ for $ x $ in the body.
If $ a $ is a pair, \Rule{LetP} applies:
the occurrences of $ x $ in the body are replaced
with a pair of variables $ (x_1, x_2) $
and each $ x_i $ is bound to $ e_i $ by new \K{let} bindings
(replacing $ x $ directly by $ (e_1, e_2) $ would duplicate expressions).
Finally, the \Rule{LetL} rule just moves a \K{let} binding out of another.

There are two rules for typecases,
by which a typecase construct $
  \Case{(x = \TypecaseExpr) \IN \TypecaseType ? e_1 : e_2}
$ can be reduced to either branch,
introducing a new binding of $ x $ to $ \TypecaseExpr $.
The rules apply only if either of $
  \Typeof(\TypecaseExpr) \leq \TypecaseType
$ or $
  \Typeof(\TypecaseExpr) \leq \lnot \TypecaseType
$ holds.
If neither holds, then the two rules do not apply,
but the \Rule{Ctx} rule can be used
to continue the evaluation of $ \TypecaseExpr $.

\subparagraph{Comparison to other presentations of call-by-need.}
These reduction rules mirror those from standard presentations of call-by-need%
~\citep{Ariola1995,Ariola1997a,Maraist1998}.
A difference is that, in \Rule{LetV} or \Rule{LetP},
we replace \emph{all} occurrences of $ x $ in $ E \HoleUnbound{x} $ at once,
whereas in the cited presentations
only the occurrence in the hole is replaced:
for example, in \Rule{LetV}
they reduce to $ E \HoleUnbound{v} $
instead of $ (E \HoleUnbound{x}) [\nicefrac{v}{x}] $.
Our \Rule{LetV} rule is mentioned as a variant in \cite[p. 38]{Maraist1998}.
We use it because it simplifies the proof of subject reduction
while maintaining an equivalent semantics.


\subparagraph{Non-determinism in the rules.}
The semantics is not deterministic.
There are two sources of non-determinism, both related to typecases.
One is that the contexts $ F $
include both $ (F, \TypecaseExpr) $ and $ (\TypecaseExpr, F) $
and thereby impose no constraint on the order with which pairs are examined.

The second source of non-determinism is that the contexts for typecases
allow us to reduce the bindings of variables in the checked expression
even when we can already apply \Rule{Case1} or \Rule{Case2}.
For example, take $
  \Let{x = e \IN \Case{(y = (\K{3}, x)) \IN (\Int \timesBot \Any) ? e_1 : e_2}}
$.
It can be immediately reduced to $
  \Let{x = e \IN \Let{y = (\K{3}, x) \IN e_1}}
$ by applying \Rule{Ctx} and \Rule{Case1},
because $
  \Typeof((\K{3}, x)) = \BasicTypeOfConstant{\K{3}} \times \Any
  \leq \Int \timesBot \Any
$.
However, we can also use \Rule{Ctx} to reduce $ e $, if it is reducible:
we do so by writing the expression as $
  \Let{x = e \IN E \HoleUnbound{x}}
$, where $ E $ is $
  \Case{(y = (\K{3}, \Hole{})) \IN (\Int \timesBot \Any) ? e_1 : e_2}
$.
To model a lazy implementation more faithfully,
we should forbid this reduction
and state that $ \Case{(x = F) \IN \TypecaseType ? e : e} $
is a context only if it cannot be reduced by \Rule{Case1} or \Rule{Case2}.

In both cases, we have chosen a non-deterministic semantics
because it is less restrictive:
as a consequence, the soundness result will also hold for semantics
which fix an order.

\section{Type system}
\label{sec:typing}

We define two typing relations
for the source language and the internal language.

A \emph{type environment} $ \Gamma $
is a finite mapping of type variables to types.
We write $ \emptyset $ for the empty environment.
We say that a type environment $ \Gamma $ is \emph{well-formed} if, for all $
  (x \colon t) \in \Gamma
$, we have $ t \not\simeq \Empty $.
Since we want to ensure that the empty type is never derivable,
we will only consider well-formed type environments
in the soundness proof.

\subsection{Type system for the source language}

\begin{figure}[t]\vspace{-5mm}
  \input{figures/typing-source}\vspace{-7mm}
  \caption{Typing rules for the source language.\vspace{-3mm}}
  \label{fig-typing-source}
\end{figure}

Figure~\ref{fig-typing-source} presents the typing rules
for source language.
The subsumption rule \Rule{S-Subsum} is used to apply subtyping.
Notably, it allows
expressions with surely converging types
(like a pair with type $ \Int \times \Bool $)
to be used where diverging types are expected:
$ t \leq \WithBot{t} $ holds for every $ t $
(since $
  \Inter{t} \subseteq \Inter{t} \cup \Set{\bot} =
  \Inter{t \lor \bot} = \Inter{\WithBot{t}}
$).
The rules \Rule{S-Var} and \Rule{S-Const} for variables and constants are standard.
The \Rule{S-Abstr} rule for functions is also straightforward.
Function interfaces have the form $
  \bigwedge_{i \in I} T'_i \toBot T_i
$, that is, $
  \bigwedge_{i \in I} \WithBot{T'_i} \to \WithBot{T_i}
$ (expanding the definition of $ \toBot $).
To type a function $ \RAAbstr{f: \Interface. x. \SourceExpr} $,
we check that it has all the arrow types in $\Interface$.
Namely, for every arrow $ T'_i \toBot T_i $ (i.e., $ \WithBot{T'_i} \to \WithBot{T_i} $),
we assume that $ x $ has type $ \WithBot{T'_i} $
and that the recursion variable $ f $ has type $ \Interface $, and we check
that the body has type $ \WithBot{T_i} $.

The \Rule{S-Appl} rule is the first one
that deals with $ \bot $ in a non-trivial way.
In call-by-value semantic subtyping systems,
to type an application $ \Appl{\SourceExpr_1}{\SourceExpr_2} $
with a type $ t $,
the standard \emph{modus ponens} rule (e.g., the one from the simply-typed \textlambda-calculus) is used:
$ \SourceExpr_1 $ must have type $ t' \to t $
and $ \SourceExpr_2 $ must have type $ t' $.
Here, instead, we allow the function to have the type $ \WithBot{t' \to t} $
(i.e., $ (t' \to t) \lor \bot $)
to allow the application also when $ \SourceExpr_1 $ might diverge.
We use $ \WithBot{t} $ as the type of the whole application,
signifying that it might diverge.
As anticipated,
we do not try to predict whether applications will converge.
The rule \Rule{S-Pair} for pairs is standard;
\Rule{S-Proj} handles $ \bot $ as in applications.

\Rule{S-Case} is the most complex one,
but it corresponds closely to that of \citet{Frisch2008}.
To type $
  \Case{(x = \SourceExpr_0) \IN \TypecaseType ? \SourceExpr_1 : \SourceExpr_2}
$ we first type $ \SourceExpr_0 $ with some type $ \WithBot{t'} $.
Then, we type the two branches $ \SourceExpr_1 $ and $ \SourceExpr_2 $.
We do not always have to type both
(note the conditions ``if \dots then'')
but for now assume that we do.
While typing either branch, we extend the environment with a binding for $ x $.
For the first branch, the type for $ x $ is $ t' \land \tau $,
a subtype of $ \WithBot{t'} $:
this type is sound because the first branch is only evaluated
if $ \SourceExpr_0 $ evaluates to an answer
(meaning we can remove the union with $ \bot $ in $ \WithBot{t'} $)
and if this answer has type $ \TypecaseType $.
Conversely, for the second branch, $ x $ is given type
$ t' \setminus \TypecaseType $, that is, $ t' \land \lnot \TypecaseType $.
Finally, if the branches have type $ t $,
the whole typecase is given type $ \WithBot{t} $
since its evaluation may diverge
in case $ \SourceExpr_0 $ diverges.

Now let us consider the conditions ``if \dots then''.
We need to type the first branch only when $ t' \not\leq \lnot \TypecaseType $;
if, conversely, $ t' \leq \lnot \TypecaseType $,
we know that the first branch can never be selected
(an expression of type $ \lnot \TypecaseType $
cannot reduce to a result of type $ \TypecaseType $)
and thus we don't need to type it.
The reasoning for the second branch is analogous.
The two conditions are pivotal to type
overloaded functions defined by typecases.
For example, a negation function implemented as $
  \RAAbstr{f: \Interface. x.
    \Case{(y = x) \IN \BasicTypeOfConstant{\K{true}} ? \K{false} : \K{true}} }
$,
with $
  \Interface =
    (\BasicTypeOfConstant{\K{true}} \to \BasicTypeOfConstant{\K{false}})
    \land
    (\BasicTypeOfConstant{\K{false}} \to \BasicTypeOfConstant{\K{true}})
$,
could not be typed without these conditions.

In the syntax
we have restricted the type $ \TypecaseType $ in typecases
requiring $ \TypecaseType \not\simeq \Any $
and $ \TypecaseType \not\simeq \Empty $.
Typecases where these conditions do not hold are uninteresting,
since they do not actually check anything.
The rule \Rule{S-Case} would be unsound for them
because these typecases can reduce to one branch
even if $ \SourceExpr_0 $ is a diverging expression
that does not evaluate to an answer.
For instance,
if $ \bar{\SourceExpr} $ has type $ \bot $ (that is, $ \WithBot{\Empty} $),
then $
  \Case{(x = \bar{\SourceExpr}) \in \Int ? \K{1} : \K{2}}
$ could be given any type, including unsound ones like $ \WithBot{\Bool} $.
%
%
To allow these typecases, we could add the side condition ``$
  \TypecaseType \not\simeq \Any \text{ and } \TypecaseType \not\simeq \Empty
$'' to \Rule{S-Case}
and give two specialized rules as follows:
\begin{mathpar}
  \Infer
    {
      \Typing{\Gamma |- \SourceExpr_0: t'} \\
      \Typing{\Gamma, x\colon t' |- \SourceExpr_1: t}
    }
    {
      \Typing{\Gamma |-
        \big(
          \Case{(x = \SourceExpr_0) \in \TypecaseType ?
            \SourceExpr_1 : \SourceExpr_2}
        \big):
        \WithBot{t}}
    }
    {\TypecaseType \simeq \Any}
  \and
  \Infer
    {
      \Typing{\Gamma |- \SourceExpr_0: t'} \\
      \Typing{\Gamma, x\colon t' |- \SourceExpr_2: t}
    }
    {
      \Typing{\Gamma |-
        \big(
          \Case{(x = \SourceExpr_0) \in \TypecaseType ?
            \SourceExpr_1 : \SourceExpr_2}
        \big):
        \WithBot{t}}
    }
    {\TypecaseType \simeq \Empty}
\end{mathpar}

\subsection{Type system for the internal language}

\begin{figure}[t]\vspace{-10mm}
  \input{figures/typing}\vspace{-7mm}
  \caption{Typing rules for the internal language.\vspace{-3mm}}
  \label{fig-typing}
\end{figure}

Figure~\ref{fig-typing} presents the typing rules for the internal language.
These include a new rule for \K{let} expressions
and a modified rule for \textlambda-abstractions;
the other rules are the same as those for the source language
(except for the different syntax of typecases).

The \Rule{S-Abstr} rule for the source language
derived the type $ \Interface $ for $ \RAAbstr{f: \Interface. x. \SourceExpr} $.
The rule for the internal language, instead,
allows us to derive a subtype of $ \Interface $
of the form $ \Interface \land t $,
where $ t $ is an intersection of negations of arrow types.
The arrows in $ t $ can be chosen freely
providing that the intersection $ \Interface \land t $ remains non-empty.
This rule (directly taken from \citet{Frisch2008}) can look surprising.
For example, it allows us to type $
  \RAAbstr{f: (\Int \toBot \Int). x. x}
$ as $
  (\Int \toBot \Int) \land \lnot (\Bool \to \Bool)
$ even though, disregarding the interface,
the function does map booleans to booleans.
But the language is explicitly typed,
and thus we can't ignore interfaces
(indeed, the function does not have type $ \Bool \to \Bool $).
The purpose of the rule is to ensure that,
given any function and any type $ t $,
either the function has type $ t $ or it has type $ \lnot t $.
This property matches the intuitive view of types as sets of values
that underpins semantic subtyping.
While in our system we do not really interpret types as sets of values
(since $ \bot $ is non-empty and yet uninhabited by values),
the property is still needed to prove subject reduction.
A consequence of this property is that a value
(i.e., a constant or a $\lambda$-abstraction)
of type $ t_1 \lor t_2 $ has always either type $ t_1 $ or type $ t_2 $.
(In the case of constants, this is obtained directly by reasoning on subtyping,
so we don't need a rule to assign negation types to them.)



The \Rule{Let} rule
combines a standard rule for (monomorphic) binders
with a union disjunction rule:
it lets us decompose the type of $ e_1 $ as a union
and type the body of the \K{let} once for each summand in the union.
The purpose of this rule was hinted at in the Introduction
and will be discussed again in Section~\ref{sec:propofsystems}, where we show that this rule
-- combined with the property on union types above -- is central to this work: it is the key technical feature that ensures the soundness of the system (see in particular Lemma~\ref{lem:text-product-decomposition} later on).
For the time being, just note that the type of $ e_1 $ can be decomposed
in arbitrarily complex ways by applying subsumption.
For example, if $ e_1 $ is a pair of type $
  (\Int \lor \Bool) \times (\Int \lor \Bool)
$, by applying \Rule{Subsum} we can type it as $
  (\Int \times \Int) \lor (\Int \times \Bool) \lor
  (\Bool \times \Int) \lor (\Bool \times \Bool)
$ and then type $ e_2 $ once for each of the four summands.

The \Rule{Abstr} and \Rule{Let} rules
introduce non-determinism
in the choice of the negations to introduce
and of how to decompose types as unions.
This would not complicate a practical implementation,
since a typechecker would only need to check the source language.

\subsection{Properties of the type system}\label{sec:propofsystems}

Full results about the type system, including proofs,
are available
\ifarxiv
  in the Appendix.
\else
  in the extended version~\cite{extendedversion}.
\fi
Here we report the main results
and describe the technical difficulties we met to obtain them.

First, we can easily show by induction that
compilation from the source language to the internal language preserves typing.

\begin{Proposition}
  \label{pro:compilation-preserves-typing}
  If $ \Typing{\Gamma |- \SourceExpr: t} $,
  then $ \Typing{\Gamma |- \Compile{\SourceExpr}: t} $.
\end{Proposition}

We show the soundness property for our type system
(``well-typed programs do not go wrong''),
following the well-known syntactic approach of Wright and Felleisen~%
  \citet{Wright1994},
by proving the two properties
of \emph{progress} and \emph{subject reduction}
for the internal language.

\begin{Theorem}[Progress]
  Let $ \Gamma $ be a well-formed type environment.
  Let $ e $ be an expression that is well-typed in $ \Gamma $
  (that is, $ \Typing{\Gamma |- e: t} $ holds for some $ t $).
  Then $ e $ is an answer, or $ e $ is of the form $ E \HoleUnbound{x} $,
  or $ \exists e'. \: \Reduces{e ~> e'} $.
\end{Theorem}

\begin{Theorem}[Subject reduction]
  Let $ \Gamma $ be a well-formed type environment.
  If $ \Typing{\Gamma |- e: t} $ and $ \Reduces{e ~> e'} $,
  then $ \Typing{\Gamma |- e': t} $.
\end{Theorem}

The statement of progress is adapted to call-by-need:
it applies also to expressions that are typed in a non-empty $ \Gamma $
and it allows a well-typed expression to have the form $ E \HoleUnbound{x} $.


As a corollary of these results, we obtain the following statement for soundness.

\begin{Corollary}[Type soundness]
  Let $ e $ be a well-typed, closed expression
  (that is, $ \Typing{\emptyset |- e: t} $ holds for some $ t $).
  If $ e \ReducesArrow^* e' $ and $ e' $ cannot reduce,
  then $ e' $ is an answer and $ \Typing{\emptyset |- e': t} $.
\end{Corollary}

The soundness result for the internal language
implies soundness for the source language.

\begin{Corollary}[Type soundness for the source language]
  Let $ \SourceExpr $ be a well-typed, closed source language expression
  (that is, $ \Typing{\emptyset |- \SourceExpr: t} $ holds for some $ t $).
  If $ \Compile{\SourceExpr} \ReducesArrow^* e' $ and $ e' $ cannot reduce,
  then $ e' $ is an answer and $ \Typing{\emptyset |- e': t} $.
\end{Corollary}

We summarize here some of the crucial properties
required to derive the results above.
We also resume the discussion of the motivations behind our choice of call-by-need.

We introduced the $ \bot $ type for diverging expressions
because assigning the type $ \Empty $ to any expression
causes unsoundness.
We must hence ensure
that no expression can be assigned the type $ \Empty $.
In well-formed type environments, we can prove this easily by induction.

\begin{Lemma}
  Let $ \Gamma $ be a well-formed type environment.
  If $ \Typing{\Gamma |- e: t} $, then $ t \not\simeq \Empty $.
\end{Lemma}

\subparagraph{Call-by-name and call-by-need.}
In the Introduction,
we have given two reasons for our choice of call-by-need rather than call-by-name.
One is that the system is only sound for call-by-name
if we make assumptions on the semantics
that might not hold in an extended language:
for example,
introducing an expression
that can reduce non-deterministically either to an integer or to a boolean
would break soundness.
The other reason is that, even when these assumptions hold
(and when presumably call-by-name and call-by-need
are observationally equivalent),
call-by-need is better suited to the soundness proof.

Let us review the example from the Introduction.
Consider the function $
  \RAAbstr{f: \Interface. x. (x, x)}
$ in the source language, where $
  \Interface =
  (\Int \toBot \Int \timesBot \Int) \land (\Bool \toBot \Bool \timesBot \Bool)
$.
It is well-typed with type $ \Interface $.
By subsumption, it also has the type
$
  (\Int \lor \Bool) \toBot (\Int \timesBot \Int) \lor (\Bool \timesBot \Bool)
$,
which is a supertype of $ \Interface $:
in general we have $
  (t'_1 \to t_1) \land (t'_2 \to t_2) \leq
  (t'_1 \lor t'_2) \to (t_1 \lor t_2)
$ and therefore $
  (t'_1 \toBot t_1) \land (t'_2 \toBot t_2) \leq
  (t'_1 \lor t'_2) \toBot (t_1 \lor t_2)
$.

Therefore, if $ \bar{\SourceExpr} $ has type $ \Int \lor \Bool \lor \bot $,
the application $
  \Appl{(\RAAbstr{f: \Interface. x. (x, x)})}{\bar{\SourceExpr}}
$ is well-typed with type $
  (\Int \timesBot \Int) \lor (\Bool \timesBot \Bool) \lor \bot
$.
Assume that $ \bar{\SourceExpr} $ can reduce
either to an integer or to a boolean:
for instance, assume
that both $ \bar{\SourceExpr} \ReducesArrow \K{3} $
and $ \bar{\SourceExpr} \ReducesArrow \K{true} $
can occur.

With call-by-name, $
  \Appl{(\RAAbstr{f: \Interface. x. (x, x)})}{\bar{\SourceExpr}}
$ reduces to $ (\bar{\SourceExpr}, \bar{\SourceExpr}) $;
then, the two occurrences of $ \bar{\SourceExpr} $ reduce independently.
It is intuitively unsound to type it as $
  (\Int \timesBot \Int) \lor (\Bool \timesBot \Bool) \lor \bot
$: there is no guarantee that the two components of the pair
will be of the same type once they are reduced.
We can find terms that break subject reduction.
Assume for example that there exists
a boolean ``$ \textsf{and} $'' operation;
then this typecase is well-typed (as $ \WithBot{\Bool} $)
but unsafe:
\[
  \Case{(y = \Appl{(\RAAbstr{f: \Interface. x. (x, x)})}{\bar{\SourceExpr}})
    \IN (\Int \timesBot \Int) ? \K{true} :
    (\ProjFst{y} \mathrel{\mathsf{and}} \ProjSnd{y})}
  \: .
\]
Since the application has type $
  \WithBot{ (\Int \timesBot \Int) \lor (\Bool \timesBot \Bool) }
$, to type the second branch of the typecase
we can assume that $ y $ has the type $
  ( (\Int \timesBot \Int) \lor (\Bool \timesBot \Bool) )
  \setminus (\Int \timesBot \Int)
  $,
which is a subtype of $
  \Bool \timesBot \Bool
$
(it is actually equivalent to $
  (\Bool \timesBot \Bool) \setminus (\bot \times \bot)
$).
Therefore, both $ \ProjFst{y} $ and $ \ProjSnd{y} $ have type $ \WithBot{\Bool} $.
We deduce then that $
  (\ProjFst{y} \mathrel{\mathsf{and}} \ProjSnd{y})
$ has type $ \WithBot{\Bool} $ as well
(we assume that ``$ \mathsf{and} $'' is defined
so as to handle arguments of type $ \bot $ correctly).

A possible reduction in a call-by-name semantics would be the following:
\begin{align*}
  &
  \Case{(y = \Appl{(\RAAbstr{f: \Interface. x. (x, x)})}{\bar{\SourceExpr}})
    \IN (\Int \timesBot \Int) ? \K{true} :
    (\ProjFst{y} \mathrel{\mathsf{and}} \ProjSnd{y})}
  \\ \ReducesArrow {} &
  \Case{(y = (\bar{\SourceExpr}, \bar{\SourceExpr}))
    \IN (\Int \timesBot \Int) ? \K{true} :
    (\ProjFst{y} \mathrel{\mathsf{and}} \ProjSnd{y})}
  \intertext{
    (the typecase must force the evaluation of
    $ (\bar{\SourceExpr}, \bar{\SourceExpr}) $
    to know which branch should be selected)}
  \ReducesArrow^* {} &
  \Case{(y = (\K{true}, \bar{\SourceExpr}))
    \IN (\Int \timesBot \Int) ? \K{true} :
    (\ProjFst{y} \mathrel{\mathsf{and}} \ProjSnd{y})}
  \intertext{
    (now we know that the first branch is impossible, so the second is chosen)}
  \ReducesArrow {} &
    \ProjFst{(\K{true}, \bar{\SourceExpr})} \mathrel{\mathsf{and}}
    \ProjSnd{(\K{true}, \bar{\SourceExpr})}
  \: \ReducesArrow \:
    \K{true} \mathrel{\mathsf{and}} \bar{\SourceExpr}
  \: \ReducesArrow \:
    \bar{\SourceExpr}
  \: \ReducesArrow \:
    \K{3}
\end{align*}
The integer $ \K{3} $ is not a $ \Bool $:
this disproves subject reduction for call-by-name
if the language contains expressions like $ \bar{\SourceExpr} $.
No such expressions exist in our current language,
but they could be introduced
if we extended it with non-deterministic constructs
like $ \mathsf{rnd}(t) $ from \citet{Frisch2008}.

Since we use a call-by-need semantics, instead,
expressions such as $ \bar{\SourceExpr} $ do not pose problems for soundness.
With call-by-need, $
  \Appl{(\RAAbstr{f: \Interface. x. (x, x)})}{\bar{\SourceExpr}}
$ reduces to $
  \Let{f = \RAAbstr{f: \Interface. x. (x, x)} \IN
    \Let{x = \bar{\SourceExpr} \IN (x, x)}}
$.
The occurrences of $ x $ in the pair are only substituted
when $ \bar{\SourceExpr} $ has been reduced to an answer,
so they cannot reduce independently.

To ensure subject reduction,
we allow the rule for \K{let} bindings to split unions in the type of the bound term.
This means that the following derivation is allowed.
\begin{mathpar}
  \Infer
    {
      \Typing{\Gamma |- \bar{\SourceExpr}: \Int \lor \Bool} \\
      \Typing{\Gamma, x\colon \Int |- (x, x): \Int \timesBot \Int} \\
      \Typing{\Gamma, x\colon \Bool |- (x, x): \Bool \timesBot \Bool}
    }
    {
      \Typing{\Gamma |- \Let{x = \bar{\SourceExpr} \IN (x, x)}:
        (\Int \timesBot \Int) \lor (\Bool \timesBot \Bool)}
    }
    {}
\end{mathpar}

\subparagraph{Proving subject reduction: main lemmas.}
While the typing rule for \K{let} bindings is simple to describe,
proving subject reduction for the reduction rules \Rule{LetV} and \Rule{LetP}
(those that actually perform substitutions)
is challenging.
For the reduction $
  \Reduces{
    \Let{x = v \IN E\HoleUnbound{x}} ~>
    (E\HoleUnbound{x})[\nicefrac{v}{x}]
  }
$, we show the following results.

\begin{Lemma}
  Let $ v $ be a value that is well-typed in $ \Gamma $
  (i.e., $ \Typing{\Gamma |- v: t'} $ holds for some $ t' $).
  Then, for every type $ t $,
  we have either $ \Typing{\Gamma |- v: t} $
  or $ \Typing{\Gamma |- v: \lnot t} $.
\end{Lemma}

\begin{Corollary}
  If $ \Typing{\Gamma |- v: \bigvee_{i \in I} t_i} $,
  then there exists an $ i_0 \in I $
  such that $ \Typing{\Gamma |- v: t_{i_0}} $.
\end{Corollary}

Consider for example the reduction $
  \Reduces{ \Let{x = v \IN (x, x)} ~> (v, v) }
$.
If $ v $ has type $ \Int \lor \Bool $,
then $ \Let{x = v \IN (x, x)} $ has type
$ (\Int \timesBot \Int) \lor (\Bool \timesBot \Bool) $
as in the derivation above.
Without this corollary,
for $ (v, v) $ we could only derive the type $
  (\Int \lor \Bool) \times (\Int \lor \Bool)
$, which is not a subtype of the type deduced for the redex.
Applying the corollary, we deduce that $ v $ has either type $ \Int $
or $ \Bool $;
in both cases $ (v, v) $ can be given the type
$ (\Int \timesBot \Int) \lor (\Bool \timesBot \Bool) $.

These results are also needed in semantic subtyping for strict languages
to prove subject reduction for applications.
To ensure them, following \citet{Frisch2008},
we have added in the type system for the internal language
the possibility of typing functions with negations of arrow types.

The reduction $
  \Reduces{
    \Let{x = (e_1, e_2) \IN E\HoleUnbound{x}} ~>
    \Let{x_1 = e_1 \IN \Let{x_2 = e_2 \IN
      (E\HoleUnbound{x})[\nicefrac{(x_1, x_2)}{x}]}} }
$, instead, is dealt with by the following lemma.

\begin{Lemma}
  \label{lem:text-product-decomposition}
  If $ \Typing{\Gamma |- (e_1, e_2): \bigvee_{i \in I} t_i} $,
  then there exist two types
  $ \bigvee_{j \in J} t_j $ and $ \bigvee_{k \in K} t_k $
  such that
  $
    \Typing{\Gamma |- e_1: \textstyle\bigvee_{j \in J} t_j}
  $, $
    \Typing{\Gamma |- e_2: \textstyle\bigvee_{k \in K} t_k}
  $, and $
    \forall j \in J . \: \forall k \in K . \: \exists i \in I . \:
      t_j \times t_k \leq t_i
  $.
\end{Lemma}

This is the result we need for the proof:
$
  \Let{x = (e_1, e_2) \IN E\HoleUnbound{x}}
$ is typed by assigning a union type to $ (e_1, e_2) $
and then typing $ E\HoleUnbound{x} $ once for every $ t_i $ in the union,
while the reduct $
  \Let{x_1 = e_1 \IN \Let{x_2 = e_2 \IN
    (E\HoleUnbound{x})[\nicefrac{(x_1, x_2)}{x}]}}
$ must be typed by typing $ e_1 $ and $ e_2 $ with two union types
and then typing the substituted expression
with every product $ t_j \times t_k $.
Showing that each $ t_j \times t_k $ is a subtype of a $ t_i $
ensures that the substituted expression is well-typed.
The proof consists in recognizing that the union $
  \bigvee_{i \in I} t_i
$ must be a decomposition into a union
of some type $ t_1 \times t_2 $
and that therefore $ t_1 $ and $ t_2 $
can be decomposed separately into two unions.

All these results rely
on the distinction between types that contain $ \bot $ and those that do not:
they would not hold if we assumed that every type implicitly contains $ \bot $.

Despite some technical difficulties,
call-by-need seems quite suited to the soundness proof.
Hence, it would probably be best to use it for the proof
even if we assumed explicitly
that the language does not include problematic expressions like $ \mathsf{rnd}(t) $.
Soundness would then also hold for a call-by-name semantics
that it is observationally equivalent to call-by-need.

\section{A discussion on the interpretation of types}
\label{sec:changing-subtyping}

We have shown in the previous sections that a set-theoretic interpretation of types, adapted to take into account divergence (Definition \ref{def:interpretation-of-types}), can be the basis for designing a sound type system for languages with lazy evaluation. In this section, we analyze the relation between such interpretation and the expressions that are actually definable in the language.

Let us first recap some notions from \cite{Frisch2008}.
The initial intuition which guides semantic subtyping is to
 see a type
as the set of values of that type in the language we consider:
for example, to see $ \Int \to \Bool $
as the set of \textlambda-abstractions of type $ \Int \to \Bool $.
However, we cannot directly define the interpretation of a type $ t $ as the set $
  \Setc{ v \given \Typing{\emptyset |- v: t} }
$, because the typing relation  $ \Typing{\emptyset |- v: t} $ depends on the definition of subtyping,
which depends in turn on the interpretation of types.
Frisch, Castagna and Benzaken~\citet{Frisch2008}
avoid this circularity by giving an interpretation $ \Inter{} $ of types
as subsets of an interpretation domain
where finite relations replace \textlambda-abstractions.

This interpretation
(like ours except that there is no $ \bot $)
is used to define subtyping
and the typing relation.
Then, the following result is shown:\\[1.5mm]
\centerline{
  \label{eq:model-of-value}\(
  \forall t_1, t_2 . \:
    \Inter{t_1} \subseteq \Inter{t_2}
    \iff
    \Inter{t_1}_\mathcal{V} \subseteq \Inter{t_2}_\mathcal{V}
  \qquad
  \text{where }
  \Inter{t}_\mathcal{V} \eqdef \Setc{ v \given \Typing{\emptyset |- v: t} }
\)}\\[1.5mm]
This result states that
a type $ t_1 $ is a subtype of a type $ t_2 $
($ t_1 \leq t_2 $, which is defined as $ \Inter{t_1} \subseteq \Inter{t_2} $)
if and only if
every value $ v $ that can be assigned the type $ t_1 $
can also be assigned the type $ t_2 $.
Showing the result above implies that, once the type system is defined,
we can indeed reason on subtyping by reasoning on inclusion between sets of values.\footnote{The circularity is avoided since
  the typing relation in $ \Setc{ v \given \Typing{\emptyset |- v: t}}$ is defined using $\Inter{}$ and not $\Inter{}_{\mathcal V}$.}

This result is useful in practice, since, when typechecking fails
because a subtyping judgment $ t_1 \leq t_2 $ does not hold,
we know that there exists a value $ v $
such that $ \Typing{\emptyset |- v: t_1} $ holds
while $ \Typing{\emptyset |- v: t_2} $ does not.
This value $ v $ can be shown
as a witness to the unsoundness of the program while reporting the error.%
  \footnote{In case of a type error, the \CDuce{} compiler shows to the programmer a default value for the type $t_1\setminus t_2$. Some heuristics are used to build a value in which only the part relevant to the error is detailed.}
Moreover, at a more foundational level, the result nicely formalizes the intuition
that types statically approximate computations, in the sense that a type $ t $ corresponds to the set of all possible values of expressions of type $ t $.

In the following we discuss how an analogous result could hold with a non-strict semantics.
%
First of all, clearly
the correspondence cannot be between interpretations of types and sets of values as in \citeauthor{Frisch2008}, since then we would identify $ \bot $ with $ \Empty$.
Hence we should consider, rather than values, sets of ``results'' of some kind,
including (a representation of) divergence.

However, whichever notion of result we consider, it is hard to define an interpretation domain of types such that the desired correspondence holds, that is, such that a type $ t $ corresponds to the set of all possible results of expressions of type $ t $.
As the reader can expect, the key challenge is to provide an interpretation where an arrow type $ t_1 \to t_2 $ corresponds, as it seems sensible,
to the set of \textlambda-abstractions $
  \Setc{ (\RAAbstr{f: \Interface. x. e}) \given
    \Typing{\emptyset |- (\RAAbstr{f: \Interface. x. e}): t_1 \to t_2} }
$.
For instance,  our proposed definition of $ \Inter{} $ is sound with respect to this correspondence, but not complete, that is, not precise enough.
We devote the rest of this section to explain why and to discuss the possibility of obtaining a complete definition.
Consider the type $ \Int \to \Empty $.
By Definition~\ref{def:interpretation-of-types},
we have\vspace{-1.5mm}
\begin{align*}
  \Inter{\Int \to \Empty} & =
    \Setc{R \in \PsetFin(\Domain \times \Domain_\Omega) \given
      \forall (d, d') \in R. \: d \in \Inter{\Int} \implies d' \in \Inter{\Empty}}
  \\[-1mm] & =
  \Setc{R \in \PsetFin(\Domain \times \Domain_\Omega) \given
    \forall (d, d') \in R. \: d \notin \Inter{\Int}}\\[-7.6mm]
\end{align*}
(since $ \Inter{\Empty} = \emptyset $,
the implication can only be satisfied if $ d \notin \Inter{\Int} $).
This type is not empty, therefore,
if a result similar to that of \citet{Frisch2008} held,
we would expect to be able to find a function $ \RAAbstr{f: \Interface. x. e} $
such that $
  \Typing{\emptyset |- (\RAAbstr{f: \Interface. x. e}): \Int \to \Empty}
$.
  Alas, no such function can be defined in our language.
This is easy to check:
interfaces must include $ \bot $ in the codomain of every arrow
(since they use the $ \toBot $ form),
so no interface can be a subtype of $ \Int \to \Empty $.
Lifting this syntactic restriction to allow any arrow type in interfaces
would not solve the problem:
for a function to have type $ \Int \to \Empty $,
its body must have type $ \Empty $, which is impossible,
and indeed \emph{must} be impossible for the system to be sound.
It is therefore to be expected that $ \Int \to \Empty $
is uninhabited in the language.
This means that our current definition of $ \Inter{\Int \to \Empty} $
as a non-empty type is imprecise.

Changing $ \Inter{} $ to make the types of the form $ t \to \Empty $ empty
is easy,
but it does not solve the problem in general.
Using intersection types we can build more challenging examples:
for instance, consider the type $
  (\Int{\lor}\Bool \to \Int) \land (\Int{\lor}\K{String} \to \Bool)
$. While neither codomain is empty, and neither arrow should be empty,
the whole intersection should:
no function, when given an $ \Int $ as argument,
can return a result which is both an $ \Int $ and a $ \Bool $.

In the call-by-value case,
it makes sense to have $ \Int \to \Empty $ and the intersection type above
be non-empty,
because they are inhabited by functions that diverge on integers.
This is because divergence is not represented in the types
(or, to put it differently, because it is represented by the type $ \Empty $).
A type like $ t_1 \to t_2 $ is interpreted
as a specification of \emph{partial correctness}:
a function of this type, when given an argument in $ t_1 $,
either diverges or returns a result in $ t_2 $.
In our system, we have introduced a separate non-empty type for divergence.
Hence, we should see a type as specifying \emph{total} correctness,
where divergence is allowed only for functions whose codomain includes $ \bot $.

Let us look again at the current interpretation of arrow types.
\[
  \Inter{t_1 \to t_2} =
    \Setc{R \in \PsetFin(\Domain \times \Domain_\Omega) \given
      \forall (d, d') \in R. \: d \in \Inter{t_1} \implies d' \in \Inter{t_2}}
\]
An arrow type is seen as a set of finite relations:
we represent functions extensionally
and approximate them with all their finite representations. We use relations instead of functions to account for non-determinism.
Within a relation,
a pair $ (d, d') $ means that the function
returns the output $ d' $ on the input $ d $;
a pair $ (d, \Omega) $ that the function crashes on $ d $;
divergence is represented simply by the absence of a pair.
In this way, as said above, a function diverging on some element of $\Inter{t_1}$ could erroneously belong to the set
even if $ \Inter{t_2} $ does not contain $ \bot $.

To formalize the requirement of totality on the domain, we could modify the definition in this way:
\[
  \Inter{t_1 \to t_2} =
    \Setc{R \in \PsetFin(\Domain \times \Domain_\Omega) \given
      \Dom(R) \supseteq \Inter{t_1}
      \textup{ and }
      \forall (d, d') \in R. \: d \in \Inter{t_1} \implies d' \in \Inter{t_2}}
\]
 (where $ \Dom(R) = \Setc{ d \given \exists d'{\in}\Domain . \: (d, d') \in R } $).

However, if we consider only finite relations as above, the definition makes no sense, since $ \Inter{t_1} \subseteq \Dom(R) $
can hold only when $ \Inter{t_1} $ is finite,
whereas types can have infinite interpretations.
On the contrary, if we allowed relations to be infinite, then the set $ \Domain $
would have to satisfy the equality $
  \Domain =
    \Constants \uplus (\Domain \times \Domain) \uplus
    \Pset(\Domain \times \Domain_\Omega)
$ (where $\uplus$ denotes disjoint union), but no such set exists:
the cardinality of $ \Pset(\Domain \times \Domain_\Omega) $
is always strictly greater than that of $ \Domain $.

Frisch, Castagna and Benzaken~\citet{Frisch2008} point out this problem
and use finite relations in the domain to avoid it.
They motivate this choice with the observation
that,
while finite relations are not really appropriate to describe functions
in a language (since these might have an infinite domain),
they are suitable to describe types
as far as subtyping is concerned.
Indeed, we do not really care what the elements in the interpretation
of a type are,
but only how they are related to those in the interpretations of other types.
It can be shown that
\[
  \forall t_1, t_1', t_2, t_2' . \enspace
  \Inter{t'_1 \to t_1} \subseteq \Inter{t'_2 \to t_2}
  \iff
  (\Inter{t'_1} \rightharpoonup \Inter{t_1}) \subseteq
  (\Inter{t'_2} \rightharpoonup \Inter{t_2})
\]
where $
  X \rightharpoonup Y \eqdef
    \Setc{R \in \Pset(\Domain \times \Domain_\Omega) \given
      \forall (d, d') \in R. \: d \in X \implies d' \in Y}
$ builds the set of possibly infinite relations.
This can be generalized to more complex types:
\[
  \textstyle
  \Inter[\big]{ \bigwedge_{i \in P} t'_i \to t_i } \subseteq
  \Inter[\big]{ \bigvee_{i \in N} t'_i \to t_i }
  \iff
  \textstyle
  \bigcap_{i \in P} \big( \Inter{t'_i} \rightharpoonup \Inter{t_i} \big) \subseteq
  \bigcup_{i \in N} \big( \Inter{t'_i} \rightharpoonup \Inter{t_i} \big)
  \: .
\]

In \citet{Frisch2008}, the authors argue
that the restriction to finite relations
does not compromise the precision of subtyping.
For reasons of space we do not elaborate further on this,
and we direct the interested reader to their work
and the notions of \emph{extensional interpretation} and of \emph{model} therein.

Let us try to proceed analogously in our case: that is, find a new interpretation of types
that matches the behaviour of possibly infinite relations that are total on their domain, while introducing an approximation to ensure that the domain is definable.
The latter point means, notably,
that functions must be represented as finite objects.
The following definition of a \emph{model}
specifies the properties that such an interpretation should satisfy.

\begin{Definition}[Model]
  A function $ \InterAlt{} : \Types \to \Pset(\DomainAlt) $
  is a \emph{model}
  if the following hold:
  \begin{itemize}
    \item
    the set $ \DomainAlt $
    satisfies $
      \DomainAlt =
        \Set{\bot} \uplus \Constants \uplus (\DomainAlt \times \DomainAlt)
        \uplus \DomainAlt^\mathsf{fun}
    $ for some set $ \DomainAlt^\mathsf{fun} $;
    \item
    for all $ b $, $ t $, $ t_1 $, and $ t_2 $,
    \begin{gather*}
      \InterAlt{\bot} = \Set{\bot} \qquad
      \InterAlt{b} = \ConstantsInBasicType(b) \qquad
      \InterAlt{t_1 \times t_2} = \InterAlt{t_1} \times \InterAlt{t_2} \qquad
      \InterAlt{t_1 \to t_2} \subseteq
        \InterAlt{\Empty \to \Any} = \DomainAlt^\mathsf{fun} \\
      \InterAlt{t_1 \lor t_2} = \InterAlt{t_1} \cup \InterAlt{t_2} \qquad
      \InterAlt{\lnot t} = \DomainAlt \setminus \InterAlt{t} \qquad
      \InterAlt{\Empty} = \emptyset
    \end{gather*}
    \item
    for every finite, non-empty intersection $ \bigwedge_{i \in P} t'_i \to t_i $
    and every finite union $ \bigvee_{i \in N} t'_i \to t_i $,
  \[
    \textstyle
    \InterAlt{ \bigwedge_{i \in P} t'_i \to t_i } \subseteq
    \InterAlt{ \bigvee_{i \in N} t'_i \to t_i }
    \iff
    \textstyle
    \bigcap_{i \in P} \big( \InterAlt{t'_i} \boldto \InterAlt{t_i} \big) \subseteq
    \bigcup_{i \in N} \big( \InterAlt{t'_i} \boldto \InterAlt{t_i} \big)
  \]
  where $
    X \boldto Y \eqdef
      \Setc{R \in \Pset(\DomainAlt \times \DomainAlt) \given
        \Dom(R) \supseteq X
        \textup{ and }
        \forall (d, d') \in R. \: d \in X \implies d' \in Y}
  $.
  \end{itemize}
\end{Definition}

We set three conditions for an interpretation of types $
  \InterAlt{} : \Types \to \Pset(\DomainAlt)
$ to be a model.
The first constrains $ \DomainAlt $
to have the same structure as $ \Domain $,
except that we do not fix the subset $ \DomainAlt^\mathsf{fun} $
in which arrow types are interpreted.
The second condition fixes the definition of $ \InterAlt{} $ completely
except for arrow types.
The third condition ensures that subtyping on arrow types
behaves as set containment between the sets of relations
that are total on the domains of the arrow types.%
  \footnote{%
    We do not use the error element $ \Omega $
    in the definition of $ X \boldto Y $,
    because the totality requirement makes it unnecessary:
    errors on a given input can be represented in a relation
    by the absence of a pair.
  }

An interesting result is that, even though we do not know whether an interpretation of types which is a model can be actually found, we can
compare such hypothetical model
with the interpretation $ \Inter{} $ defined in Section~\ref{sec:types}.
Indeed $ \Inter{} $ turns out to be a sound approximation
of every model;
that is, the subtyping relation $ \leq $ defined in Definition~\ref{def:subtyping} from $ \Inter{} $
is contained in every subtyping relation $ \leq_{\llparenthesis\,\rrparenthesis} $
defined from some model $ \InterAlt{} $.
We have proven that this holds for non-recursive types:

\begin{Proposition}
  Let $ \InterAlt{} : \Types \to \Pset(\DomainAlt) $ be a model.
  Let $ t_1 $ and $ t_2 $ be two finite (i.e., non-recursive) types.
  If $ \Inter{t_1} \subseteq \Inter{t_2} $,
  then $ \InterAlt{t_1} \subseteq \InterAlt{t_2} $.
\end{Proposition}

We conjecture that the result holds for recursive types too,
but this proof is left for future work.

Showing that $ \InterTot{} $ exists
would be important to understand the connection between our types
and the semantics.
To use $ \InterTot{} $ to define subtyping for the use of a typechecker, though,
we would also need to show that the resulting definition is decidable.
Otherwise, $ \Inter{} $ would remain the definition used
in a practical implementation
since it is sound and decidable, though less precise.

\section{Conclusion}
\label{sec:conclusion}

We have shown
how to adapt the framework of semantic subtyping~\cite{Frisch2008}
to languages with non-strict semantics.
Our type system
uses the subtyping relation from \cite{Frisch2008} unchanged
(except for the addition of $ \bot $),
while the typing rules are reworked
to avoid the pathological behaviour of semantic subtyping on empty types.
Notably, typing rules for constructs
like application and projection must handle $ \bot $ explicitly.
This ensures soundness for call-by-need.

This approach ensures
that the subtyping relation still behaves set-theoretically:
we can still see union, intersection, and negation in types
as the corresponding operations on sets.
We can still use intersection types to express overloading.

The type $ \bot $ we introduce
has no analogue in well-known type systems
like the simply typed \textlambda-calculus or Hindley-Milner typing.
However, $ \bot $ never appears explicitly in programs
(it does not appear in types of the forms $ T $ and $ \tau $
given at the beginning of Section~\ref{pag:types-big-t-and-tau}).
Hence, programmers do not need to use it and to consider the difference
between terminating and non-terminating types
while writing function interfaces or typecases.
Still, sub-expressions of a program can have types with explicit $ \bot $
(e.g., the type $ \Int \lor \bot $).
Such types are not expressible in the grammar of types visible to the programmer.
Accordingly, error reporting is required to be more elaborated, to avoid mentioning
internal types that are unknown to the programmer.

A different approach to use semantic subtyping with non-strict languages
would be to change the interpretation of types
(and, as a result, the definition of subtyping)
to avoid the pathological behaviour on $ \Empty $,
and then to use standard typing rules.

We have explored this alternative approach,
but we have not found it promising.
A modified subtyping relation loses important properties
-- especially results on the decomposition of product types --
that we need to prove soundness via subject reduction.
The approach we have adopted here is more suited to this technical work.
However, a modified subtyping relation could yield an alternative type system
for the source language,
provided that we can relate it to the current system for the internal language.

We also plan to study more expressive typing rules
that can track termination with some precision.
For example, we could change the application rule
so that it does not always introduce $ \bot $.
In function interfaces, some arrows could include $ \bot $ and some could not:
then, overloaded function types would express
that a function behaves differently on terminating or diverging arguments.
For example, the function $ \Abstr{x. x + \K{1}} $
could have type $ (\Int \to \Int) \land (\bot \to \bot) $,
while $ \Abstr{x. \K{3}} $ could have type $ \Any \to \Int $:
the first diverges on diverging arguments, the other always terminates.
It would be interesting for future work to explore forms of termination analysis to obtain greater precision.
The difficulty is to ensure that the type $ \Empty $ remains uninhabited
and that all diverging expressions still have types that include $ \bot $.
This is trivial in the current system,
but it is no longer straightforward with more refined typing rules.

A further direction for future work is
to extend the language and the type system we have considered
with more features.
Notably, polymorphism, gradual typing, and record types
are needed to be able to type effectively the Nix Expression Language,
which was the starting inspiration for our work.



\bibliography{main}

\ifarxiv
  \newpage
  \appendix
  \section{Appendix}
  \subsection{Properties of subtyping and type decompositions}

\begin{Lemma}
  \label{lem:subtyping-decomposition-arrows}
  \begin{multline*}
    \bigwedge_{i \in I} t'_i \to t_i \leq \bigvee_{j \in J} t'_j \to t_j
    \iff \\
    \exists j_0 \in J . \enspace
    \Bigg( t'_{j_0} \leq \bigvee_{i \in I} t'_i \Bigg)
    \land
    \Bigg(
      \forall I' \subsetneq I . \enspace
        \Big( t'_{j_0} \leq \bigvee_{i \in I'} t'_i \Big)
        \lor
        \Big( \bigwedge_{i \in I \setminus I'} t_i \leq t_{j_0} \Big)
    \Bigg)
  \end{multline*}
\end{Lemma}

\begin{Proof}
Consequence of Lemma~6.8 of \citet{Frisch2008}.
\end{Proof}

\begin{Corollary}
  \label{lem:subtyping-decomposition-arrows-disjoint}
  Let $
    \bigwedge_{i \in I} t'_i \to t_i
  $ (with $ |I| > 0 $)
  be such that $ t'_{i_1} \land t'_{i_2} \simeq \Empty $ for $ i_1 \neq i_2 $.
  Then:
  \[
    \bigwedge_{i \in I} t'_i \to t_i \leq t' \to t
    \implies
    \Big( t' \leq \bigvee_{i \in I} t'_i \Big)
    \land
    \Big(
      \forall i \in I . \enspace
        \big( t'_i \land t' \not\simeq \Empty \big)
        \implies
        \big( t_i \leq t \big)
    \Big)
  \]
\end{Corollary}

\begin{Proof}
  By applying Lemma~\ref{lem:subtyping-decomposition-arrows}, we get
  \[
    \big( t' \leq \textstyle\bigvee_{i \in I} t'_i \big)
    \land
    \big(
      \forall I' \subsetneq I . \enspace
        ( t' \leq \textstyle\bigvee_{i \in I'} t'_i )
        \lor
        ( \textstyle\bigwedge_{i \in I \setminus I'} t_i \leq t )
    \big)
    \: .
  \]
  Now consider an arbitrary $ i_0 \in I $ such that $
    t'_{i_0} \land t' \not\simeq \Empty
  $; we must show $ t_{i_0} \leq t $.
  Instantiating the quantifier above with $ I' = I \setminus \Set{i_0} $
  we get $
    (t' \leq \textstyle\bigvee_{i \in I \setminus \Set{i_0}} t'_i)
    \lor
    (\textstyle\bigwedge_{i \in I \setminus (I \setminus \Set{i_0})} t_i \leq t)
  $.
  We show $
    t' \not\leq \textstyle\bigvee_{i \in I \setminus \Set{i_0}} t'_i
  $, which concludes the proof
  since the second term of the union is $ t_{i_0} \leq t $.

  By contradiction, assume $
    t' \leq \textstyle\bigvee_{i \in I \setminus \Set{i_0}} t'_i
  $.
  Note that $
    t'_{i_0} \land \textstyle\bigvee_{i \in I \setminus \Set{i_0}} t'_i
    \simeq \Empty
  $ (because the $ t'_i $ are disjoint);
  therefore we would also have $
    t'_{i_0} \land t \simeq \Empty
  $, which is false by hypothesis.
\end{Proof}

\begin{Corollary}
\label{cor:subtyping-negative-arrows}
  Let $
    \bar{t} =
    (\bigwedge_{i \in I} t'_i \to t_i) \land
    (\bigwedge_{j \in J} \lnot (t'_j \to t_j))
  $. If $ \bar{t} \not\simeq \Empty $
  and $ \bar{t} \leq t' \to t $, then $
    (\bigwedge_{i \in I} t'_i \to t_i) \leq t' \to t
  $.
\end{Corollary}

\begin{Proof}
  By definition of subtyping, we have
  \begin{align*}
      \bar{t} \not\simeq \Empty
    & \iff
      (\textstyle\bigwedge_{i \in I} t'_i \to t_i) \land
      (\textstyle\bigwedge_{j \in J} \lnot (t'_j \to t_j))
      \not\leq \Empty
    \iff
      \textstyle\bigwedge_{i \in I} t'_i \to t_i
      \not\leq
      \textstyle\bigvee_{j \in J} t'_j \to t_j
    \\
      \bar{t} \leq t' \to t
    & \iff
      \textstyle\bigwedge_{i \in I} t'_i \to t_i
      \leq
      (\textstyle\bigvee_{j \in J} t'_j \to t_j)
      \lor (t' \to t)
  \end{align*}
  Let $ \bar{j} $ be such that $ \bar{j} \notin J $
  and let $ t'_{\bar{j}} = t' $ and $ t_{\bar{j}} = t $.
  By Lemma~\ref{lem:subtyping-decomposition-arrows},
  we derive
  \begin{gather*}
    \forall j_0 \in J . \enspace
    \lnot \Big(
    \big( t'_{j_0} \leq \textstyle\bigvee_{i \in I} t'_i \big)
    \land
    \big(
      \forall I' \subsetneq I . \enspace
        ( t'_{j_0} \leq \textstyle\bigvee_{i \in I'} t'_i )
        \lor
        ( \textstyle\bigwedge_{i \in I \setminus I'} t_i \leq t_{j_0} )
    \big)
    \Big)
    \\
    \exists j_0 \in J \cup \Set{\bar{j}} . \enspace
    \big( t'_{j_0} \leq \textstyle\bigvee_{i \in I} t'_i \big)
    \land
    \big(
      \forall I' \subsetneq I . \enspace
        ( t'_{j_0} \leq \textstyle\bigvee_{i \in I'} t'_i )
        \lor
        ( \textstyle\bigwedge_{i \in I \setminus I'} t_i \leq t_{j_0} )
    \big)
  \end{gather*}
  where clearly the existentially quantified proposition
  must be true for $ \bar{j} $,
  which allows us to conclude.
\end{Proof}

\begin{Lemma}
  \label{lem:union-of-all-subsets-is-any}
  For every set $ J $
  and every set $ \Setc{t_j \given j \in J} $,
  \[
    \textstyle\bigvee_{J' \subseteq J} \big(
      \textstyle\bigwedge_{j \in J'} t_j \land
      \textstyle\bigwedge_{j \in J \setminus J'} \lnot t_j
    \big)
    \simeq \Any
  \]
  (with the convention that an intersection over an empty set is $ \Any $).
\end{Lemma}

\begin{Proof}
  We prove this by induction on $ |J| $.
  If $ |J| = 0 $, then the only $ J' $ is $ J $ itself,
  and the equivalence holds.
  If $ |J| > 0 $, consider an arbitrary $ j_0 \in J $
  and let $ \bar{J} = J \setminus \Set{j_0} $.
  We have
  \begin{align*}
    & \textstyle\bigvee_{J' \subseteq J} \big(
      \textstyle\bigwedge_{j \in J'} t_j \land
      \textstyle\bigwedge_{j \in J \setminus J'} \lnot t_j
    \big)
    \\ \simeq {} &
    \textstyle\bigvee_{J' \subseteq \bar{J}} \big(
      \textstyle\bigwedge_{j \in J'} t_j \land
      \textstyle\bigwedge_{j \in \bar{J} \setminus J'} \lnot t_j
      \land \lnot t_{j_0}
    \big) \lor
    \textstyle\bigvee_{J' \subseteq \bar{J}} \big(
      t_{j_0} \land
      \textstyle\bigwedge_{j \in J'} t_j \land
      \textstyle\bigwedge_{j \in \bar{J} \setminus J'} \lnot t_j
    \big)
    \\ \simeq {} &
    \Big( \lnot t_{j_0} \land
    \textstyle\bigvee_{J' \subseteq \bar{J}} \big(
      \textstyle\bigwedge_{j \in J'} t_j \land
      \textstyle\bigwedge_{j \in \bar{J} \setminus J'} \lnot t_j
    \big) \Big) \lor
    \Big( t_{j_0} \land
    \textstyle\bigvee_{J' \subseteq \bar{J}} \big(
      \textstyle\bigwedge_{j \in J'} t_j \land
      \textstyle\bigwedge_{j \in \bar{J} \setminus J'} \lnot t_j
    \big) \Big)
    \\ \simeq {} &
    (\lnot t_{j_0} \land \Any) \lor (t_{j_0} \land \Any)
    \intertext{(by the induction hypothesis)}
    \simeq {} & \Any
    \: .
    \qedhere
  \end{align*}
\end{Proof}

\begin{Lemma}
\label{lem:interface-disjoint}
  Let $
    \Interface = \bigwedge_{i \in I} t_i' \to t_i
  $ (with $ |I| > 0 $) be a type.
  Then:
  \[
    \Interface \simeq
    \textstyle\bigwedge_{\emptyset \subsetneq I' \subseteq I} s_{I'} \to u_{I'}
    \qquad \text{where\:\:}
    s_{I'} \eqdef \textstyle\bigwedge_{i \in I'} t_i' \land
      \textstyle\bigwedge_{i \in I \setminus I'} \lnot t_i'
    \text{\:\:and\:\:}
    u_{I'} \eqdef \textstyle\bigwedge_{i \in I'} t_i
  \]
  (with the convention: $
    \bigwedge_{i \in \emptyset} \lnot t_i' = \Any
  $).
\end{Lemma}

\begin{Proof}
  We first show $
    \Interface \leq
    \bigwedge_{\emptyset \subsetneq I' \subseteq I} s_{I'} \to u_{I'}
  $.
  To do this, we show that, for every $ I' $ such that $
    \emptyset \subsetneq I' \subseteq I
  $, we have $
    \Interface \leq s_{I'} \to u_{I'}
  $, that is,
  \[
    \Interface \leq
      \big( \textstyle\bigwedge_{i \in I'} t_i' \land
      \textstyle\bigwedge_{i \in I \setminus I'} \lnot t_i' \big)
      \to
      \big( \textstyle\bigwedge_{i \in I'} t_i \big)
    \: .
  \]
  We have
  \begin{align*}
    \Interface & = \textstyle\bigwedge_{i \in I} t_i' \to t_i \\
      & \leq \textstyle\bigwedge_{i \in I'} t_i' \to t_i \\
      & \leq \big( \textstyle\bigwedge_{i \in I'} t_i' \big) \to
        \big( \textstyle\bigwedge_{i \in I'} t_i \big)
        & \text{by definition of subtyping for arrows} \\
      & \leq \big( \textstyle\bigwedge_{i \in I'} t_i' \land
          \textstyle\bigwedge_{i \in I \setminus I'} \lnot t_i' \big) \to
        \big( \textstyle\bigwedge_{i \in I'} t_i \big)
      \: .
  \end{align*}

  We now consider the opposite direction.
  To show $
    \bigwedge_{\emptyset \subsetneq I' \subseteq I} s_{I'} \to u_{I'} \leq
    \Interface
  $,
  we show that, for every $ i \in I $, we have $
    \bigwedge_{\emptyset \subsetneq I' \subseteq I} s_{I'} \to u_{I'}
    \leq t_i' \to t_i
  $.
  Consider an arbitrary $ i_0 \in I $ and let $
    \overline{I} = I \setminus \Set{i_0}
  $.
  We have
  \begin{align*}
    \textstyle\bigwedge_{\emptyset \subsetneq I' \subseteq I} s_{I'} \to u_{I'}
    & \leq
    \textstyle\bigwedge_{
      \substack{\emptyset \subsetneq I' \subseteq I \\ i_0 \in I'}}
      s_{I'} \to u_{I'}
    \\ & \simeq
    \textstyle\bigwedge_{\substack{I' \subseteq \overline{I}}}
      s_{(I' \cup \Set{i_0})} \to u_{(I' \cup \Set{i_0})}
    \\ & \simeq
    \textstyle\bigwedge_{\substack{I' \subseteq \overline{I}}}
      \Big(
      \big( t'_{i_0} \land \bigwedge_{i \in I'} t'_i
        \land \bigwedge_{i \in \overline{I} \setminus I'} \lnot t'_i
      \big)
      \to
      \big( t_{i_0} \land \bigwedge_{i \in I'} t_i \big)
      \Big)
    \\ & \leq
      \Big(
        \textstyle\bigvee_{\substack{I' \subseteq \overline{I}}}
        \big( t'_{i_0} \land \bigwedge_{i \in I'} t'_i
        \land \bigwedge_{i \in \overline{I} \setminus I'} \lnot t'_i
      \big)
      \Big)
      \to
      \Big(
        \textstyle\bigvee_{\substack{I' \subseteq \overline{I}}}
        \big( t_{i_0} \land \bigwedge_{i \in I'} t_i \big)
      \Big)
    \\ & \simeq
      \Big(
      t'_{i_0} \land
        \textstyle\bigvee_{\substack{I' \subseteq \overline{I}}}
        \big( \bigwedge_{i \in I'} t'_i
        \land \bigwedge_{i \in \overline{I} \setminus I'} \lnot t'_i
      \big)
      \Big)
      \to
      \Big(
      t_{i_0} \land
        \textstyle\bigvee_{\substack{I' \subseteq \overline{I}}}
        \big(\bigwedge_{i \in I'} t_i \big)
      \Big)
    \\ & \simeq
      t'_{i_0}
      \to
      \Big(
      t_{i_0} \land
        \textstyle\bigvee_{\substack{I' \subseteq \overline{I}}}
        \big(\bigwedge_{i \in I'} t_i \big)
      \Big)
    \intertext{(by Lemma~\ref{lem:union-of-all-subsets-is-any})}
    & \leq
    t'_{i_0} \to t_{i_0} \: .
    \qedhere
  \end{align*}
\end{Proof}

\begin{Definition}[Product decomposition]
  \label{def:product-decomposition}
  A \emph{product decomposition} $ \Pi $
  is a finite set of \emph{product atoms}, that is,
  of types of the form $ t_1 \times t_2 $.

  We say that a product decomposition $
    \Pi = \Setc{ t^1_i \times t^2_i \given i \in I }
  $ is \emph{fully disjoint} if $
    \forall i \in I . \: t^1_i \times t^2_i \not\simeq \Empty
  $ and if the following conditions hold
  for all $ i_1 \neq i_2 \in I $:
  \begin{itemize}
    \item $
      (t^1_{i_1} \land t^1_{i_2} \simeq \Empty) \lor
      (t^1_{i_1} \simeq t^1_{i_2}) $;
    \item $
      (t^2_{i_1} \land t^2_{i_2} \simeq \Empty) \lor
      (t^2_{i_1} \simeq t^2_{i_2}) $.
  \end{itemize}
\end{Definition}

\begin{Lemma}
  \label{lem:product-decomposition-exists}
  For every type $ t $ such that $ t \leq \Any \times \Any $,
  there exists a product decomposition $ \Pi $
  such that $ t \simeq \bigvee_{t_1 \times t_2 \in \Pi} t_1 \times t_2 $.
\end{Lemma}

\begin{Lemma}
  \label{lem:product-decomposition-disjoint}
  For every product decomposition $ \Pi $,
  there exists a product decomposition $ \Pi' $
  such that $ \Pi' $ is fully disjoint,
  that $
    \bigvee_{t \in \Pi} t \simeq \bigvee_{t' \in \Pi'} t'
  $, and that $
    \forall t' \in \Pi . \: \exists t \in \Pi . \: t' \leq t
  $.
\end{Lemma}

\begin{Proof}
  Let $ \Pi = \Setc{ t^1_i \times t^2_i \given i \in I } $.
  When $ i \in I $ and $ I', I_1, I_2 \subseteq I $, and $ k \in \Set{1, 2} $,
  we define
  \[
    \textstyle
    \mathbb{T}^k(i, I') = t^k_i \land \bigwedge_{j \in I'} t^k_j
      \land \bigwedge_{j \in I \setminus \Set{i} \setminus I'} \lnot t^k_j
    \qquad
    \mathbb{T}(i, I_1, I_2) = \mathbb{T}^1(i, I_1) \times \mathbb{T}^2(i, I_2)
  \]
  and we consider the product decomposition
  \[
    \textstyle
    \Pi' = \bigcup_{i \in I}
    \Setc{ \mathbb{T}(i, I_1, I_2) \given
      I_1 \subseteq I \setminus \Set{i},
      I_2 \subseteq I \setminus \Set{i},
      \mathbb{T}^1(i, I_1) \not\simeq \Empty,
      \mathbb{T}^2(i, I_2) \not\simeq \Empty }
    \: .
  \]

  We first show that $ \Pi' $ is fully disjoint.
  First, consider an arbitrary element of $ \Pi' $,
  $ \mathbb{T}(i, I_1, I_2) = \mathbb{T}^1(i, I_1) \times \mathbb{T}^2(i, I_2) $.
  We must show $ \mathbb{T}(i, I_1, I_2) \not\simeq \Empty $, which holds
  because we explicitly require both $ \mathbb{T}^k(i, I_k) $ to be non-empty.
  Now, we consider two arbitrary elements of $ \Pi' $:
  \[
    \mathbb{T}(i, I_1, I_2) = \mathbb{T}^1(i, I_1) \times \mathbb{T}^2(i, I_2)
    \qquad
    \mathbb{T}(i', I_1', I_2') =
    \mathbb{T}^1(i', I_1') \times \mathbb{T}^2(i', I_2')
  \]
  and we must prove:
  \begin{gather*}
    \big( \mathbb{T}^1(i, I_1) \land \mathbb{T}^1(i', I_1') \simeq \Empty \big)
    \lor \big( \mathbb{T}^1(i, I_1) \simeq \mathbb{T}^1(i', I_1') \big)
    \\
    \big( \mathbb{T}^2(i, I_2) \land \mathbb{T}^2(i', I_2') \simeq \Empty \big)
    \lor \big( \mathbb{T}^2(i, I_2) \simeq \mathbb{T}^2(i', I_2') \big)
    \: .
  \end{gather*}
  We prove the first (the second is proved identically).
  Note that if $
    \Set{i} \cup I_1 = \Set{i'} \cup I_1'
  $, then $ \mathbb{T}^1(i, I_1) $ and $ \mathbb{T}^1(i', I_1') $
  are the same up to reordering of the intersections:
  therefore $ \mathbb{T}^1(i, I_1) \simeq \mathbb{T}^1(i', I_1') $ holds.
  Otherwise, assume without loss of generality that there exists an $ i_0 $
  such that $ i_0 \in \Set{i} \cup I_1 $ but $ i_0 \notin \Set{i'} \cup I_1' $.
  Then, we have $ \mathbb{T}^1(i, I_1) \leq t^1_{i_0} $
  and $ \mathbb{T}^1(i', I_1') \leq \lnot t^1_{i_0} $.
  Then, $
    \mathbb{T}^1(i, I_1) \land \mathbb{T}^1(i', I_1')
    \leq t^1_{i_0} \land \lnot t^1_{i_0} \leq \Empty
  $.

  Now we show that, for every $
    \mathbb{T}(i, I_1, I_2) = \mathbb{T}^1(i, I_1) \times \mathbb{T}^2(i, I_2)
  $ in $ \Pi' $,
  there exists a $ i' \in I $
  such that $ \mathbb{T}(i, I_1, I_2) \leq t^1_{i'} \times t^2_{i'} $.
  We simply take $ i' = i $,
  since both $ \mathbb{T}^1(i, I_1) \leq t^1_i $
  and $ \mathbb{T}^2(i, I_2) \leq t^2_i $ always hold.

  Finally, we show that $
    \bigvee_{i \in I} t^1_i \times t^2_i \simeq \bigvee_{t \in \Pi'} t
  $. We do so by showing that, for every $ i \in I $,
  \[
    t^1_i \times t^2_i \simeq
    \textstyle\bigvee_{
      i \in I,
      I_1 \subseteq I \setminus \Set{i},
      I_2 \subseteq I \setminus \Set{i},
      \mathbb{T}^1(i, I_1) \not\simeq \Empty,
      \mathbb{T}^2(i, I_2) \not\simeq \Empty
    } \mathbb{T}(i, I_1, I_2)
    \: .
  \]
  Note that we can show this by showing
  \[
    t^1_i \times t^2_i \simeq
    \textstyle\bigvee_{
      i \in I,
      I_1 \subseteq I \setminus \Set{i},
      I_2 \subseteq I \setminus \Set{i}
    } \mathbb{T}(i, I_1, I_2)
    \: ,
  \]
  without the conditions of non-emptiness
  (we have more summands in the union, but they are empty).
  We have
  \begin{align*}
    & \textstyle
    \bigvee_{ i \in I,
      I_1 \subseteq I \setminus \Set{i},
      I_2 \subseteq I \setminus \Set{i}
    } \mathbb{T}(i, I_1, I_2)
    \\
    = {} &
    \textstyle
    \bigvee_{
      i \in I,
      I_1 \subseteq I \setminus \Set{i},
      I_2 \subseteq I \setminus \Set{i}
    } \mathbb{T}^1(i, I_1) \times \mathbb{T}^2(i, I_2)
    \\ \simeq {} &
    \textstyle
    \bigvee_{ i \in I, I_1 \subseteq I \setminus \Set{i} } \big(
      \bigvee_{ I_2 \subseteq I \setminus \Set{i} }
      \mathbb{T}^1(i, I_1) \times \mathbb{T}^2(i, I_2)
    \big)
    \\ \simeq {} &
    \textstyle
    \bigvee_{ i \in I, I_1 \subseteq I \setminus \Set{i} }
    \mathbb{T}^1(i, I_1) \times \big(
      \bigvee_{ I_2 \subseteq I \setminus \Set{i} }
      \mathbb{T}^2(i, I_2)
    \big)
    \intertext{(subtyping of product types satisfies $
      \bigvee_{i \in I} (t \times t_i)
      \simeq t \times (\bigvee_{i \in I} t_i)
    $)}
    \simeq {} &
    \textstyle
    \bigvee_{ i \in I, I_1 \subseteq I \setminus \Set{i} }
    \mathbb{T}^1(i, I_1) \times \Big(
      \bigvee_{ I_2 \subseteq I \setminus \Set{i} }
      \big(
        t^2_i \land \bigwedge_{j \in I_2} t^2_j \land
        \bigwedge_{j \in I \setminus \Set{i} \setminus I_2} \lnot t^2_j
      \big)
    \Big)
    \\ \simeq {} &
    \textstyle
    \bigvee_{ i \in I, I_1 \subseteq I \setminus \Set{i} }
    \mathbb{T}^1(i, I_1) \times \Big(
       t^2_i \land \bigvee_{ I_2 \subseteq I \setminus \Set{i} }
      \big(
        \bigwedge_{j \in I_2} t^2_j \land
        \bigwedge_{j \in I \setminus \Set{i} \setminus I_2} \lnot t^2_j
      \big)
    \Big)
    \\ \simeq {} &
    \textstyle
    \bigvee_{ i \in I, I_1 \subseteq I \setminus \Set{i} }
    \mathbb{T}^1(i, I_1) \times \big(
       t^2_i \land \Any \big)
    \intertext{(by Lemma~\ref{lem:union-of-all-subsets-is-any})}
    \simeq {} &
    \textstyle
    \bigvee_{ i \in I, I_1 \subseteq I \setminus \Set{i} }
    \mathbb{T}^1(i, I_1) \times t^2_i
    \\ \simeq {} &
    t^1_i \times t^2_i
  \end{align*}
  (proceeding as above).
\end{Proof}

\begin{Lemma}
  \label{lem:product-decomposition-disjoint-split}
  Let $ \Pi = \Setc{ t^1_i \times t^2_i \given i \in I } $
  be a fully disjoint product decomposition
  and let $ t^1 $ and $ t^2 $ be two types
  such that $ t^1 \times t^2 \simeq \bigvee_{i \in I} t^1_i \times t^2_i $.
  Then, $ t^1 \simeq \bigvee_{i \in I} t^1_i $,
  $ t^2 \simeq \bigvee_{i \in I} t^2_i $,
  and $
    \forall i_1, i_2 \in I . \: \exists i \in I . \:
    t^1_{i_1} \times t^2_{i_2} \leq t^1_i \times t^2_i
  $.
\end{Lemma}

\begin{Proof}
  We have
  \[
    t^1 \times t^2
    \simeq
    \textstyle\bigvee_{i \in I} t^1_i \times t^2_i
    \leq
    \big( \textstyle\bigvee_{i \in I} t^1_i \big) \times
      \big( \textstyle\bigvee_{i \in I} t^2_i \big)
  \]
  and therefore $ t^1 \leq \bigvee_{i \in I} t^1_i $
  and $ t^2 \leq \textstyle\bigvee_{i \in I} t^2_i $,
  since all $ t^1_i $ and $ t^2_i $ are non-empty.

  Since $
    \textstyle\bigvee_{i \in I} t^1_i \times t^2_i
    \leq t^1 \times t^2
  $, we have, for all $ i \in I $, $
    t^1_i \times t^2_i \leq t^1 \times t^2
  $ and hence
  (by definition of subtyping, since $ t^1_i $ and $ t^2_i $ are non-empty)
  $ t^1_i \leq t^1 $ and $ t^2_i \leq t^2 $.
  Hence, we also have $
    \bigvee_{i \in I} t^1_i \leq t^1
  $ and $
    \bigvee_{i \in I} t^2_i \leq t^2
  $.
  This yields $
    t^1 \simeq \bigvee_{i \in I} t^1_i
  $ and $
    t^2 \simeq \bigvee_{i \in I} t^2_i
  $.

  To prove $
    \forall i_1, i_2 \in I . \: \exists i \in I . \:
    t^1_{i_1} \times t^2_{i_2} \leq t^1_i \times t^2_i
  $, we consider arbitrary $ i_1 $ and $ i_2 $ in $ I $;
  we must show $
    \exists i \in I . \: t^1_{i_1} \times t^2_{i_2} \leq t^1_i \times t^2_i
  $.
  Note that $ t^1_{i_1} \times t^2_{i_2} \leq t^1 \times t^2 $.
  Hence, we have
  \begin{align*}
    t^1_{i_1} \times t^2_{i_2}
    & \simeq
    (t^1_{i_1} \times t^2_{i_2}) \land (t^1 \times t^2)
    \\ & \simeq
    (t^1_{i_1} \times t^2_{i_2}) \land
      (\textstyle\bigvee_{i \in I} t^1_i \times t^2_i)
    \\ & \simeq
    \textstyle\bigvee_{i \in I}
      \big( (t^1_{i_1} \times t^2_{i_2}) \land (t^1_i \times t^2_i) \big)
    \\ & \simeq
    \textstyle\bigvee_{i \in I}
      \big( (t^1_{i_1} \land t^1_i) \times (t^2_{i_2} \land t^2_i) \big)
    \: .
  \end{align*}
  Since $ t^1_{i_1} \times t^2_{i_2} $ is not empty,
  there must exist an $ i_0 \in I $ such that $
    (t^1_{i_1} \land t^1_{i_0}) \times (t^2_{i_2} \land t^2_{i_0})
  $ is not empty, that is, an $ {i_0} $ such that
  $ t^1_{i_1} \land t^1_{i_0} \not\simeq \Empty $
  and $ t^2_{i_2} \land t^2_{i_0} \not\simeq \Empty $.
  Since the decomposition is fully disjoint, we have
  $ t^1_{i_1} \simeq t^1_{i_0} $ and $ t^2_{i_2} \simeq t^2_{i_0} $:
  therefore $ t^1_{i_1} \times t^2_{i_2} \simeq t^1_{i_0} \times t^2_{i_0} $.
\end{Proof}

\subsection{Syntactic meta-theory of the type system of the internal language}

\begin{Lemma}[Weakening]
\label{lem:typing-weakening}
  Let $ \Gamma $ and $ \Gamma' $ be two type environments
  such that, whenever $ \Gamma(x) = t $,
  $ x \in \Dom(\Gamma') $ and $ \Gamma'(x) \leq t $.
  If $ \Typing{\Gamma |- e: t} $, then $ \Typing{\Gamma' |- e: t} $.
\end{Lemma}

\begin{Proof}
  For every $ \Gamma $ and $ \Gamma' $, we define
  \[
    \Gamma \leq \Gamma'
    \iffdef
    \forall x \in \Dom(\Gamma) . \:
      (x \in \Dom(\Gamma')) \land (\Gamma'(x) \leq \Gamma(x))
    \: .
  \]

  We prove that, if $ \Typing{\Gamma |- e: t} $ and $ \Gamma' \leq \Gamma $,
  then $ \Typing{\Gamma' |- e: t} $.
  We proceed by induction on the derivation of $ \Typing{\Gamma |- e: t} $
  and by case on the last rule applied.

  If the last rule is \Rule{Var},
  we show the result by applying \Rule{Var} and \Rule{Subsum}.
  For \Rule{Const}, the result is straightforward.
  For \Rule{Abstr}, we can assume by \textalpha-renaming
  that $ f $ and $ x $ do not appear in $ \Gamma $ and $ \Gamma' $;
  then, we have (for all $ i $) $
    (\Gamma', f\colon \Interface, x\colon \WithBot{T'_i})
    \leq
    (\Gamma, f\colon \Interface, x\colon \WithBot{T'_i})
  $ and we apply the induction hypothesis to conclude.
  We proceed similarly for \Rule{Case} and \Rule{Let}.
  For \Rule{Appl}, \Rule{Pair}, \Rule{Proj}, and \Rule{Subsum},
  we directly apply the induction hypothesis.
\end{Proof}

\begin{Lemma}[Intersection introduction]
\label{lem:typing-intersection-in}
  If $ \Typing{\Gamma |- e: t_1} $ and $ \Typing{\Gamma |- e: t_2} $,
  then $ \Typing{\Gamma |- e: t_1 \land t_2} $.
\end{Lemma}

\begin{Proof}
  By induction on the derivations
  of $ \Typing{\Gamma |- e: t_1} $ and of $ \Typing{\Gamma |- e: t_2} $.
  As a measure we use the sum of the depth of the two derivations.

  If the last rule applied in the derivation of $ \Typing{\Gamma |- e: t_1} $
  is \Rule{Subsum},
  we have $ \Typing{\Gamma |- e: t_1'} $ and $ t_1' \leq t_1 $.
  We apply the induction hypothesis
  to $ \Typing{\Gamma |- e: t_1'} $ and $ \Typing{\Gamma |- e: t_2} $
  to derive $ \Typing{\Gamma |- e: t_1' \land t_2} $
  and then apply \Rule{Subsum} since $ t_1' \land t_2 \leq t_1 \land t_2 $.
  If the last rule applied for $ e_1 $ is not \Rule{Subsum},
  and that for $ e_2 $ is, we do the reverse.

  Having dealt with the cases where the last rule applied
  in one derivation at least is \Rule{Subsum},
  we can assume for the remainder that the derivations end with the same rule:
  every derivation for $ e $
  must end with the application of the rule corresponding to the form of $ e $,
  possibly followed by applications of \Rule{Subsum}.

  If the last rule applied is \Rule{Var} or \Rule{Const},
  we have $ t_1 = t_2 $
  and we can derive $ \Typing{\Gamma |- x: t_1 \land t_2} $ by subsumption.

  If the last rule applied is \Rule{Abstr}, we have
  \[
    t_1 = \Interface \land (\textstyle\bigwedge_{j \in J} \lnot (t'_j \to t_j))
    \qquad
    t_2 = \Interface \land (\textstyle\bigwedge_{k \in K} \lnot (t'_k \to t_k))
  \]
  and we want to derive $ t_1 \land t_2 $.
  We do so by applying \Rule{Abstr} to derive $
    \Interface \land (\textstyle\bigwedge_{j \in J} \lnot (t'_j \to t_j))
    \land (\textstyle\bigwedge_{k \in K} \lnot (t'_k \to t_k))
  $
  (which is non-empty by Corollary~\ref{cor:subtyping-negative-arrows}
  since $ t_1 $ and $ t_2 $ are both non-empty),
  followed by \Rule{Subsum}.

  If the last rule applied is \Rule{Appl}, we have $ e = \Appl{e_1}{e_2} $ and
  \begin{gather*}
    \Typing{\Gamma |- e_1: \WithBot{t_1' \to t_1''}} \qquad
    \Typing{\Gamma |- e_2: t_1'} \qquad
    t_1 = \WithBot{t_1''} \\
    \Typing{\Gamma |- e_1: \WithBot{t_2' \to t_2''}} \qquad
    \Typing{\Gamma |- e_2: t_2'} \qquad
    t_2 = \WithBot{t_2''}
  \end{gather*}
  and, by induction, we derive
  \[
    \Typing{\Gamma |- e_1:
      \WithBot{t_1' \to t_1''} \land \WithBot{t_2' \to t_2''}}
    \qquad
    \Typing{\Gamma |- e_2: t_1' \land t_2'}
    \: .
  \]
  We have
  \begin{multline*}
    \WithBot{t_1' \to t_1''} \land \WithBot{t_2' \to t_2''} =
    ((t_1' \to t_1'') \lor \bot) \land ((t_2' \to t_2'') \lor \bot) \\ \simeq
    ((t_1' \to t_1'') \land (t_2' \to t_2'')) \lor \bot \leq
    \WithBot{(t_1' \land t_2') \to (t_1'' \land t_2'')}
  \end{multline*}
  and can therefore conclude by applying \Rule{Subsum} and \Rule{Appl}.

  The cases for \Rule{Pair} and \Rule{Proj} are similar.
  They rely on the following properties of subtyping:
  \begin{gather*}
    (t_1^1 \land t_2^1) \times (t_1^2 \land t_2^2)
    \simeq
    (t_1^1 \times t_2^1) \land (t_1^2 \times t_2^2)
    \\
    \WithBot{t_1^1 \times t_1^2} \land \WithBot{t_2^1 \times t_2^2}
    \leq
    \WithBot{(t_1^1 \land t_2^1) \times (t_1^2 \land t_2^2)}
  \end{gather*}

  For \Rule{Case}, we have
  \begin{flalign*}
    &
    \Typing{\Gamma |- \big(\Case{(x = \TypecaseExpr) \in \TypecaseType ? e_1 : e_2}\big):
      \WithBot{t_1}}
    \qquad
    \Typing{\Gamma |- \TypecaseExpr: \WithBot{t'_1}}
    \\ & \qquad
    \big( \text{if } t'_1 \nleq \lnot \TypecaseType \text{ then }
    \Typing{\Gamma, x\colon (t'_1 \land \TypecaseType) |- e_1: t_1} \big)
    \qquad
    \big( \text{if } t'_1 \nleq \TypecaseType \text{ then }
    \Typing{\Gamma, x\colon (t'_1 \setminus \TypecaseType) |- e_2: t_1} \big)
    \\ &
    \Typing{\Gamma |- \big(\Case{(x = \TypecaseExpr) \in \TypecaseType ? e_1 : e_2}\big):
      \WithBot{t_2}}
    \qquad
    \Typing{\Gamma |- \TypecaseExpr: \WithBot{t'_2}}
    \\ & \qquad
    \big( \text{if } t'_2 \nleq \lnot \TypecaseType \text{ then }
    \Typing{\Gamma, x\colon (t'_2 \land \TypecaseType) |- e_1: t_2} \big)
    \qquad
    \big( \text{if } t'_2 \nleq \TypecaseType \text{ then }
    \Typing{\Gamma, x\colon (t'_2 \setminus \TypecaseType) |- e_2: t_2} \big)
  \end{flalign*}
  and we derive $
    \Typing{\Gamma |- \big(\Case{(x = \TypecaseExpr) \in \TypecaseType ? e_1 : e_2}\big):
      \WithBot{t_1 \land t_2}}
  $ from the premises
  \begin{gather*}
    \Typing{\Gamma |- \TypecaseExpr: \WithBot{t'_1 \land t'_2}}
    \\
    \big( \text{if } t'_1 \land t'_2 \nleq \lnot \TypecaseType \text{ then }
    \Typing{\Gamma, x\colon ((t'_1 \land t'_2) \land \TypecaseType) |- e_1:
      t_1 \land t_2} \big)
    \\
    \big( \text{if } t'_1 \land t'_2 \nleq \TypecaseType \text{ then }
    \Typing{\Gamma, x\colon ((t'_1 \land t'_2) \setminus \TypecaseType) |- e_2:
      t_1 \land t_2} \big)
  \end{gather*}
  The first premise can be derived by applying the induction hypothesis
  and then subsumption,
  since $ \WithBot{t'_1} \land \WithBot{t'_2} \simeq \WithBot{t'_1 \land t'_2} $.
  For the second premise, note that, when $
    t'_1 \land t'_2 \nleq \lnot \TypecaseType
  $, we have $ t'_1 \nleq \lnot \TypecaseType $ and $ t'_2 \nleq \lnot \TypecaseType $
  and therefore we have
  \[
    \Typing{\Gamma, x\colon (t'_1 \land \TypecaseType) |- e_1: t_1}
    \qquad
    \Typing{\Gamma, x\colon (t'_2 \land \TypecaseType) |- e_1: t_2}
  \]
  and, by weakening (Lemma~\ref{lem:typing-weakening}),
  \[
    \Typing{\Gamma, x\colon ((t'_1 \land t'_2) \land \TypecaseType) |- e_1: t_1}
    \qquad
    \Typing{\Gamma, x\colon ((t'_1 \land t'_2) \land \TypecaseType) |- e_1: t_2}
    \: .
  \]
  Hence, we derive the premise by the induction hypothesis.
  The third premise is derived analogously to the second.
  Finally, we apply subsumption since $
    \WithBot{t_1 \land t_2} \simeq \WithBot{t_1} \land \WithBot{t_2}
  $.

  For \Rule{Let}, we have
  \begin{flalign*}
    &
    \Typing{\Gamma |- \Let{x = e_1 \IN e_2}: t_1}
    \qquad
    \Typing{\Gamma |- e_1: \textstyle\bigvee_{i \in I} t_i}
    \qquad
    \forall i \in I . \enspace \Typing{\Gamma, x\colon t_i |- e_2: t_1}
    \\ &
    \Typing{\Gamma |- \Let{x = e_1 \IN e_2}: t_2}
    \qquad
    \Typing{\Gamma |- e_1: \textstyle\bigvee_{j \in J} t_j}
    \qquad
    \forall j \in J . \enspace \Typing{\Gamma, x\colon t_j |- e_2: t_2}
  \end{flalign*}
  By the induction hypothesis we derive $
    \Typing{\Gamma |- e_1:
      (\textstyle\bigvee_{i \in I} t_i) \land (\textstyle\bigvee_{j \in J} t_j)}
  $; by subsumption we obtain $
    \Typing{\Gamma |- e_1:
      \textstyle\bigvee_{(i, j) \in I \times J} (t_i \land t_j)}
  $ since the two types are equivalent.
  Then, for every $ (i, j) \in I \times J $, we want to show $
    \Typing{\Gamma, x\colon (t_i \land t_j) |- e_2: t_1 \land t_2}
  $, which we show by applying Lemma~\ref{lem:typing-weakening}
  and the induction hypothesis.
\end{Proof}

\begin{Lemma}
\label{lem:typing-never-empty}
  Let $ \Gamma $ be a well-formed type environment.
  If $ \Typing{\Gamma |- e: t} $, then $ t \not\simeq \Empty $.
\end{Lemma}

\begin{Proof}
  By induction on the derivation of $ \Typing{\Gamma |- e: t} $
  and by case on the last typing rule applied.

  If the last rule is \Rule{Var},
  the conclusion follows from the assumption on $ \Gamma $.
  If the last rule is
  \Rule{Const}, \Rule{Abstr}, \Rule{Appl}, \Rule{Proj}, or \Rule{Case},
  then $ t $ is surely non-empty.
  For the remaining rules (\Rule{Pair}, \Rule{Let}, and \Rule{Subsum})
  we apply the induction hypothesis.
  In particular, for \Rule{Pair},
  we have by definition of subtyping that,
  if $ t_1 $ and $ t_2 $ are both non-empty,
  $ t_1 \times t_2 $ is non-empty as well.
  For \Rule{Let}, by the induction hypothesis we derive that the type $
    \bigvee_{i \in I} t_i
  $ of $ e_1 $ is non-empty;
  therefore, there exists an $ i_0 \in I $ such that $ t_{i_0} $ is non-empty.
  The corresponding environment $ \Gamma, x\colon t_{i_0} $
  is well-formed (whereas other $ \Gamma, x\colon t_i $ might not be)
  and thus we can apply the induction hypothesis to it
  to conclude that $ t $ is non-empty.
\end{Proof}

\begin{Lemma}[Generation]
\label{lem:typing-answer-inversion}
  Let $ \Gamma $ be a well-formed type environment
  and let $ a $ be an answer such that $ \Typing{\Gamma |- a: t} $ holds.
  Then:
  \begin{itemize}
    \item if $ t = \WithBot{t_1 \to t_2} $,
      then $ a $ is either of the form $ \RAAbstr{f: \Interface. x. e} $
      or of the form $ \Let{x = e \IN a'} $;
    \item if $ t = \WithBot{t_1 \times t_2} $,
      then $ a $ is either of the form $ (e_1, e_2) $
      or of the form $ \Let{x = e \IN a'} $.
  \end{itemize}
\end{Lemma}

\begin{Proof}
  The typing derivation $ \Typing{\Gamma |- a: t} $
  must end with the application of the rule corresponding to the form of $ a $
  (for an answer, one of \Rule{Const}, \Rule{Abstr}, \Rule{Pair}, or \Rule{Let})
  possibly followed by applications of \Rule{Subsum}.
  Therefore, if $ a = c $, then we must have $ \BasicTypeOfConstant{c} \leq t $:
  by definition of subtyping, this excludes both
  $ t = \WithBot{t_1 \to t_2} $ and $ t = \WithBot{t_1 \times t_2} $.
  Similarly, a non-empty intersection of the form $
    (\bigwedge_{i \in I} t'_i \to t_i) \land
    (\bigwedge_{j \in J} \lnot (t'_j \to t_j))
  $ cannot be a subtype of $ \WithBot{t_1 \times t_2} $,
  nor can a type $ t_1 \times t_2 $ be a subtyping of $ \WithBot{t_1 \to t_2} $,
  except if it is empty
  (which is impossible by Lemma~\ref{lem:typing-never-empty}).
\end{Proof}

\begin{Lemma}
\label{lem:typing-typeof-derivable}
  If $ \TypecaseExpr $ is well-typed in an environment $ \Gamma $
  (i.e. if $ \Typing{\Gamma |- \TypecaseExpr: t} $ holds for some $ t $),
  then $ \Typing{\Gamma |- \TypecaseExpr: \Typeof(\TypecaseExpr)} $.
\end{Lemma}

\begin{Proof}
  By induction on $ \TypecaseExpr $.
  If it is a variable, a constant, or a function,
  the result is straightforward
  (note that $ \Empty \to \Any $ is greater than any functional type).
  If it is a pair, we apply the induction hypothesis and use rule \Rule{Pair}.
\end{Proof}

\begin{Lemma}
\label{lem:typing-typeof-novar}
  Let $ \SimpleTypecaseExpr $ be an expression generated by the grammar $
    \SimpleTypecaseExpr \Coloneqq c \mid \RAAbstr{f: \Interface. x. e}
      \mid (\SimpleTypecaseExpr, \SimpleTypecaseExpr)
  $.
  Then, for every $ \TypecaseType $,
  either $ \Typeof(\SimpleTypecaseExpr) \leq \TypecaseType $
  or $ \Typeof(\SimpleTypecaseExpr) \leq \lnot \TypecaseType $.
\end{Lemma}

\begin{Proof}
  By induction on the pair $ (\SimpleTypecaseExpr, \TypecaseType) $
  and by case on $ \SimpleTypecaseExpr $ and $ \TypecaseType $.

  If $ \TypecaseType = \Empty $,
  we have $ \Typeof(\SimpleTypecaseExpr) \leq \lnot \TypecaseType $.

  If $ \TypecaseType = \TypecaseType_1 \lor \TypecaseType_2 $,
  we apply the induction hypothesis to both $ \TypecaseType_i $.
  If $ \Typeof(\SimpleTypecaseExpr) \leq \TypecaseType_1 $
  or $ \Typeof(\SimpleTypecaseExpr) \leq \TypecaseType_2 $,
  then $ \Typeof(\SimpleTypecaseExpr) \leq \TypecaseType_1 \lor \TypecaseType_2 $.
  Otherwise, we must have
  $ \Typeof(\SimpleTypecaseExpr) \leq \lnot \TypecaseType_1 $
  and $ \Typeof(\SimpleTypecaseExpr) \leq \lnot \TypecaseType_2 $;
  hence, $
    \Typeof(\SimpleTypecaseExpr)
    \leq \lnot \TypecaseType_1 \land \lnot \TypecaseType_2
    \simeq \lnot (\TypecaseType_1 \lor \TypecaseType_2)
  $.

  If $ \TypecaseType = \lnot \TypecaseType' $,
  we apply the induction hypothesis to $ \TypecaseType' $.
  If $ \Typeof(\SimpleTypecaseExpr) \leq \TypecaseType' $,
  then $ \Typeof(\SimpleTypecaseExpr) \leq \lnot \TypecaseType $.
  Conversely, if $ \Typeof(\SimpleTypecaseExpr) \leq \lnot \TypecaseType' $,
  then $ \Typeof(\SimpleTypecaseExpr) \leq \TypecaseType $.

  If $ \TypecaseType = b $ and $ \SimpleTypecaseExpr = c $,
  then $ \Typeof(\SimpleTypecaseExpr) = b_c $.
  Since $ \Inter{b_c} = \Set{c} $,
  either $ \Typeof(\SimpleTypecaseExpr) \leq b $
  or $ \Typeof(\SimpleTypecaseExpr) \leq \lnot b $ holds.
  If instead $ \SimpleTypecaseExpr $ is not a constant,
  then $ \Typeof(\SimpleTypecaseExpr) \leq \lnot b $.

  If $ \TypecaseType = \TypecaseType_1 \timesBot \TypecaseType_2 $
  and $ \SimpleTypecaseExpr = (\SimpleTypecaseExpr_1, \SimpleTypecaseExpr_2) $,
  we apply the induction hypothesis
  to $ (\SimpleTypecaseExpr_1, \TypecaseType_1) $
  and $ (\SimpleTypecaseExpr_2, \TypecaseType_2) $.
  If $ \Typeof(\SimpleTypecaseExpr_1) \leq \TypecaseType_1 $
  and $ \Typeof(\SimpleTypecaseExpr_2) \leq \TypecaseType_2 $,
  then $
    \Typeof((\SimpleTypecaseExpr_1, \SimpleTypecaseExpr_2)) \leq
    \TypecaseType_1 \timesBot \TypecaseType_2
  $.
  Otherwise, $
    \Typeof((\SimpleTypecaseExpr_1, \SimpleTypecaseExpr_2))
  $ must be subtype of one of the following types: $
    \lnot \TypecaseType_1 \times \TypecaseType_2
  $, $
    \TypecaseType_1 \times \lnot \TypecaseType_2
  $, or $
    \lnot \TypecaseType_1 \times \lnot \TypecaseType_2
  $.
  We also have $
    \Typeof((\SimpleTypecaseExpr_1, \SimpleTypecaseExpr_2))
    \leq \lnot \bot \times \lnot \bot
  $.
  The intersection of any of the three types above with $
    \lnot \bot \times \lnot \bot
  $ is a subtype of $ \lnot (\TypecaseType_1 \timesBot \TypecaseType_2) $.
  Finally, if $ \SimpleTypecaseExpr $ is not a pair,
  then $
    \Typeof(\SimpleTypecaseExpr) \leq
    \lnot (\TypecaseType_1 \timesBot \TypecaseType_2)
  $.

  If $ \TypecaseType = \Empty \to \Any $,
  then if $ \SimpleTypecaseExpr = \RAAbstr{f: \Interface. x. e} $,
  we have $
    \Typeof(\SimpleTypecaseExpr) \leq \TypecaseType
  $;
  otherwise, we have $
    \Typeof(\SimpleTypecaseExpr) \leq \lnot \TypecaseType
  $.
\end{Proof}

\begin{Lemma}
\label{lem:typing-values-disjunction}
  Let $ v $ be a value that is well-typed in $ \Gamma $
  (i.e. $ \Typing{\Gamma |- v: t'} $ holds for some $ t' $).
  Then, for every $ t $,
  we have either $ \Typing{\Gamma |- v: t} $
  or $ \Typing{\Gamma |- v: \lnot t} $.
\end{Lemma}

\begin{Proof}
  A value $ v $ is either a constant $ c $
  or an abstraction $ \RAAbstr{f: \Interface. x. e} $.
  If $ v = c $, then $ \Typing{\Gamma |- v: b_c} $ holds.
  By subsumption, we have $ \Typing{\Gamma |- v: t} $
  whenever $ b_c \leq t $.
  By definition, $ b_c \leq t $ is equivalent to $ c \in \Inter{t} $.
  Since $ c \in \Domain $, for every type $ t $,
  either $ c \in \Inter{t} $
  or $ c \in \Inter{\lnot t} = \Domain \setminus \Inter{t} $ must hold.
  Hence, either $ \Typing{\Gamma |- v: t} $ or $ \Typing{\Gamma |- v: \lnot t} $
  is derivable.

  Consider now $ v = \RAAbstr{f: \Interface. x. e} $.
  Note that, since $ v $ is well-typed, we know by inversion of the typing rules
  that $ \Typing{\Gamma |- v: \Interface} $ holds.
  We prove the result by induction on $ t $.

  If $ t = \bot $, $ t = b $, $ t = t_1 \times t_2 $, or $ t = \Empty $,
  we have $ \Interface \leq \lnot t $
  and hence $ \Typing{\Gamma |- v: \lnot t} $ holds.

  If $ t = t_1 \to t_2 $, either $ \Interface \leq t_1 \to t_2 $ holds or not.
  If it holds,
  we can derive $ \Typing{\Gamma |- v: t_1 \to t_2} $ by subsumption.
  If it does not hold we have (by definition of subtyping)
  $ \Interface \land \lnot (t_1 \to t_2) \not\simeq \Empty $.
  We can therefore derive $
    \Typing{\Gamma |- v: \Interface \land \lnot (t_1 \to t_2)}
  $ and, by subsumption, $
    \Typing{\Gamma |- v: \lnot (t_1 \to t_2)}
  $.

  If $ t = t_1 \lor t_2 $,
  we apply the induction hypothesis to $ t_1 $ and $ t_2 $.
  If either $ \Typing{\Gamma |- v: t_1} $ or $ \Typing{\Gamma |- v: t_2} $ hold,
  $ \Typing{\Gamma |- v: t_1 \lor t_2} $ holds by subsumption.
  Otherwise, we must have both
  $ \Typing{\Gamma |- v: \lnot t_1} $ and $ \Typing{\Gamma |- v: \lnot t_2} $.
  Then, by Lemma~\ref{lem:typing-intersection-in}, we have
  $ \Typing{\Gamma |- v: (\lnot t_1) \land (\lnot t_2)} $
  and, by subsumption, $ \Typing{\Gamma |- v: \lnot (t_1 \lor t_2)} $
  since $ (\lnot t_1) \land (\lnot t_2) \simeq \lnot (t_1 \lor t_2) $.

  If $ t = \lnot t' $, by the induction hypothesis we have either
  $ \Typing{\Gamma |- v: t'} $ or $ \Typing{\Gamma |- v: \lnot t'} $.
  In the former case, we have $ \Typing{\Gamma |- v: \lnot t} $
  since $ t' \simeq \lnot \lnot t' = \lnot t $;
  in the latter, we have $ \Typing{\Gamma |- v: t} $.
\end{Proof}

\begin{Corollary}
\label{cor:typing-values-disjunction-n}
  If $ \Typing{\Gamma |- v: \bigvee_{i \in I} t_i} $,
  then there exists an $ i_0 \in I $
  such that $ \Typing{\Gamma |- v: t_{i_0}} $.
\end{Corollary}

\begin{Proof}
  By induction on $ |I| $.
  If $ |I| = 1 $, the result is straightforward.

  If $ |I| = 2 $, that is, if $
    \Typing{\Gamma |- v: t_1 \lor t_2}
  $, either $ \Typing{\Gamma |- v: t_1} $ holds or not.
  In the former case, the result holds.
  In the latter, by Lemma~\ref{lem:typing-values-disjunction}, we must have $
    \Typing{\Gamma |- v: \lnot t_1}
  $. Hence, by Lemma~\ref{lem:typing-intersection-in}, we have $
    \Typing{\Gamma |- v: (t_1 \lor t_2) \land \lnot t_1}
  $, and $
    (t_1 \lor t_2) \land \lnot t_1 \simeq
    (t_1 \land \lnot t_1) \lor (t_2 \land \lnot t_1) \leq t_2
  $, so we can derive $
    \Typing{\Gamma |- v: t_2}
  $ by subsumption.

  If $ |I| = n > 2 $, we have $
    \Typing{\Gamma |- v: (t_1 \lor \dots \lor t_{n-1}) \lor t_n}
  $. We apply the induction hypothesis to conclude.
\end{Proof}

\begin{Lemma}
\label{lem:function-type-disjoint}
  Let $
    \Interface = \bigwedge_{i \in I} t_i' \to t_i
  $ (with $ |I| > 0 $) be a type.
  There exists a type $
    \Interface' = \bigwedge_{k \in K} t_k' \to t_k
  $ (with $ |K| > 0 $) such that:
  \begin{itemize}
    \item $ \Interface \simeq \Interface' $;
    \item $ \forall k_1 \neq k_2 \in K . \:
      t_{k_1} \land t_{k_2} \simeq \Empty $;
    \item if $ \Typing{\Gamma |- (\RAAbstr{f: \Interface. x. e}): \Interface} $,
      then $
        \forall k \in K . \enspace
        \Typing{\Gamma, f\colon \Interface, x\colon t_k' |- e: t_k}
      $.
      \qedhere
  \end{itemize}
\end{Lemma}

\begin{Proof}
  Given $ \Interface $, we take
  \[
    \Interface' =
    \textstyle\bigwedge_{\emptyset \subsetneq I' \subseteq I} s_{I'} \to u_{I'}
    \qquad \text{where\:\:}
    s_{I'} \eqdef \textstyle\bigwedge_{i \in I'} t_i' \land
      \textstyle\bigwedge_{i \in I \setminus I'} \lnot t_i'
    \text{\:\:and\:\:}
    u_{I'} \eqdef \textstyle\bigwedge_{i \in I'} t_i
  \]
  (defining $ \bigwedge_{i \in \emptyset} \lnot t_i' $ to be $ \Any $).
  By Lemma~\ref{lem:interface-disjoint}, we have $
    \Interface \simeq \Interface'
  $.
  To prove that the domains are pairwise disjoint,
  let $ I'_1 $ and $ I'_2 $ be two non-empty, arbitrary subsets of $ I $;
  if $ I'_1 \neq I'_2 $, then there exists an $ i_0 \in I $
  which is in one set and not in the other.
  Assume, without loss of generality, $ i_0 \in I'_1 $ and $ i_0 \notin I'_2 $.
  Then:
  \begin{align*}
    s_{I'_1} \land s_{I'_2}
    & =
    \big(
      \textstyle\bigwedge_{i \in I'_1} t_i' \land
      \textstyle\bigwedge_{i \in I \setminus I'_1} \lnot t_i'
    \big) \land
    \big(
      \textstyle\bigwedge_{i \in I'_2} t_i' \land
      \textstyle\bigwedge_{i \in I \setminus I'_2} \lnot t_i'
    \big)
    \\ & \simeq
    \big(
      t'_{i_0} \land
      \textstyle\bigwedge_{i \in I'_1 \setminus \Set{i_0}} t_i' \land
      \textstyle\bigwedge_{i \in I \setminus I'_1} \lnot t_i'
    \big) \land
    \big(
      \lnot t'_{i_0} \land
      \textstyle\bigwedge_{i \in I'_2} t_i' \land
      \textstyle\bigwedge_{i \in I \setminus \Set{i_0} \setminus I'_2}
        \lnot t_i'
    \big)
    \\ & \simeq
    t'_{i_0} \land
    \lnot t'_{i_0} \land
    \big(
      \textstyle\bigwedge_{i \in I'_1 \setminus \Set{i_0}} t_i' \land
      \textstyle\bigwedge_{i \in I \setminus I'_1} \lnot t_i'
    \big) \land
    \big(
      \textstyle\bigwedge_{i \in I'_2} t_i' \land
      \textstyle\bigwedge_{i \in I \setminus \Set{i_0} \setminus I'_2}
        \lnot t_i'
    \big)
    \\ & \simeq \Empty \: .
  \end{align*}

  To prove the third condition, note that $
    \Typing{\Gamma |- (\RAAbstr{f: \Interface. x. e}): \Interface}
  $ implies that, for every $ i \in I $, we can derive $
    \Typing{\Gamma, f\colon \Interface, x\colon t_i' |- e: t_i}
  $. Now consider an arbitrary $ I' $ such that $
    \emptyset \subsetneq I' \subseteq I
  $. We must show $
    \Typing{\Gamma, f\colon \Interface, x\colon s_{I'} |- e: u_{I'}}
  $.
  Note that $ s_{I'} \leq t_i' $ for every $ i \in I' $.
  Hence, by Lemma~\ref{lem:typing-weakening}, we have (for all $ i \in I' $) $
    \Typing{\Gamma, f\colon \Interface, x\colon s_{I'} |- e: t_i}
  $. By Lemma~\ref{lem:typing-intersection-in}, we have $
    \Typing{\Gamma, f\colon \Interface, x\colon s_{I'} |- e: u_{I'}}
  $ since $ u_{I'} = \bigwedge_{i \in I'} t_i $.
\end{Proof}

\begin{Lemma}[Expression substitution]
\label{lem:typing-substitution}
  If $ \Typing{\Gamma, x\colon t' |- e: t} $
  and $ \Typing{\Gamma |- e': t'} $,
  then $ \Typing{\Gamma |- e [\nicefrac{e'}{x}]: t} $.
\end{Lemma}

\begin{Proof}
  By induction on the derivation of $ \Typing{\Gamma, x\colon t' |- e: t} $
  and by case on the last rule applied.
\end{Proof}

\begin{Lemma}
\label{lem:typing-product-decompose}
  If $ \Typing{\Gamma |- (e_1, e_2): \bigvee_{i \in I} t_i} $,
  then there exist two types
  $ \bigvee_{j \in J} t_j $ and $ \bigvee_{k \in K} t_k $
  such that
  \[
    \Typing{\Gamma |- e_1: \textstyle\bigvee_{j \in J} t_j} \qquad
    \Typing{\Gamma |- e_2: \textstyle\bigvee_{k \in K} t_k} \qquad
    \forall j \in J . \: \forall k \in K . \: \exists i \in I . \:
      t_j \times t_k \leq t_i
    \: .
  \]
\end{Lemma}

\begin{Proof}
  Since $ \Typing{\Gamma |- (e_1, e_2): \bigvee_{i \in I} t_i} $,
  by inversion of the typing derivation, we have
  $ \Typing{\Gamma |- e_1: t^1} $, $ \Typing{\Gamma |- e_2: t^2} $,
  and $ t^1 \times t^2 \leq \bigvee_{i \in I} t_i $.

  If $ t^1 \simeq \Empty $ or $ t^2 \simeq \Empty $,
  then we choose $ \bigvee_{j \in J} t_j = t^1 $
  and $ \bigvee_{k \in K} t_k = t^2 $
  (i.e. $ |J| = |K| = 1 $),
  which ensures the result.

  Now we assume $ t^1 \not\simeq \Empty $ and $ t^2 \not\simeq \Empty $.
  We have $
    t^1 \times t^2 \simeq (\bigvee_{i \in I} t_i) \land (t^1 \times t^2)
      \simeq \bigvee_{i \in I} (t_i \land (t^1 \times t^2))
  $.
  For every $ i $, we have $
    t_i \land (t^1 \times t^2) \leq \Any \times \Any
  $; therefore, by Lemma~\ref{lem:product-decomposition-exists},
  we can find a product decomposition $ \Pi_i $ such that $
    t_i \land (t^1 \times t^2) \simeq
    \bigvee_{(t_1, t_2) \in \Pi_i} t_1 \times t_2
  $.
  Then, $ \Pi = \bigcup_{i \in I} \Pi_i $ is itself a product decomposition,
  such that $
    \bigvee_{(t_1, t_2) \in \Pi} t_1 \times t_2 \simeq t^1 \times t^2
  $.

  By Lemma~\ref{lem:product-decomposition-disjoint},
  there exists a fully disjoint product decomposition $ \Pi' $ such that
  \[
    \textstyle
    \bigvee_{(t_1, t_2) \in \Pi} t_1 \times t_2 \simeq
    \bigvee_{(t_1, t_2) \in \Pi'} t_1 \times t_2
    \qquad
    \forall (t_1', t_2') \in \Pi' . \: \exists (t_1, t_2) \in \Pi . \:
    t_1' \times t_2' \leq t_1 \times t_2
    \: .
  \]

  Since $
    t^1 \times t^2 \simeq
    \bigvee_{(t_1, t_2) \in \Pi'} t_1 \times t_2
  $, by Lemma~\ref{lem:product-decomposition-disjoint-split}
  we have
  \begin{gather*}
    \textstyle
    t^1 \simeq \bigvee_{(t_1, t_2) \in \Pi'} t_1
    \qquad
    t^2 \simeq \bigvee_{(t_1, t_2) \in \Pi'} t_2
    \\
    \forall (t_1', t_2'), (t_1'', t_2'') \in \Pi' \: .
      \exists (t_1, t_2) \in \Pi' . \:
      t_1' \times t_2'' \leq t_1 \times t_2
    \: .
  \end{gather*}

  We take the two decompositions above for $ t^1 $ and $ t^2 $:
  by subsumption, we have
  \[
    \textstyle
    \Typing{\Gamma |- e_1: \bigvee_{(t_1, t_2) \in \Pi'} t_1}
    \qquad
    \Typing{\Gamma |- e_2: \bigvee_{(t_1, t_2) \in \Pi'} t_2}
  \]

  It remains to prove that $
    \forall (t_1', t_2'), (t_1'', t_2'') \in \Pi' \: .
      \exists i \in I . \:
      t_1' \times t_2'' \leq t_i
  $.
  Consider two arbitrary $ (t_1', t_2') $ and $ (t_1'', t_2'') $ in $ \Pi' $.
  There exists a $ (t_1, t_2) \in \Pi' $
  such that $ t_1' \times t_2'' \leq t_1 \times t_2 $.
  Therefore, there exists also a $ (t_1, t_2) \in \Pi $
  such that $ t_1' \times t_2'' \leq t_1 \times t_2 $.
  This $ (t_1, t_2) \in \Pi $ belongs to some $ \Pi_i $
  and therefore $ t_1 \times t_2 \leq t_i $,
  implying also $ t_1' \times t_2'' \leq t_i $.
\end{Proof}

\begin{Lemma}
  \label{lem:context-with-variable-needs-gamma}
  If $ \Typing{\Gamma |- E \HoleUnbound{x}: t} $,
  then $ x \in \Dom(\Gamma) $.
\end{Lemma}

\begin{Proof}
  By induction on the derivation of $ \Typing{\Gamma |- E \HoleUnbound{x}: t} $
  and by case on the last rule applied.

  \begin{itemize}
    \item \Rule{Var}:
      we have $ E \HoleUnbound{x} = x $
      and $ x \in \Dom(\Gamma) $.
    \item \Rule{Const}, \Rule{Abstr}, \Rule{Pair}:
      impossible, since $ E \HoleUnbound{x} $ cannot be
      a constant, a function, or a pair.
    \item \Rule{Appl}:
      we have $ E \HoleUnbound{x} = \Appl{e_1}{e_2} $,
      therefore $ E = \Appl{E'}{e_2} $
      and $ e_1 = E' \HoleUnbound{x} $;
      we conclude by applying the induction hypothesis to the derivation of
      $ \Typing{\Gamma |- e_1: \WithBot{t' \to t}} $.
    \item \Rule{Proj}:
      we have $ E \HoleUnbound{x} = \ProjIth{e} $,
      therefore $ E = \ProjIth{E'} $ and $ e = E' \HoleUnbound{x} $;
      we conclude by applying the induction hypothesis to the derivation of
      $ \Typing{\Gamma |- e: \WithBot{t_1 \times t_2}} $.
    \item \Rule{Case}:
      we have $
        E \HoleUnbound{x} =
        \big( \Case{(y = \TypecaseExpr) \IN \TypecaseType ? e_1 : e_2} \big)
      $,
      therefore $
        E = \big( \Case{(y = F) \IN \TypecaseType ? e_1 : e_2} \big)
      $ and $ \TypecaseExpr = F \HoleUnbound{x} $
      (hence, $ x \neq y $);
      we also have that $ \TypecaseExpr $ is well-typed in $ \Gamma $,
      so we can conclude by showing, by induction on $ F $,
      that $ \Typing{\Gamma |- F \HoleUnbound{x}: t} $
      implies $ x \in \Dom(\Gamma) $.
    \item \Rule{Let}:
      since $
        E \HoleUnbound{x} = \big( \Let{y = e_1 \IN e_2} \big)
      $ we have
      either $ E = \big( \Let{y = e_1 \IN E'} \big) $
      and $ e_2 = E' \HoleUnbound{x} $
      or $ E = \big( \Let{y = E' \IN E'' \HoleUnbound{y}} \big) $
      and $ e_1 = E' \HoleUnbound{x} $;
      in both cases we have a derivation for $ E' \HoleUnbound{x} $
      and, by the induction hypothesis, we derive $ x \in \Dom(\Gamma) $
      (in the first case,
      the derivation is in an environment $ (\Gamma, y\colon t_i) $,
      but we have $ x \neq y $).
    \item \Rule{Subsum}: we conclude by applying the induction hypothesis.
      \qedhere
  \end{itemize}
\end{Proof}

\subsection{Type soundness for the internal language}

\begin{Theorem}[Progress]
  Let $ \Gamma $ be a well-formed type environment.
  Let $ e $ be an expression that is well-typed in $ \Gamma $
  (that is, $ \Typing{\Gamma |- e: t} $ holds for some $ t $).
  Then $ e $ is an answer, or $ e $ is of the form $ E \HoleUnbound{x} $,
  or $ \exists e'. \: \Reduces{e ~> e'} $.
\end{Theorem}

\begin{Proof}
  By induction on the derivation of $ \Typing{\Gamma |- e: t} $
  and by case on the last typing rule applied.

  \begin{description}
    \item[Case \Rule{Var}]
      In this case $ e = x $, and therefore $ e $ has the form $ E \HoleUnbound{x} $.

    \item[Cases \Rule{Const}, \Rule{Abstr}, and \Rule{Pair}]
      In all cases, $ e $ is an answer.

    \item[Case \Rule{Appl}]
      We have
      \[
        e = \Appl{e_1}{e_2} \qquad
        \Typing{\Gamma |- e: \WithBot{t}} \qquad
        \Typing{\Gamma |- e_1: \WithBot{t' \to t}} \qquad
        \Typing{\Gamma |- e_2: t'}
        \: .
      \]
      We apply the induction hypothesis to $ e_1 $.
      If $ e_1 $ reduces, then $ e $ reduces by the rule \Rule{Ctx}.
      If $ e_1 $ is of the form $ E \HoleUnbound{x} $,
      then $ e $ is of the form $ E' \HoleUnbound{x} $ with $ E' = \Appl{E}{e_2} $.

      If $ e_1 $ is an answer, then by Lemma~\ref{lem:typing-answer-inversion}
      it is either of the form $ \RAAbstr{f: \Interface. x. e'} $
      or of the form $ \Let{x = e'' \IN a} $.
      Therefore, $ e $ reduces by \Rule{Appl} or \Rule{ApplL}.

    \item[Case \Rule{Proj}]
      We have
      \[
        e = \ProjIth{e'} \qquad
        \Typing{\Gamma |- e: \WithBot{t_i}} \qquad
        \Typing{\Gamma |- e': \WithBot{t_1 \times t_2}}
        \: .
      \]
      We apply the induction hypothesis to $ e' $.
      If $ e' $ reduces, then $ e $ reduces by the rule \Rule{Ctx}.
      If it is of the form $ E \HoleUnbound{x} $,
      then $ e $ is of the form $ E' \HoleUnbound{x} $ with $ E' = \ProjIth{E} $.

      If $ e' $ is an answer, then by Lemma~\ref{lem:typing-answer-inversion}
      it is either of the form $ (e_1, e_2) $
      or of the form $ \Let{x = e'' \IN a} $.
      Then $ e $ reduces by \Rule{Proj} or \Rule{ProjL}.

    \item[Case \Rule{Case}]
      We have
      \begin{gather*}
        e = (\Case{(x = \TypecaseExpr) \in \TypecaseType ? e_1 : e_2}) \qquad
        \Typing{\Gamma |- e: \WithBot{t}} \qquad
        \Typing{\Gamma |- \TypecaseExpr: \WithBot{t'}} \\
        t' \leq \lnot \TypecaseType \text{\: or \:}
        \Typing{\Gamma, x\colon (t' \land \TypecaseType) |- e_1: t}
        \qquad
        t' \leq \TypecaseType \text{\: or \:}
        \Typing{\Gamma, x\colon (t' \setminus \TypecaseType) |- e_2: t}
        \: .
      \end{gather*}
      We apply the induction hypothesis to $ \TypecaseExpr $.
      If it reduces, then $ e $ reduces by \Rule{Ctx}.
      If it is of the form $ E \HoleUnbound{y} $,
      then we must have $ \TypecaseExpr = y $ and $ E = \Hole{} $
      because all other productions in the grammar for $ E $
      do not appear in the grammar for $ \TypecaseExpr $.
      Then, we have $ \TypecaseExpr = F \HoleUnbound{y} $
      (with $ F = \Hole{} $),
      and hence $ e $ is of the form $ E \HoleUnbound{y} $
      with $ E = (\Case{(x = \Hole{}) \in \TypecaseType ? e_1 : e_2}) $.

      If $ \TypecaseExpr $ is an answer,
      it is either generated by the restricted grammar $
        \SimpleTypecaseExpr \Coloneqq c \mid \RAAbstr{f: \Interface. x. e}
          \mid (\SimpleTypecaseExpr, \SimpleTypecaseExpr)
      $ (i.e. it does not contain variables except under abstractions)
      or not.
      In the latter case, $ \TypecaseExpr $ is of the form $ F \HoleUnbound{y} $
      for some $ F $ and $ y $,
      and hence $ e $ is of the form $ E \HoleUnbound{y} $.
      In the former case, by Lemma~\ref{lem:typing-typeof-novar},
      either $ \Typeof(\TypecaseExpr) \leq \TypecaseType $
      or $ \Typeof(\TypecaseExpr) \leq \lnot \TypecaseType $.
      Then $ e $ reduces by \Rule{Case1} or \Rule{Case2}.

    \item[Case \Rule{Let}]
      We have
      \[
        e = (\Let{x = e_1 \IN e_2}) \qquad
        \Typing{\Gamma |- e: t} \qquad
        \Typing{\Gamma |- e_1: \textstyle\bigvee_{i \in I} t_i} \qquad
        \forall i \in I  . \: \Typing{\Gamma, x\colon t_i |- e_2: t}
        \: .
      \]
      Since $ \Gamma $ is well-formed, by Lemma~\ref{lem:typing-never-empty},
      we know that $
        \bigvee_{i \in I} t_i
      $ is not empty.
      As a consequence, at least one of the $ t_i $ is non-empty,
      and hence at least one of the environments $
        (\Gamma, x\colon t_i)
      $ is well-formed,
      and we can apply the induction hypothesis to it.

      We derive that
      $ e_2 $ is an answer, or it has the form $ E \HoleUnbound{y} $, or it reduces.
      If $ e_2 $ is an answer, then $ e $ is an answer as well.
      If $ e_2 $ reduces, then $ e $ reduces by \Rule{Ctx}.
      If $ e_2 $ is of the form $ E \HoleUnbound{y} $
      for some context $ E $ and variable $ y $,
      then either $ x = y $ or not.
      In the latter case, $ e $ is of the form $ E' \HoleUnbound{y} $ too.

      If $ x = y $, we apply the induction hypothesis to $ e_1 $.
      If $ e_1 $ is of the form $ E'' \HoleUnbound{z} $
      for some context $ E'' $ and variable $ z $,
      then $ e $ is of such form as well.
      If $ e_1 $ reduces, then $ e $ reduces by \Rule{Ctx}.
      If $ e_1 $ is an answer,
      then $ e $ reduces by \Rule{LetV}, \Rule{LetP}, or \Rule{LetL}.

    \item[Case \Rule{Subsum}]
      We apply the induction hypothesis to the premise and conclude.
      \qedhere
  \end{description}
\end{Proof}

\begin{Theorem}[Subject reduction]
  Let $ \Gamma $ be a well-formed type environment.
  If $ \Typing{\Gamma |- e: t} $ and $ \Reduces{e ~> e'} $,
  then $ \Typing{\Gamma |- e': t} $.
\end{Theorem}

\begin{Proof}
  By induction on the derivation of $ \Typing{\Gamma |- e: t} $
  and by case on the last typing rule applied.

  \begin{description}
    \item[Cases \Rule{Var}, \Rule{Const}, \Rule{Abstr}, and \Rule{Pair}]
      These cases do not occur,
      because $ \Reduces{e ~> e'} $ cannot hold
      when $ e $ is a variable, a constant, an abstraction, or a pair.

    \item[Case \Rule{Appl}]
      We have
      \[
        e = \Appl{e_1}{e_2} \qquad
        \Typing{\Gamma |- e: \WithBot{t}} \qquad
        \Typing{\Gamma |- e_1: \WithBot{t' \to t}} \qquad
        \Typing{\Gamma |- e_2: t'}
        \: .
      \]
      If $ \Reduces{\Appl{e_1}{e_2} ~> e'} $ occurs by the rule \Rule{Ctx},
      then $ e' = \Appl{e_1'}{e_2} $ and, by the induction hypothesis,
      $ \Typing{\Gamma |- e_1': \WithBot{t' \to t}} $:
      we apply \Rule{Appl} again to type $ e' $.

      If the reduction occurs by \Rule{Appl},
      we have
      \[
        e = \Appl{(\RAAbstr{f: \Interface. x. e_3})}{e_2} \qquad
        e' = \big( \Let{f = (\RAAbstr{f: \Interface. x. e_3}) \IN
          \Let{x = e_2 \IN e_3}} \big)
      \]
      and we must show $ \Typing{\Gamma |- e': \WithBot{t}} $.
      Let $
        \Interface =
        \bigwedge_{i \in I} \InterfaceType_i' \toBot \InterfaceType_i
      $.
      The typing derivation for $ e $ is
      \[
        \Infer[Appl]
          {
            \Infer[Subsum]
              {
                \Infer[Abstr]
                  {
                    \forall i \in I . \enspace
                    \Typing{
                      \Gamma, f\colon \Interface,
                      x\colon \WithBot{\InterfaceType'_i} |-
                      e_3: \WithBot{\InterfaceType_i}}
                  }
                  {\Typing{\Gamma |- (\RAAbstr{f: \Interface. x. e_3}):
                    \Interface \land
                    \textstyle\bigwedge_{j \in J} \lnot (t_j' \to t_j)}}
                  {}
              }
              {\Typing{\Gamma |- (\RAAbstr{f: \Interface. x. e_3}):
                \WithBot{t' \to t}}}
              {}
            \\
            \Typing{\Gamma |- e_2: t'}
          }
          {\Typing{\Gamma |- \Appl{(\RAAbstr{f: \Interface. x. e_3})}{e_2}:
            \WithBot{t}}}
          {}
      \]
      The side conditions of \Rule{Abstr} and \Rule{Subsum} ensure
      \[
        \Interface \land \textstyle\bigwedge_{j \in J} \lnot (t_j' \to t_j)
          \not\simeq \Empty
        \qquad
        \Interface \land \textstyle\bigwedge_{j \in J} \lnot (t_j' \to t_j)
          \leq \WithBot{t' \to t}
      \]
      from which we have (by definition of subtyping) $
        \Interface \land \textstyle\bigwedge_{j \in J} \lnot (t_j' \to t_j)
          \leq t' \to t
      $ and, by Corollary~\ref{cor:subtyping-negative-arrows} $,
        \Interface \leq t' \to t
      $.

      By Lemma~\ref{lem:function-type-disjoint}, we find a type $
        \Interface' = \bigwedge_{k \in K} t_k' \to t_k
      $ such that $ \Interface \simeq \Interface' $,
      that $ t'_{k_1} \land t'_{k_2} \simeq \Empty $ when $ k_1 \neq k_2 $,
      and that, for all $ k \in K $, $
        \Typing{\Gamma, f\colon \Interface, x\colon t_k' |- e_3: t_k}
      $.
      Since $ \Interface \leq t' \to t $,
      we also have $ \Interface' \leq t' \to t $.
      By Corollary~\ref{lem:subtyping-decomposition-arrows-disjoint}, we have $
        t' \leq \bigvee_{k \in K} t'_k
      $.
      Let $
        \bar{K} = \Setc{k \in K \given t' \land t'_k \not\simeq \Empty}
      $.
      We have $ t' \leq \bigvee_{k \in \bar{K}} t'_k $.
      By Corollary~\ref{lem:subtyping-decomposition-arrows-disjoint},
      we also have $ \bigvee_{k \in \bar{K}} t_k \leq t $
      and therefore $ \bigvee_{k \in \bar{K}} t_k \leq \WithBot{t} $.

      We build the typing derivation for $ e' $ as follows:
      \[
        \Infer
          {
            \Infer
              {\Typing{\Gamma, f\colon \Interface |- e_2: t'}}
              {\Typing{\Gamma, f\colon \Interface |- e_2:
                \textstyle\bigvee_{k \in \bar{K}} t'_k}}
              {}
            \\
            \forall k \in \bar{K} . \enspace
            \Infer
              {
                \Infer
                  {\Typing{\Gamma, f\colon \Interface, x\colon t'_k |- e_3:
                    t_k}}
                  {\Typing{\Gamma, f\colon \Interface, x\colon t'_k |- e_3:
                    \textstyle\bigvee_{k \in \bar{K}} t_k}}
                  {}
              }
              {\Typing{\Gamma, f\colon \Interface, x\colon t'_k |- e_3:
                \WithBot{t}}}
              {}
          }
          {\Typing{\Gamma, f\colon \Interface |-
            (\Let{x = e_2 \IN e_3}): \WithBot{t}}}
          {}
      \]
      \[
        \Infer
          {
            \Typing{\Gamma |- (\RAAbstr{f: \Interface. x. e_3}): \Interface} \\
            \Typing{\Gamma, f\colon \Interface |-
                (\Let{x = e_2 \IN e_3}): \WithBot{t}}
          }
          {\Typing{\Gamma |-
            \big( \Let{f = (\RAAbstr{f: \Interface. x. e_3}) \IN
            \Let{x = e_2 \IN e_3}} \big): \WithBot{t}}}
          {}
      \]

      If the reduction occurs by the rule \Rule{ApplL}, we have
      \[
        e = \Appl{(\Let{x = e_1' \IN a})}{e_2} \qquad
        e' = (\Let{x = e_1' \IN \Appl{a}{e_2}})
      \]
      and we must show $ \Typing{\Gamma |- e': \WithBot{t}} $.
      The typing derivation for $ e $ (collapsing the use of \Rule{Subsum}) is
      \[
        \Infer
          {
            \Infer
              {
                \Typing{\Gamma |- e_1': \textstyle\bigvee_{i \in I} t_i} \\
                \forall i \in I . \enspace
                \Typing{\Gamma, x\colon t_i |- a: t''}
              }
              {\Typing{\Gamma |- (\Let{x = e_1' \IN a}): \WithBot{t' \to t}}}
              {t'' \leq \WithBot{t' \to t}}
            \\
            \Typing{\Gamma |- e_2: t'}
          }
          {\Typing{\Gamma |- \Appl{(\Let{x = e_1' \IN a})}{e_2}: \WithBot{t}}}
          {}
      \]
      from which we build the needed derivation as follows:
      \[
        \Infer
          {
            \Typing{\Gamma |- e_1': \textstyle\bigvee_{i \in I} t_i} \\
            \forall i \in I . \enspace
            \Infer
              {
                \Infer
                  {\Typing{\Gamma, x\colon t_i |- a: t''}}
                  {\Typing{\Gamma, x\colon t_i |- a: \WithBot{t' \to t}}}
                  {t'' \leq \WithBot{t' \to t}}
                \\
                \Typing{\Gamma |- e_2: t'}
              }
              {\Typing{\Gamma, x\colon t_i |- \Appl{a}{e_2}: \WithBot{t}}}
              {}
          }
          {\Typing{\Gamma |- (\Let{x = e_1' \IN \Appl{a}{e_2}}): \WithBot{t}}}
          {}
      \]

    \item[Case \Rule{Proj}]
      We have
      \[
        e = \ProjIth{e''} \qquad
        \Typing{\Gamma |- e: \WithBot{t_i}} \qquad
        \Typing{\Gamma |- e'': \WithBot{t_1 \times t_2}}
        \: .
      \]
      If $ e $ reduces by rule \Rule{Ctx},
      we obtain the result from the induction hypothesis.
      Otherwise, the reduction must occur
      by rule \Rule{Proj} or rule \Rule{ProjL}.

      If \Rule{Proj} applies, we have $
        e = \ProjIth{(e_1, e_2)}
      $ and $
        e' = e_i
      $. We must show $
        \Typing{\Gamma |- e_i: \WithBot{t_i}}
      $.
      Note that we have $
        \Typing{\Gamma |- (e_1, e_2): \WithBot{t_1 \times t_2}}
      $: by inversion of the typing derivation, we have $
        \Typing{\Gamma |- e_1: t_1'}
      $, $
        \Typing{\Gamma |- e_2: t_2'}
      $, and $ t_1' \times t_2' \leq \WithBot{t_1 \times t_2} $.
      By Lemma~\ref{lem:typing-never-empty},
      we know $ t_1' \not\simeq \Empty $ and $ t_2' \not\simeq \Empty $;
      hence, by definition of subtyping
      we have $ t_1' \leq t_1 $ and $ t_2' \leq t_2 $.
      Therefore we can derive $ \Typing{\Gamma |- e_i: \WithBot{t_i}} $
      by \Rule{Subsum}.

      If \Rule{ProjL} applies, we have $
        e = \ProjIth{(\Let{x = e''' \IN a})}
      $ and $
        e' = (\Let{x = e''' \IN \ProjIth{a}})
      $.
      The typing derivation for $ e $ (collapsing the use of \Rule{Subsum}) is
      \[
        \Infer
          {
            \Infer
              {
                \Typing{\Gamma |- e''': \textstyle\bigvee_{i \in I} t_i} \\
                \forall i \in I . \enspace
                \Typing{\Gamma, x\colon t_i |- a: t}
              }
              {\Typing{\Gamma |- (\Let{x = e''' \IN a}): t}}
              {}
          }
          {\Typing{\Gamma |- \ProjIth{(\Let{x = e''' \IN a})}: \WithBot{t_i}}}
          {t \leq \WithBot{t_1 \times t_2}}
      \]
      from which we build the derivation for $
        \Typing{\Gamma |- e': \WithBot{t_i}}
      $ as follows:
      \[
        \Infer
          {
            \Typing{\Gamma |- e''': \textstyle\bigvee_{i \in I} t_i} \\
            \forall i \in I . \enspace
            \Infer
              {\Typing{\Gamma, x\colon t_i |- a: t}}
              {\Typing{\Gamma, x\colon t_i |- \ProjIth{a}: \WithBot{t_i}}}
              {t \leq \WithBot{t_1 \times t_2}}
          }
          {\Typing{\Gamma |- (\Let{x = e''' \IN \ProjIth{a}}): \WithBot{t_i}}}
          {}
      \]

    \item[Case \Rule{Case}]
      We have
      \begin{gather*}
        e = (\Case{(x = \TypecaseExpr) \in \TypecaseType ? e_1 : e_2}) \qquad
        \Typing{\Gamma |- e: \WithBot{t}} \qquad
        \Typing{\Gamma |- \TypecaseExpr: \WithBot{t'}} \\
        t' \leq \lnot \TypecaseType \text{\: or \:}
        \Typing{\Gamma, x\colon (t' \land \TypecaseType) |- e_1: t}
        \qquad
        t' \leq \TypecaseType \text{\: or \:}
        \Typing{\Gamma, x\colon (t' \setminus \TypecaseType) |- e_2: t}
        \: .
      \end{gather*}
      If $ e $ reduces by rule \Rule{Ctx},
      we obtain the result from the induction hypothesis.
      Otherwise, it reduces by either \Rule{Case1} or \Rule{Case2}.

      If \Rule{Case1} applies, we have $
        e' = (\Let{x = \TypecaseExpr \IN e_1})
      $ and $
        \Typeof(\TypecaseExpr) \leq \TypecaseType
      $.
      Note that $ \TypecaseExpr $ cannot be a variable:
      if it were, we would have $ \Typeof(\TypecaseExpr) = \Any $,
      but this would require $ \TypecaseType \simeq \Any $,
      which is forbidden by the syntax of typecases.
      Since $ \TypecaseExpr $ is not a variable,
      we have $ \Typeof(\TypecaseExpr) \leq \lnot \bot $.
      Hence, we also have
      $ \Typeof(\TypecaseExpr) \leq \TypecaseType \land \lnot \bot $.
      By Lemma~\ref{lem:typing-typeof-derivable}, we can derive $
        \Typing{\Gamma |- \TypecaseExpr: \Typeof(\TypecaseExpr)}
      $; then we can derive $
        \Typing{\Gamma |- \TypecaseExpr: \TypecaseType \land \lnot \bot}
      $ by \Rule{Subsum} and $
        \Typing{\Gamma |- \TypecaseExpr:
          \WithBot{t'} \land \TypecaseType \land \lnot \bot}
      $ by Lemma~\ref{lem:typing-intersection-in};
      again by \Rule{Subsum}, we derive $
        \Typing{\Gamma |- \TypecaseExpr: t' \land \TypecaseType}
      $ because $
        \WithBot{t'} \land \TypecaseType \land \lnot \bot \leq
        t' \land \TypecaseType
      $.
      If $ \Typing{\Gamma, x\colon (t' \land \TypecaseType) |- e_1: t} $
      holds, we can derive $ \Typing{\Gamma |- e': \WithBot{t}} $
      by applying \Rule{Let} and \Rule{Subsum}.
      If $ \Typing{\Gamma, x\colon (t' \land \TypecaseType) |- e_1: t} $
      does not hold, by hypothesis we would have $
        t' \leq \lnot \TypecaseType
      $: we show that this cannot occur.
      If we had $
        t' \leq \lnot \TypecaseType
      $, we could derive $
        \Typing{\Gamma |- \TypecaseExpr:
          \lnot \TypecaseType \land \TypecaseType}
      $ and $
        \Typing{\Gamma |- \TypecaseExpr: \Empty}
      $ by subsumption.
      This is impossible by Lemma~\ref{lem:typing-never-empty}.

      If \Rule{Case2} applies, we proceed similarly.
      We have $
        e' = (\Let{x = \TypecaseExpr \IN e_2})
      $ and $
        \Typeof(\TypecaseExpr) \leq \lnot \TypecaseType
      $.
      We have $ \Typeof(\TypecaseExpr) \leq \lnot \bot $
      because $ \TypecaseExpr $ cannot be a variable
      (since in that case we would have $ \TypecaseType \simeq \Empty $,
      which is forbidden by the syntax).
      We can derive $
        \Typing{\Gamma |- \TypecaseExpr:
          t' \land \lnot \TypecaseType}
      $, and, if $
        \Typing{\Gamma, x\colon (t' \setminus \TypecaseType) |- e_2: t}
      $ holds, $ \Typing{\Gamma |- e': \WithBot{t}} $.
      As before, we can show that $
        \Typing{\Gamma, x\colon (t' \setminus \TypecaseType) |- e_2: t}
      $ must always hold by showing that the alternative, $
        t' \leq \TypecaseType
      $, cannot occur:
      if we had $ t' \leq \TypecaseType $,
      we would have $
        \Typing{\Gamma |- \TypecaseExpr:
          \TypecaseType \land \lnot \TypecaseType}
      $, which is impossible.

    \item[Case \Rule{Let}]
      We have
      \[
        e = (\Let{x = e_1 \IN e_2}) \qquad
        \Typing{\Gamma |- e: t} \qquad
        \Typing{\Gamma |- e_1: \textstyle\bigvee_{i \in I} t_i} \qquad
        \forall i \in I  . \: \Typing{\Gamma, x\colon t_i |- e_2: t}
        \: .
      \]
      If $ e $ reduces by rule \Rule{Ctx},
      we obtain the result from the induction hypothesis.
      Otherwise, $ e $ reduces by \Rule{LetV}, \Rule{LetP}, or \Rule{LetL}.

      If \Rule{LetV} applies, we have $
        e = (\Let{x = v \IN E \HoleUnbound{x}})
      $ and $
        e' = (E\HoleUnbound{x})[\nicefrac{v}{x}]
      $. We must show $ \Typing{\Gamma |- e': t} $.
      Since $ \Typing{\Gamma |- v: \textstyle\bigvee_{i \in I} t_i} $,
      by Corollary~\ref{cor:typing-values-disjunction-n}
      we have $ \Typing{\Gamma |- v: t_{i_0}} $ for some $ i_0 \in I $.
      We also have $ \Typing{\Gamma, x\colon t_{i_0} |- E \HoleUnbound{x}: t} $.
      By Lemma~\ref{lem:typing-substitution},
      we derive $ \Typing{\Gamma |- (E\HoleUnbound{x})[\nicefrac{v}{x}]: t} $.

      If \Rule{LetP} applies, we have
      \[
        e = \big( \Let{x = (e_1', e_1'') \IN E \HoleUnbound{x}} \big) \qquad
        e' = \big( \Let{x' = e_1' \IN \Let{x'' = e_1'' \IN
          (E\HoleUnbound{x})[\nicefrac{(x', x'')}{x}]}} \big)
      \]
      and must show $ \Typing{\Gamma |- e': t} $.
      Since $ \Typing{\Gamma |- (e_1', e_1''): \bigvee_{i \in I} t_i} $,
      by Lemma~\ref{lem:typing-product-decompose} we can find two types
      $ \bigvee_{j \in J} t_j $ and $ \bigvee_{k \in K} t_k $
      such that
      \[
        \Typing{\Gamma |- e_1': \textstyle\bigvee_{j \in J} t_j} \qquad
        \Typing{\Gamma |- e_1'': \textstyle\bigvee_{k \in K} t_k} \qquad
        \forall j \in J . \: \forall k \in K . \: \exists i \in I . \:
          t_j \times t_k \leq t_i
        \: .
      \]
      We show $ \Typing{\Gamma |- e': t} $ by showing
      \[
        \forall j \in J . \: \forall k \in K . \quad
          \Typing{\Gamma, x'\colon t_j, x''\colon t_k |-
            (E\HoleUnbound{x})[\nicefrac{(x', x'')}{x}]: t}
      \]
      which can be derived by Lemma~\ref{lem:typing-substitution} from
      \[
        \forall j \in J . \: \forall k \in K . \enspace
        \begin{cases}
          \Typing{\Gamma, x'\colon t_j, x''\colon t_k |- (x', x''):
            t_j \times t_k} \\
          \Typing{\Gamma, x'\colon t_j, x''\colon t_k, x\colon t_j \times t_k
            |- E\HoleUnbound{x}: t}
        \end{cases}
      \]
      For every $ j $ and $ k $, the second derivation
      is obtained from
      \[
        \forall i \in I  . \: \Typing{\Gamma, x\colon t_i |- E \HoleUnbound{x}: t}
      \]
      by weakening (Lemma~\ref{lem:typing-weakening}), because $
        t_j \times t_k \leq t_i
      $ for some $ i \in I $.

      If \Rule{LetL} applies, we have
      \[
        e = \big( \Let{x = (\Let{y = e'' \IN a}) \IN E \HoleUnbound{x}} \big) \qquad
        e' = \big( \Let{y = e'' \IN \Let{x = a \IN E \HoleUnbound{x}}} \big)
        \: .
      \]
      The typing derivation for $ e $ (collapsing the use of \Rule{Subsum}) is
      \[
        \Infer
          {
            \Infer
            {
              \Typing{\Gamma |- e'': \textstyle\bigvee_{j \in J} t_j} \\
              \forall j \in J . \enspace
              \Typing{\Gamma, y\colon t_j |- a: t'}
            }
            {\Typing{\Gamma |- (\Let{y = e'' \IN a}):
              \textstyle\bigvee_{i \in I} t_i}}
            {t' \leq \textstyle\bigvee_{i \in I} t_i} \\
            \forall i \in I . \enspace
              \Typing{\Gamma, x\colon t_i |- E \HoleUnbound{x}: t}
          }
          {\Typing{\Gamma |-
            \big( \Let{x = (\Let{y = e'' \IN a}) \IN E \HoleUnbound{x}} \big): t}}
          {}
      \]
      We show $ \Typing{\Gamma |- e': t} $ as follows:
      \[
        \Infer
          {
            \Typing{\Gamma |- e'': \textstyle\bigvee_{j \in J} t_j} \\
            \forall j \in J . \enspace
              \Infer
                {
                  \Typing{\Gamma, y\colon t_j |- a:
                    \textstyle\bigvee_{i \in I} t_i} \\
                  \forall i \in I . \enspace
                  \Typing{\Gamma, y\colon t_j, x\colon t_i |-
                    E \HoleUnbound{x}: t}
                }
                { \Typing{\Gamma, y\colon t_j |-
                  (\Let{x = a \IN E \HoleUnbound{x}}): t}}
                {}
          }
          {\Typing{\Gamma |-
            \big( \Let{y = e'' \IN \Let{x = a \IN E \HoleUnbound{x}}} \big): t}}
          {}
      \]
      The premise for the typing of $ E \HoleUnbound{x} $
      is derived by weakening (Lemma~\ref{lem:typing-weakening}):
      we can assume $ y \notin \Dom(\Gamma) $  by \textalpha-renaming.

    \item[Case \Rule{Subsum}]
      We have $ \Typing{\Gamma |- e: t'} $ for some $ t' \leq t $.
      By the induction hypothesis, we derive $ \Typing{\Gamma |- e': t'} $,
      and we apply \Rule{Subsum} to conclude.
      \qedhere
  \end{description}
\end{Proof}

\begin{Corollary}[Type soundness]
  Let $ e $ be a well-typed, closed expression
  (that is, $ \Typing{\emptyset |- e: t} $ holds for some $ t $).
  If $ e \ReducesArrow^* e' $ and $ e' $ cannot reduce,
  then $ e' $ is an answer and $ \Typing{\emptyset |- e': t} $.
\end{Corollary}

\subsection{Results on the source language}

\begin{Proposition}
  If $ \Typing{\Gamma |- \SourceExpr: t} $,
  then $ \Typing{\Gamma |- \Compile{\SourceExpr}: t} $.
\end{Proposition}

\begin{Corollary}[Type soundness for the source language]
  Let $ \SourceExpr $ be a well-typed, closed source language expression
  (that is, $ \Typing{\emptyset |- \SourceExpr: t} $ holds for some $ t $).
  If $ \Compile{\SourceExpr} \ReducesArrow^* e' $ and $ e' $ cannot reduce,
  then $ e' $ is an answer and $ \Typing{\emptyset |- e': t} $.
\end{Corollary}

\subsection{Interpreting types as sets of values}

\begin{Lemma}
  \label{lem:model-arrow-decomposition}
  Let $ \InterTot{} : \Types \to \Pset(\DomainAlt) $ be a model.
  Let $ P $ and $ N $ be finite sets of types of the form $ t_1 \to t_2 $,
  with $ P \neq \emptyset $.
  Then:
  \begin{multline*}
    \exists t_1' \to t_2' \in N . \enspace
    \textstyle
    \InterTot{ t_1' \setminus \bigvee_{t_1 \to t_2\in P} t_1 } = \emptyset
    \mathrel{\mathsf{and}}
    \\
    \textstyle
    \big(
      \forall P' \subsetneq P . \enspace
        \InterTot{ t_1' \setminus \bigvee_{t_1 \to t_2\in P'} t_1 } = \emptyset
        \mathrel{\mathsf{or}}
        \InterTot{ \bigwedge_{t_1 \to t_2\in P \setminus P'} t_2 \setminus t_2' }
          = \emptyset
    \big)
    \\
    \implies
    \textstyle
    \bigcap_{t_1 \to t_2 \in P} \InterTot{ t_1 \to t_2 } \subseteq
    \bigcup_{t_1 \to t_2 \in N} \InterTot{ t_1 \to t_2 }
  \end{multline*}
\end{Lemma}

\begin{Proof}
  We define
  \begin{align*}
    \mathsf{Tot}(X) & \eqdef
      \Setc{R \in \Pset(\DomainAlt \times \DomainAlt) \given
        \Dom(R) \supseteq X
      }
    \\
    X \rightharpoonup Y & \eqdef
      \Setc{R \in \Pset(\DomainAlt \times \DomainAlt) \given
        \forall (d, d') \in R. \: d \in X \implies d' \in
      }
    \\
    X \boldto Y & \eqdef
      \Setc{R \in \Pset(\DomainAlt \times \DomainAlt) \given
        \Dom(R) \supseteq X
        \textup{ and }
        \forall (d, d') \in R. \: d \in X \implies d' \in Y
      }
  \end{align*}
  and therefore we have $
    X \boldto Y = \mathsf{Tot}(X) \cap (X \rightharpoonup Y)
  $.
  We also have $
    X \rightharpoonup Y =
    \Pset(
    \overline{X \times \overline{Y}^{\DomainAlt}}^{\DomainAlt \times \DomainAlt})
  $, using the notation $ \overline{A}^B $ for $ B \setminus A $.

  To show
  $
    \textstyle
    \bigcap_{t_1 \to t_2 \in P} \InterTot{ t_1 \to t_2 } \subseteq
    \bigcup_{t_1 \to t_2 \in N} \InterTot{ t_1 \to t_2 }
  $,
  by the definition of model, it suffices to show
  $
    \textstyle
    \bigcap_{t_1 \to t_2 \in P} \InterTot{t_1} \boldto \InterTot{t_2}
    \subseteq
    \bigcup_{t_1 \to t_2 \in N} \InterTot{t_1} \boldto \InterTot{t_2}
  $.
  We will actually show $
    \textstyle
    \bigcap_{t_1 \to t_2 \in P} \InterTot{t_1} \boldto \InterTot{t_2}
    \subseteq
    \InterTot{t_1'} \boldto \InterTot{t_2'}
  $, which is enough to conclude.

  To show it, we first show $
    \textstyle
    \bigcap_{t_1 \to t_2 \in P} \InterTot{t_1} \rightharpoonup \InterTot{t_2}
    \subseteq
    \InterTot{t_1'} \rightharpoonup \InterTot{t_2'}
  $, without the requirement of totality.

  The premise
  \[
    \textstyle
    \InterTot{ t_1' \setminus \bigvee_{t_1 \to t_2\in P} t_1 } = \emptyset
    \mathrel{\mathsf{and}}
    \textstyle
    \big(
      \forall P' \subsetneq P . \enspace
        \InterTot{ t_1' \setminus \bigvee_{t_1 \to t_2\in P'} t_1 } = \emptyset
        \mathrel{\mathsf{or}}
        \InterTot{ \bigwedge_{t_1 \to t_2\in P \setminus P'} t_2 \setminus t_2' }
          = \emptyset
    \big)
  \]
  can be rewritten as
  \[
    \textstyle
    \InterTot{ t_1' } \subseteq \InterTot{ \bigvee_{t_1 \to t_2\in P} t_1 }
    \mathrel{\mathsf{and}}
    \textstyle
    \big(
      \forall P' \subsetneq P . \enspace
        \InterTot{ t_1' } \subseteq \InterTot{ \bigvee_{t_1 \to t_2\in P'} t_1 }
        \mathrel{\mathsf{or}}
        \InterTot{ \bigwedge_{t_1 \to t_2\in P \setminus P'} t_2 } \subseteq
        \InterTot{ t_2' }
    \big)
  \]
  and implies
  \[
    \textstyle
    \InterTot{ t_1' } \subseteq \bigcup_{t_1 \to t_2\in P} \InterTot{ t_1 }
    \mathrel{\mathsf{and}}
    \textstyle
    \big(
      \forall P' \subsetneq P . \enspace
        \InterTot{ t_1' } \subseteq \bigcup_{t_1 \to t_2\in P'} \InterTot{ t_1 }
        \mathrel{\mathsf{or}}
        \overline{\InterTot{ t_2' }}^{\DomainAlt} \subseteq
        \bigcup_{t_1 \to t_2\in P \setminus P'} \overline{
          \InterTot{ t_2 }
        }^{\DomainAlt}
    \big)
  \]
  and therefore implies
  \[
    \textstyle
    \forall P' \subseteq P . \enspace
      \InterTot{ t_1' } \subseteq \bigcup_{t_1 \to t_2\in P'} \InterTot{ t_1 }
      \mathrel{\mathsf{or}}
      \overline{\InterTot{ t_2' }}^{\DomainAlt} \subseteq
      \bigcup_{t_1 \to t_2\in P \setminus P'} \overline{
        \InterTot{ t_2 }
      }^{\DomainAlt}
  \]
  as well.
  We can apply Lemma~6.4 of \citet{Frisch2008} to obtain
  \[
    \textstyle
    \InterTot{t_1'} \times \overline{\InterTot{t_2'}}^{\DomainAlt}
    \subseteq
    \bigcup_{t_1 \to t_2 \in P}
      \InterTot{t_1} \times \overline{\InterTot{t_2}}^{\DomainAlt}
  \]
  whence
  \[
    \textstyle
    \overline{
      \bigcup_{t_1 \to t_2 \in P}
      \InterTot{t_1} \times \overline{\InterTot{t_2}}^{\DomainAlt}
    }^{\DomainAlt \times \DomainAlt}
    \subseteq
    \overline{
      \InterTot{t_1'} \times \overline{\InterTot{t_2'}}^{\DomainAlt}
    }^{\DomainAlt \times \DomainAlt}
  \]
  and
  \[
    \textstyle
    \bigcap_{t_1 \to t_2 \in P}
    \overline{
      \InterTot{t_1} \times \overline{\InterTot{t_2}}^{\DomainAlt}
    }^{\DomainAlt \times \DomainAlt}
    \subseteq
    \overline{
      \InterTot{t_1'} \times \overline{\InterTot{t_2'}}^{\DomainAlt}
    }^{\DomainAlt \times \DomainAlt}
  \]
  and finally
  \[
    \textstyle
    \bigcap_{t_1 \to t_2 \in P}
    \Pset \big( \overline{
      \InterTot{t_1} \times \overline{\InterTot{t_2}}^{\DomainAlt}
    }^{\DomainAlt \times \DomainAlt} \big)
    \subseteq
    \Pset \big( \overline{
      \InterTot{t_1'} \times \overline{\InterTot{t_2'}}^{\DomainAlt}
    }^{\DomainAlt \times \DomainAlt} \big)
  \]
  where note that the powerset construction obtained is equivalent
  to the definition of $ \rightharpoonup $.

  Now we have
  \[
    \textstyle
    \bigcap_{t_1 \to t_2 \in P} \InterTot{t_1} \rightharpoonup \InterTot{t_2}
    \subseteq
    \InterTot{t_1'} \rightharpoonup \InterTot{t_2'}
  \]
  and we want
  \[
    \textstyle
    \bigcap_{t_1 \to t_2 \in P} \InterTot{t_1} \boldto \InterTot{t_2}
    \subseteq
    \InterTot{t_1'} \boldto \InterTot{t_2'}
    \: ,
  \]
  that is,
  \[
    \textstyle
    \bigcap_{t_1 \to t_2 \in P}
      \Big( \mathsf{Tot}(\InterTot{t_1})
      \cap (\InterTot{t_1} \rightharpoonup \InterTot{t_2}) \Big)
    \subseteq
    \mathsf{Tot}(\InterTot{t_1'})
    \cap (\InterTot{t_1'} \rightharpoonup \InterTot{t_2'})
    \: .
  \]
  The latter is further equivalent to
  \[
    \textstyle
    \mathsf{Tot}(\InterTot{\bigvee_{t_1 \to t_2\in P} t_1})
    \cap
    \bigcap_{t_1 \to t_2 \in P}
      (\InterTot{t_1} \rightharpoonup \InterTot{t_2})
    \subseteq
    \mathsf{Tot}(\InterTot{t_1'})
    \cap (\InterTot{t_1'} \rightharpoonup \InterTot{t_2'})
    \: .
  \]
  Note that $
    \InterTot{ t_1' \setminus \bigvee_{t_1 \to t_2\in P} t_1 } = \emptyset
  $ implies $
    \InterTot{t_1'} \subseteq \InterTot{\bigvee_{t_1 \to t_2\in P} t_1}
  $.
  Therefore, we have $
    \mathsf{Tot}(\InterTot{t_1'}) \supseteq
    \mathsf{Tot}(\InterTot{\bigvee_{t_1 \to t_2\in P} t_1})
  $.
  This allows us to conclude that the containment above holds.
\end{Proof}

\begin{Proposition}
  Let $ \InterAlt{} : \Types \to \Pset(\DomainAlt) $ be a model.
  Let $ t_1 $ and $ t_2 $ be two finite (that is, non-recursive) types.
  If $ \Inter{t_1} \subseteq \Inter{t_2} $,
  then $ \InterAlt{t_1} \subseteq \InterAlt{t_2} $.
\end{Proposition}

\begin{Proof}
  First, note that $
    \Inter{t_1} \subseteq \Inter{t_2} \iff \Inter{t_1 \setminus t_2} = \emptyset
  $ and that $
    \InterTot{t_1} \subseteq \InterTot{t_2} \iff
    \InterTot{t_1 \setminus t_2} = \emptyset
  $.
  We therefore show this equivalent proposition:
  for all finite $ t $,
  if $ \Inter{t} = \emptyset $, then $ \InterTot{t} = \emptyset $.

  We define the function $ h(\cdot) $ on finite types by structural induction
  as follows:
  \begin{gather*}
    h(\bot) = h(b) = h(\Empty) = 0 \qquad
    h(t_1 \times t_2) = h(t_1 \to t_2) = \max(h(t_1), h(t_2)) + 1 \\
    h(t_1 \lor t_2) = \max(h(t_1), h(t_2)) \qquad
    h(\lnot t) = h(t)
  \end{gather*}
  That is, $ h(t) $ is the maximum number of $ \times $ and $ \to $
  constructors found on paths from the root of $ t $ to the leaves.
  We use $ h(\cdot) $ as the measure for induction.

  Now, let us consider an arbitrary finite type $ t $
  such that $ \Inter{t} = \emptyset $.
  We want to show $ \InterTot{t} = \emptyset $.

  In the following,
  we refer to a type of the form $ \bot $, $ b $, $ t_1 \times t_2 $,
  or $ t_1 \to t_2 $ as an \emph{atom}.
  We refer to a finite set of pairs of finite sets of atoms
  (that is, to a set $ \Setc{ (P_i, N_i) \given i \in I } $
  where $ I $ is finite and
  where each $ P_i $ and $ N_i $ is a finite set of atoms)
  as a \emph{disjunctive normal form}.

  We define the $ \mathsf{dnf}(\cdot) $ and $ \mathsf{dnf}^-(\cdot) $
  functions as follows
  (by mutual induction):
  \begin{align*}
    \mathsf{dnf}(t) & = \Set{(\Set{t}, \emptyset)}
      & \text{if $ t $ is an atom} \\
    \mathsf{dnf}(t_1 \lor t_2) & = \mathsf{dnf}(t_1) \cup \mathsf{dnf}(t_2) \\
    \mathsf{dnf}(\lnot t) & = \mathsf{dnf}^-(t) \\
    \mathsf{dnf}(\Empty) & = \emptyset \\
    \mathsf{dnf}^-(t) & = \Set{(\emptyset, \Set{t})}
      & \text{if $ t $ is an atom} \\
    \mathsf{dnf}^-(t_1 \lor t_2) & =
      \Setc{ (P_1 \cup P_2, N_1 \cup N_2) \given
        (P_1, N_1) \in \mathsf{dnf}^-(t_1),
        (P_2, N_2) \in \mathsf{dnf}^-(t_2) }
    \\
    \mathsf{dnf}^-(\lnot t) & = \mathsf{dnf}(t) \\
    \mathsf{dnf}^-(\Empty) & = \Set{(\emptyset, \emptyset)}
  \end{align*}

  We extend $ \Inter{} $ to disjunctive normal forms by defining
  \[
    \textstyle
    \Inter{ \Setc{ (P_i, N_i) \given i \in I } } =
    \bigcup_{i \in I} \big(
      \bigcap_{t \in P_i} \Inter{t}
      \cap
      \bigcap_{t \in N_i} ( \Domain \setminus \Inter{t} )
    \big)
  \]
  where by convention $ \bigcap_{t \in \emptyset} \Inter{t} = \Domain $.
  We can check by induction that, for all types $ t $, $
    \Inter{t} = \Inter{\mathsf{dnf}(t)} =
    \Domain \setminus \Inter{\mathsf{dnf}^-(t)}
  $.

  We extend $ \InterTot{} $ to disjunctive normal forms likewise;
  it holds that $
    \InterTot{t} = \InterTot{\mathsf{dnf}(t)} =
    \DomainAlt \setminus \InterTot{\mathsf{dnf}^-(t)}
  $.

  Let $ \Setc{ (P_i, N_i) \given i \in I } = \mathsf{dnf}(t) $.
  Since $ \Inter{t} = \emptyset $, we have
  \[
    \textstyle
    \forall i \in I . \:
      \bigcap_{t \in P_i} \Inter{t}
      \cap
      \bigcap_{t \in N_i} ( \Domain \setminus \Inter{t} )
      = \emptyset
    \: .
  \]
  We want to show
  \[
    \textstyle
    \forall i \in I . \:
      \bigcap_{t \in P_i} \InterTot{t}
      \cap
      \bigcap_{t \in N_i} ( \DomainAlt \setminus \InterTot{t} )
      = \emptyset
    \: .
  \]
  which would conclude our proof.

  Let us partition atoms into four kinds, according to their form:
  $ \bot $, $ b $, $ t_1 \times t_2 $, or $ t_1 \to t_2 $.
  Note that if $ t_1 $ and $ t_2 $ are two atoms of different kind,
  then $ \InterTot{t_1} \cap \InterTot{t_2} = \emptyset $
  (the same holds for $ \Inter{} $).

  Now, let us consider an arbitrary $ i \in I $.
  Either $ P_i $ is empty or it contains at least one atom.
  First we show that if $ P_i $ contains atoms of at least two different kinds,
  then $
    \bigcap_{t \in P_i} \InterTot{t}
    \cap
    \bigcap_{t \in N_i} ( \DomainAlt \setminus \InterTot{t} )
  $ is empty.
  This holds
  because the intersection is a subset of $ \InterTot{t_1} \cap \InterTot{t_2} $,
  where $ t_1 $ and $ t_2 $ are two atoms of different kind in $ P_i $,
  and we have remarked that atoms of different kinds have disjoint interpretations.

  There remain two cases to consider:
  $ P_i = \emptyset $ or $ P_i $ non-empty and composed of atoms of a single kind.
  We consider the case $ P_i = \emptyset $ first.
  Note that in that case
  \begin{align*}
    \textstyle
    \bigcap_{t \in P_i} \Inter{t}
    \cap
    \bigcap_{t \in N_i} ( \Domain \setminus \Inter{t} )
    & =
    \textstyle
    \Domain
    \cap
    \bigcap_{t \in N_i} ( \Domain \setminus \Inter{t} )
    \\
    & =
    \textstyle
    (\Set{\bot} \cup \Constants \cup \Inter{\Any \times \Any}
      \cup \Inter{\Empty \to \Any})
    \cap
    \bigcap_{t \in N_i} ( \Domain \setminus \Inter{t} )
  \end{align*}
  because the domain $ \Domain $ can be decomposed as a union
  of four sets corresponding to the four kinds of atoms.
  Since the intersection is empty, we have
  \begin{align*}
    \textstyle
    \Set{\bot} \cap \bigcap_{t \in N_i} ( \Domain \setminus \Inter{t} )
    & = \emptyset
    &
    \textstyle
    \Constants \cap \bigcap_{t \in N_i} ( \Domain \setminus \Inter{t} )
    & = \emptyset
    \\
    \textstyle
    \Inter{\Any \times \Any}
      \cap \bigcap_{t \in N_i} ( \Domain \setminus \Inter{t} )
    & = \emptyset
    &
    \textstyle
    \Inter{\Empty \to \Any}
      \cap \bigcap_{t \in N_i} ( \Domain \setminus \Inter{t} )
    & = \emptyset
  \end{align*}

  Setting aside the intersection with $ \Constants $ for a moment,
  observe that the others are equivalent to
  \begin{align*}
    \textstyle
    \bigcap_{t \in \Set{\bot}} \Inter{t}
    \cap \bigcap_{t \in N_i} ( \Domain \setminus \Inter{t} )
    & = \emptyset
    \\
    \textstyle
    \bigcap_{t \in \Set{\Any \times \Any}} \Inter{t}
      \cap \bigcap_{t \in N_i} ( \Domain \setminus \Inter{t} )
    & = \emptyset
    \\
    \textstyle
    \bigcap_{t \in \Set{\Empty \to \Any}} \Inter{t}
      \cap \bigcap_{t \in N_i} ( \Domain \setminus \Inter{t} )
    & = \emptyset
  \end{align*}
  so they can be treated together with the case of non-empty $ P_i $.

  As for the intersection with $ \Constants $,
  if $
    \Constants \cap \bigcap_{t \in N_i} ( \Domain \setminus \Inter{t} )
    = \emptyset
  $, then $
    \Constants \subseteq \bigcup_{b \in N_i} \ConstantsInBasicType(b)
  $ (we can ignore atoms of different kind in $ N_i $).
  But then $
    \Constants \subseteq \bigcup_{b \in N_i} \InterTot{b}
  $, and therefore $
    \Constants \subseteq \bigcup_{t \in N_i} \InterTot{t}
  $, which shows that $
    \Constants \cap \bigcap_{t \in N_i} ( \DomainAlt \setminus \InterTot{t} )
    = \emptyset
  $.

  Now, assuming
  \begin{align*}
    \textstyle
    \Set{\bot} \cap \bigcap_{t \in N_i} ( \DomainAlt \setminus \InterAlt{t} )
    & = \emptyset
    &
    \textstyle
    \Constants \cap \bigcap_{t \in N_i} ( \DomainAlt \setminus \InterAlt{t} )
    & = \emptyset
    \\
    \textstyle
    \InterAlt{\Any \times \Any}
      \cap \bigcap_{t \in N_i} ( \DomainAlt \setminus \InterAlt{t} )
    & = \emptyset
    &
    \textstyle
    \InterAlt{\Empty \to \Any}
      \cap \bigcap_{t \in N_i} ( \DomainAlt \setminus \InterAlt{t} )
    & = \emptyset
  \end{align*}
  we have
  \[
    \textstyle
    (\Set{\bot} \cup \Constants \cup \InterAlt{\Any \times \Any}
    \cup \InterAlt{\Empty \to \Any})
    \cap \bigcap_{t \in N_i} ( \DomainAlt \setminus \InterAlt{t} )
    = \emptyset
  \]
  which is
  \[
    \textstyle
    \DomainAlt
    \cap \bigcap_{t \in N_i} ( \DomainAlt \setminus \InterAlt{t} )
    = \emptyset
    \: .
  \]

  We now consider the remaining case.
  That is, we assume
  \[
    \textstyle
    \bigcap_{t \in P_i} \Inter{t}
    \cap
    \bigcap_{t \in N_i} ( \Domain \setminus \Inter{t} )
    = \emptyset
  \]
  with $ P_i $ non-empty and formed of atoms of a single kind,
  and we show
  \[
    \textstyle
    \bigcap_{t \in P_i} \InterTot{t}
    \cap
    \bigcap_{t \in N_i} ( \DomainAlt \setminus \InterTot{t} )
    = \emptyset
    \: .
  \]
  Equivalently, we assume
  $
    \textstyle
    \bigcap_{t \in P_i} \Inter{t}
    \subseteq
    \bigcup_{t \in N_i} \Inter{t}
  $
  and we show
  $
    \textstyle
    \bigcap_{t \in P_i} \InterTot{t}
    \subseteq
    \bigcup_{t \in N_i} \InterTot{t}
  $.
  In doing so, we can disregard the atoms in $ N_i $
  that are not of the same kind as those in $ P_i $.
  Therefore, we consider that $ N_i $ only contains atoms of that same kind.

  If the atoms of $ P_i $ are of the kind of $ \bot $
  (that is, if $ P_i = \Set{\bot} $)
  or if they are basic types,
  the result is immediate because the two interpretations are defined
  identically on these kinds of atoms.

  If the atoms of $ P_i $ are all products,
  then we have (by Lemmas~6.4 and 6.5 of \citet{Frisch2008}):
  \begin{multline*}
    \textstyle
    \bigcap_{t \in P_i} \Inter{t}
    \subseteq
    \bigcup_{t \in N_i} \Inter{t}
    \iff \\
    \textstyle
    \forall N \subseteq N_i . \enspace
      \Inter{
        \bigwedge_{t_1 \times t_2 \in P_i} t_1
        \land \bigwedge_{t_1 \times t_2 \in N} \lnot t_1
      } = \emptyset
      \mathrel{\mathsf{or}}
      \Inter{
        \bigwedge_{t_1 \times t_2 \in P_i} t_2
        \land \bigwedge_{t_1 \times t_2 \in N_i \setminus N} \lnot t_2
      } = \emptyset
  \end{multline*}
  (with the convention $
    \bigwedge_{t_1 \times t_2 \in \emptyset} \lnot t_i = \Domain
  $).
  Since $ \InterTot{} $ also satisfies $
    \InterTot{t_1 \times t_2} = \InterTot{t_1} \times \InterTot{t_2}
  $, we also have
  \begin{multline*}
    \textstyle
    \bigcap_{t \in P_i} \InterTot{t}
    \subseteq
    \bigcup_{t \in N_i} \InterTot{t}
    \iff \\
    \textstyle
    \forall N \subseteq N_i . \enspace
      \InterTot{
        \bigwedge_{t_1 \times t_2 \in P_i} t_1
        \land \bigwedge_{t_1 \times t_2 \in N} \lnot t_1
      } = \emptyset
      \mathrel{\mathsf{or}}
      \InterTot{
        \bigwedge_{t_1 \times t_2 \in P_i} t_2
        \land \bigwedge_{t_1 \times t_2 \in N_i \setminus N} \lnot t_2
      } = \emptyset
  \end{multline*}
  (with the convention $
    \bigwedge_{t_1 \times t_2 \in \emptyset} \lnot t_i = \DomainTot
  $).
  This allows us to conclude $
    \bigcap_{t \in P_i} \InterTot{t}
    \subseteq
    \bigcup_{t \in N_i} \InterTot{t}
  $, because we can apply the induction hypothesis to all the types $
    \bigwedge_{t_1 \times t_2 \in P_i} t_1
    \land \bigwedge_{t_1 \times t_2 \in N} \lnot t_1
  $ and $
    \bigwedge_{t_1 \times t_2 \in P_i} t_2
    \land \bigwedge_{t_1 \times t_2 \in N_i \setminus N} \lnot t_2
  $.
  Indeed, note that $ h(\cdot) $ on these types is always strictly less
  than $
    \max \Setc{ h(t_1 \times t_2) \given t_1 \times t_2 \in P_i \cup N_i }
  $, because the $ \times $ constructor has been eliminated.
  Also, $
    h(t) \geq
    \max \Setc{ h(t_1 \times t_2) \given t_1 \times t_2 \in P_i \cup N_i }
  $ because any atom $ t_1 \times t_2 $ appeared under $ t $.

  The last case to examine is that of $ P_i $ composed only of arrow types.
  In that case, by Lemma~\ref{lem:subtyping-decomposition-arrows}, we have
  \begin{multline*}
    \textstyle
    \bigcap_{t_1 \to t_2 \in P_i} \Inter{ t_1 \to t_2 } \subseteq
    \bigcup_{t_1 \to t_2 \in N_i} \Inter{ t_1 \to t_2 }
    \iff \\
    \exists t_1' \to t_2' \in N_i . \enspace
    \textstyle
    \Inter{ t_1' \setminus \bigvee_{t_1 \to t_2\in P_i} t_1 } = \emptyset
    \mathrel{\mathsf{and}}
    \\
    \textstyle
    \big(
      \forall P \subsetneq P_i . \enspace
        \Inter{ t_1' \setminus \bigvee_{t_1 \to t_2\in P} t_1 } = \emptyset
        \mathrel{\mathsf{or}}
        \Inter{ \bigwedge_{t_1 \to t_2\in P_i \setminus P} t_2 \setminus t_2' }
          = \emptyset
    \big)
  \end{multline*}
  and, by Lemma~\ref{lem:model-arrow-decomposition},
  \begin{multline*}
    \exists t_1' \to t_2' \in N_i . \enspace
    \textstyle
    \InterTot{ t_1' \setminus \bigvee_{t_1 \to t_2\in P_i} t_1 } = \emptyset
    \mathrel{\mathsf{and}}
    \\
    \textstyle
    \big(
      \forall P \subsetneq P_i . \enspace
        \InterTot{ t_1' \setminus \bigvee_{t_1 \to t_2\in P} t_1 } = \emptyset
        \mathrel{\mathsf{or}}
        \InterTot{ \bigwedge_{t_1 \to t_2\in P_i \setminus P} t_2 \setminus t_2' }
          = \emptyset
    \big)
    \\
    \implies
    \textstyle
    \bigcap_{t_1 \to t_2 \in P_i} \InterTot{ t_1 \to t_2 } \subseteq
    \bigcup_{t_1 \to t_2 \in N_i} \InterTot{ t_1 \to t_2 }
  \end{multline*}
  and we can therefore conclude by applying the induction hypothesis
  (with the same argument as before to show that $ h(\cdot) $ decreases).
\end{Proof}

\fi

\end{document}